\documentclass{aa}

\usepackage{graphicx}
\usepackage{txfonts}
\usepackage{placeins}
\usepackage{bm}                                

\usepackage[colorlinks, citecolor=blue, urlcolor=blue, linkcolor=blue]{hyperref}

\begin{document}

\title{Delayed phase mixing in the self-gravitating Galactic disc}

\author{T. Asano\inst{1,2,3}
        \and T. Antoja\inst{1,2,3}
       }

\institute{Departament de F\'isica Qu\`antica i Astrof\'isica (FQA), Universitat de Barcelona (UB), c. Mart\'i i Franqu\`es, 1, 08028 Barcelona, Spain\\
  \email{asano@fqa.ub.edu}
  \and
  Institut de Ci\`encies del Cosmos (ICCUB), Universitat de Barcelona (UB), c. Mart\'i i Franqu\`es, 1, 08028 Barcelona, Spain
  \and
  Institut d'Estudis Espacials de Catalunya (IEEC), c. Gran Capit\`a, 2-4, 08034 Barcelona, Spain
}
\date{Received 13 October 2025 / Accepted 13 February 2026}

  \abstract
	{The discovery of the phase spiral in the vertical position--velocity ($z$--$v_z$) distribution of the solar neighbourhood stars revealed the Milky Way (MW) disc disequilibrium.
	The phase spiral is considered to work as a dynamical clock for dating past perturbations, but some previous studies  neglected the disc self-gravity, which might bias estimates of the phase spiral excitation time.}
	{We revisit self-gravitating effects on the evolution of vertical phase spirals and quantify the bias introduced in estimating their excitation time when these effects are ignored.}
	{We analysed a high-resolution self-consistent $N$-body simulation of the MW interaction with the Sagittarius dwarf galaxy (Sgr), alongside four test-particle simulations in potentials constructed from the $N$-body snapshots. In the test-particle simulations, we used static and time-dependent potentials that included (or excluded) Sgr and the dark matter (DM) wake.
	In each case, we estimated the winding time of phase spirals by measuring the slope of the density contrast in the vertical angle--frequency ($\theta_z$--$\Omega_z$) space.}
	{In test-particle models, the phase spiral immediately begins to wind after the Sgr pericentric passage, and the winding time closely tracks the true elapsed time since the Sgr pericentric passage.
		Adding the DM wake yields only a modest ($<100$~Myr) reduction of the winding time relative to Sgr alone.
	By contrast, the self-consistent $N$-body simulation exhibits an initial coherent vertical oscillation lasting $\gtrsim 300$~Myr before a clear spiral forms, leading to a systematic  underestimation of the excitation times. An analytical shearing-box model with self-gravity, developed in a previous study, qualitatively reproduces this delay. This supports the hypothesis that it originates in the response of the self-gravitating disc.}
	{Assuming that self-gravity affects phase mixing in the MW to the same degree as the $N$-body model, we estimated the lag induced by self-gravity to be $\sim 0.3$~Gyr in the solar neighbourhood.
		Accounting for this delay revises the inferred age of the MW observed phase spiral to $\sim0.6$--$1.2$~Gyr, which agrees better with the Sgr pericentric passage.
	An accurate dynamical dating of past perturbations thus requires models that include the self-gravitating response of the Galactic disc.}
\keywords{Galaxy: kinematics and dynamics -- Galaxy: disk -- Methods: numerical}

\maketitle

\section{Introduction}\label{sec:intro}
The phase spiral is a spiral structure that is observed in the vertical position ($z$) versus vertical velocity ($v_z$) space \citep{2018Natur.561..360A}. It is one of the most significant discoveries of the \textit{Gaia} mission \citep{2016A&A...595A...1G}.
It provides direct evidence of the incomplete vertical phase mixing in the Milky Way (MW) disc and indicates that the MW disc is in a disequilibrium state due to strong perturbation events in the past.
Although the properties of the phase spiral have been studied both theoretically and observationally since its initial discovery in Gaia Data Release 2 \citep[DR2;][]{2018A&A...616A...1G}, its origin remains under debate (see \citealt{2025NewAR.10001721H} for a review).

Of the various proposed scenarios, those involving external perturbations such as satellite galaxies, dark matter (DM) wakes, and DM sub-haloes have been studied most extensively.
Of all potential external perturbers, the Sagittarius dwarf galaxy \citep[Sgr;][]{1994Natur.370..194I} is a strong candidate for the origin of the phase spiral, as it is the most massive satellite after the Magellanic Clouds and has a small pericentre distance of $\sim20$~kpc, but may have had several past pericentres \citep[e.g.][]{2010ApJ...714..229L, 2011Natur.477..301P, 2017ApJ...836...92D,2021MNRAS.501.2279V}.
The formation of phase spirals through Sgr-like interactions has been demonstrated by
analytical calculations \citep{2018MNRAS.481.1501B, 2021MNRAS.503..376B, 2022ApJ...935..135B, 2023ApJ...952...65B},
test-particle simulations \citep[e.g.][]{2021ApJ...911..107L, 2022ApJ...928...80G},
$N$-body simulations \citep{2019MNRAS.485.3134L, 2019MNRAS.486.1167B, 2021MNRAS.504.3168B, 2021MNRAS.508.1459H, 2022ApJ...927..131B, 2025A&A...700A.109A}, and $N$-body+hydrodynamical simulations \citep{2025MNRAS.542.1987T}.
Additionally, phase spirals have been detected in cosmological simulations \citep{2022MNRAS.510..154G, 2023MNRAS.524..801G}, although the masses and orbits of the satellites in these simulations do not necessarily correspond to those of Sgr due to the nature of cosmological modelling. 
As alternative explanations, small stochastic perturbations \citep{2023MNRAS.521..114T, 2025ApJ...980...24G, 2025MNRAS.542.1987T}, bar buckling \citep{2019A&A...622L...6K}, DM wake \citep{2023MNRAS.524..801G}, and Galactic echoes\footnote{A Galactic echo is an analogue of a plasma echo \citep{1967PhRvL..19..219G}. These echoes are generated in collisionless systems that experience two successive perturbations due to the non-linear coupling between the first and second perturbations. For further details, see \citet{2025MNRAS.543..190C}.} \citep{2025MNRAS.543..190C} have also been proposed.
Additionally, \citet{2024A&A...683A..47G} demonstrated that the dark sub-haloes, disc-halo misalignment, and anisotropic gas accretion can trigger disc bending and phase spirals by analysing cosmological simulations.

To investigate the origin of the phase spiral, it is crucial to accurately estimate the excitation time of the bending mode associated with the one-arm phase spiral.
For example, when the estimated time coincides with the timing of the pericentric passage of Sgr within the uncertainties, this supports the Sgr scenario.
When the times do not agree, the results might indicate that the phase spiral was instead excited by other perturbations.
Assuming that the perturbation is impulsive and the gravitational potential remains static, we can estimate the formation time of the phase spiral by unwinding it, or in other words, by reconstructing its evolution backwards in time.
Equivalently, this can be achieved by measuring the slopes of ridges in the vertical angle ($\theta_z$) versus vertical frequency ($\Omega_z$) space, which correspond to the phase spirals in the $z$--$v_z$ space \citep{2021MNRAS.503.1586L, 2023ApJ...955...74D, 2023MNRAS.521.5917F, 2025ApJ...988..254L}.
With these methods, the formation of the one-arm phase spiral has been estimated to have started $\sim 300\text{--}900$~Myr ago \citep{2023A&A...673A.115A, 2023ApJ...955...74D, 2023MNRAS.521.5917F, 2025ApJ...987...81F, 2025arXiv250719579W}. 

However, these estimates can be significantly biased when the disc self-gravity is neglected.
\citet{2019MNRAS.484.1050D} qualitatively compared the time evolution of phase spirals in a self-consistent $N$-body simulation with that in a test-particle simulation using a static potential similar to that of the $N$-body model. They found that the phase spiral in the $N$-body model is less tightly wound than in the test-particle model.
\citet{2023MNRAS.522..477W} analytically investigated the effect of self-gravity on two-arm and one-arm phase spirals.
In the case of a breathing perturbation, which is associated with the two-arm phase spiral, the evolution is complex; not only does the initially excited phase spiral wind up, but new phase spirals are generated by swing amplification of the surface density perturbation.
Furthermore, the study showed that static (non-winding) two-arm phase spirals can exist around massive clouds.
In contrast, for a bending perturbation, which is associated with the one-arm phase spiral, the model does not predict such a complicated evolution.
Instead, it shows that the phase spiral winds up more slowly than in kinematic (non-self-gravitating) models, as observed in the simulations by \citet{2019MNRAS.484.1050D}.

In \citet{2025A&A...700A.109A}, we performed high-resolution $N$-body simulations of the MW-Sgr system and visually examined the time evolution and spatial variation of phase spirals.
We found that two-arm phase spirals appear intermittently, while one-arm phase spirals wind up in a more orderly manner (see Fig. 12 of \citealt{2025A&A...700A.109A}).
The relatively regular evolution of the one-arm phase spiral compared to the two-arm phase spiral in the analytical model \citep{2023MNRAS.522..477W} and $N$-body simulations suggests that it can be used as a dynamical clock to date past perturbations in the MW disc.
However, to use it reliably, the effects of self-gravity must be accounted for.
As a first step in this direction, this paper aims to quantitatively evaluate the bias introduced when self-gravitating effects are ignored. 
In addition, we provide an empirical correction of the unwinding method in order to incorporate self-gravity.
We analyse the $N$-body simulation \citep{2025A&A...700A.109A}, which employs 200 million particles in the disc (and 4.9 billion particles in the DM halo).
This high-resolution $N$-body model allows us to track the long-term ($\sim 1$~Gyr) self-gravitating evolution of phase spirals.

The structure of this paper is as follows.
Section 2 describes the $N$-body and test-particle simulations and qualitatively discusses the evolution of the phase spirals in both types of simulation.
Section 3 presents our unwinding method and applies it to the simulations. We show the winding time as a function of time and phase-space position (guiding radius and azimuthal angle), and compare it between the $N$-body and test-particle models.
Section 4 compares the results with an analytical shearing-box model including self-gravity, discusses the origin of the delayed winding, and considers the implications for the MW.
Section 5 provides our conclusions.

\section{Simulations}
\subsection{\texorpdfstring{$N$}{N}-body simulation}
We used the $N$-body model of the MW-Sgr interaction from \citet{2025A&A...700A.109A}.\footnote{The simulation data are available at \url{http://galaxies.astron.s.u-tokyo.ac.jp/data/mwsgr}.}
The MW-like host galaxy consists of a classical bulge, a stellar disc, and a DM halo, which follow a Hernquist profile \citep{1990ApJ...356..359H}, a radially exponential and vertically sech$^2$ profile, and a Navarro-Frenk-White (NFW) profile \citep{1997ApJ...490..493N}, respectively.
The initial condition was  generated using the \texttt{Galactics} code \citep{1995MNRAS.277.1341K, 2005ApJ...631..838W, 2008ApJ...679.1239W}.
The model parameters are identical to those of \citet{2019MNRAS.482.1983F}'s best-fit MW model, MWa.
The masses of the components are $M_\mathrm{bulge} = 5.4 \times 10^9 M_{\odot}$, $M_\mathrm{disc} = 3.6 \times 10^{10} M_{\odot}$, and $M_\mathrm{halo} = 8.7 \times 10^{11} M_{\odot}$.
Each component consists of equal-mass particles of $\sim170M_{\odot}$, with a total number of particles of 5.1 billion.
The initial condition of the Sgr-like satellite was modelled as a combination of a stellar component with a Hernquist profile and a DM component with an NFW profile. Their initial masses were $M_\mathrm{star} = 1\times10^9M_{\odot}$ and $M_\mathrm{DM} = 5\times10^{10}M_{\odot}$, respectively.

In \citet{2025A&A...700A.109A}, we first integrated the MW model in isolation for 8~Gyr to reach a quasi-steady state, then injected the satellite at $r\sim150$~kpc and simulated their interaction for $\sim3$~Gyr.
The satellite experienced three pericentric passages with pericentre distances of $\sim$ 20, 13, and 10~kpc. Its position about 30 Myr after the second pericentric passage was close to the present-day location of Sgr (see Fig. 2 of \citealt{2025A&A...700A.109A}).
The orbit is nearly polar, with an orbital plane roughly aligned with the $x$--$z$ plane, and the satellite crosses the disc shortly before the first pericentric passage.
This disc crossing occurs at $(R,\phi) \sim (25\,\mathrm{kpc},\, 0^\circ)$. We primarily focused on the time interval between the first and second pericentric passages.
We performed the simulation on Pegasus at the Center for Computational Sciences, University of Tsukuba, using the \texttt{Bonsai} code \citep{2012JCoPh.231.2825B, 2014hpcn.conf...54B}.
Further details of the simulation can be found in \citet{2025A&A...700A.109A}.

The fundamental properties of the simulated Galactic disc, such as the rotation curve and the velocity dispersion, are consistent with those of the MW.
Therefore, we expect the effect of self-gravity on the evolution of the phase spiral to be similar to that in the real Galaxy, although additional contributions from the gaseous components can further modify the evolution \citep{2022MNRAS.515.5951T, 2025MNRAS.542.1987T}.

We computed actions, angles, and orbital frequencies of the MW stellar particles using \texttt{Agama} \citep{2019MNRAS.482.1525V} with the St\"{a}ckel fudge method \citep{2012MNRAS.426.1324B}.
To do this, we constructed an axisymmetric potential model from the $N$-body snapshot at 300~Myr before the first pericentric passage of Sgr, using a potential expansion that is described in the next subsection.

\subsection{Test-particle simulations}
We constructed potential models from the previous $N$-body snapshots with \texttt{Agama} to run test-particle simulations.
For the disc and the bulge, we constructed static axisymmetric potentials from the $N$-body snapshot taken 300~Myr before the first pericentric passage of Sgr.
At this time, Sgr is sufficiently distant from the Galactic centre, and its effect on the disc is negligible.
We applied an azimuthal harmonic expansion to the disc and a multipole expansion to the bulge.
For the DM halo of MW, we constructed both static and time-dependent potentials.
The static potential was derived from the same snapshot as we used for the stellar component through an axisymmetric multipole expansion.
The time-dependent potential was obtained by interpolating the non-axisymmetric potentials from individual $N$-body snapshots, whose output cadence of $\sim 10$~Myr is sufficiently short to resolve the time evolution of the potential.
For Sgr, we first performed multipole expansions in the Sgr-centric frame and then transformed the potential into the MW-centric frame.
We also applied temporal interpolation to the Sgr potential.
For further details of the potential models, see Appendix~\ref{appendix:potential_models}.

We performed four test-particle simulations using different combinations of these potential models.
Table~\ref{tab:tp_models} summarises the model setup.
We used snapshots of the $N$-body model as the initial conditions of the test-particle simulations.
In the model with a purely static potential TP (static), we took the $N$-body snapshot at $t=0$~Gyr, which is the time of the first Sgr pericentric passage.
In the other models including the perturber potential, we took the snapshot at $t=-300$~Myr.
In the TP (static) model, particles were perturbed only once at $t=0$.
They were then evolved in the fully static axisymmetric potential,without any further external perturbations.
As a result, the subsequent evolution of the phase spiral in this model reflects purely kinematic phase mixing in a rigid potential.
By contrast, in the other test-particle models, particles experienced continuous perturbations arising from the time-dependent gravitational effect of Sgr, the DM wake, or both, but self-gravity was absent in all of them.

\begin{table*}
        \begin{center}
                \caption{Summary of the test-particle models.}\label{tab:tp_models}
                \begin{tabular}{l l l}
                        \hline
                        \hline
                        Model & Potential & Initial condition \\
                        \hline
                        TP (static) & static bulge + static disc + static halo & $t=0$ Myr \\
                        TP (Sgr) & static bulge + static disc + static halo + time-evolving Sgr & $t=-300$ Myr \\
                        TP (wake) & static bulge + static disc + time-evolving halo & $t=-300$ Myr \\
                        TP (Sgr+wake) & static bulge + static disc + time-evolving halo + time-evolving Sgr & $t=-300$ Myr \\
                        \hline
                \end{tabular}
                \tablefoot{First column: Model names. Second column: Gravitational potential configurations. Third column: Times of $N$-body snapshots adopted for the initial conditions of the test-particle simulations. \citet{2025A&A...702A.223B} also used the TP (static) model and referred to it as test-particle rerun.}
        \end{center}
\end{table*}

Before presenting the results of the test-particle simulations, we first examine the characteristics of Sgr and the MW halo DM wake by evaluating the force field they generate.
Since our interest lies in the internal motion of disc stars rather than their overall motion in an inertial frame, the perturbing force is given by
\begin{align}
\bm{F}_\mathrm{pert} (\bm{r}, t) = -\nabla \Phi_\mathrm{pert} (\bm{r}, t) + \nabla \Phi_\mathrm{pert} (\bm{0}, t),
\end{align}
where $\Phi_\mathrm{pert}$ is the potential of the perturbers (i.e. Sgr and the DM wake).
The first term corresponds to the force in an inertial frame, and the second term represents the acceleration of the coordinate system comoving with the MW disc.
\begin{figure}
        \begin{center}
                \includegraphics[width=0.9\hsize]{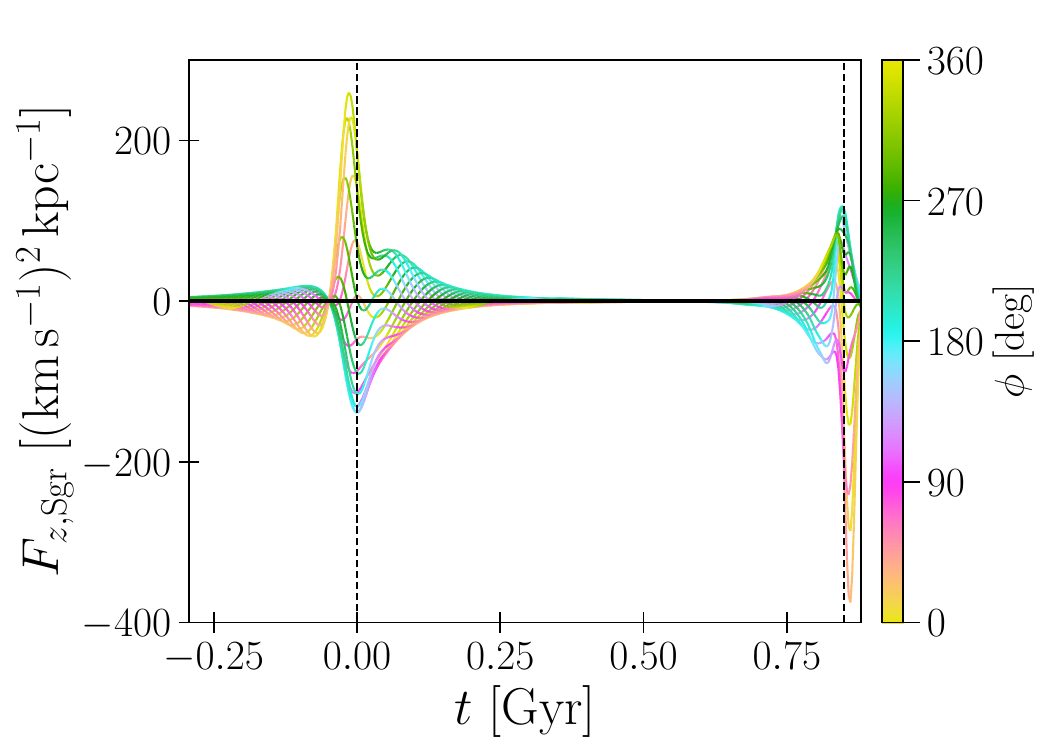}
                \includegraphics[width=0.9\hsize]{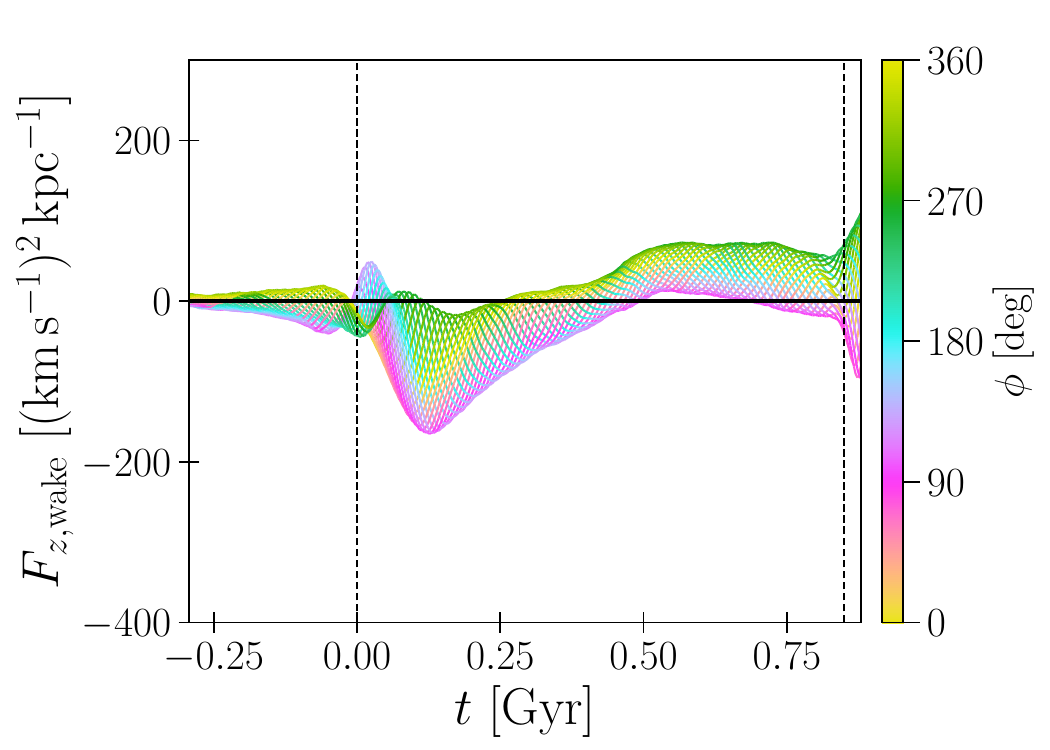}
                \caption{Vertical forces due to Sgr (\textit{upper panel}) and the DM wake (\textit{lower panel}). The forces were evaluated along circular orbits at $R=8$~kpc, and the colours of the lines indicate the azimuth $\phi=\Omega_{\mathrm{8\,kpc}} t + \phi_0$. The vertical dashed lines indicate the times of the Sgr pericentric passages.}\label{fig:Fz_8kpc}
        \end{center}
\end{figure}
Figure~\ref{fig:Fz_8kpc} shows the vertical component of the perturbing force as a function of time. The force was evaluated along circular orbits at $R=8$~kpc, and the colours of the curves indicate the azimuth $\phi=\Omega_{\mathrm{8\,kpc}} t + \phi_0$.
The upper panel shows the contribution of Sgr.
The curves show strong peaks around the times of the Sgr pericentric passages ($t=0$ and $t=0.8$~Gyr).
At the first pericentric passage, the stars at $\phi \sim 0^\circ$ and $\sim 180^\circ$ are kicked up and down (positive and negative vertical forces), respectively.
The lower panel shows the contribution of the DM wake. 
The peak of the DM wake is shifted to a later time than the Sgr pericentric passage by $\sim0.1$~Gyr.
Compared to the direct perturbation of Sgr, the wake induces a slightly weaker force at pericentre, but a more continuous force is present all the time.

We note that this behaviour contrasts with the results of the cosmological simulation by \citet{2023MNRAS.524..801G}, who found that the gravitational torque induced by the DM wake exceeds that from the satellite itself. This difference likely reflects variations in the adopted model setups. The strength of the DM wake is expected to depend on the satellite mass and orbit, as well as on the structure of the host halo, and the relative importance of the satellite and wake can also vary with time.
        For example, in the $N$-body simulation of the MW-Sgr system by \citet{2018MNRAS.481..286L}, the DM wake dominates the torque during the first pericentric passage, whereas the direct contribution from Sgr becomes comparable to or stronger than that of the wake in subsequent passages (see \citealt{2021MNRAS.507.2825G} for an analysis of the force fields in that simulation).
        In the simulations by \citet{2023MNRAS.524..801G} and \citet{2018MNRAS.481..286L} and also in ours, the peaks of the DM wake perturbation occur close to or with only a slight delay ($\lesssim 100$~Myr) relative to the peaks of the satellite perturbation.
Regardless of which component dominates the instantaneous perturbation amplitude, the onset of vertical phase mixing is therefore expected to occur close to the pericentric passage time. As discussed below, the subsequent evolution of the phase spiral is more strongly affected by the disc self-gravity.

\begin{figure*}
        \begin{center}
                \includegraphics[width=0.33\hsize]{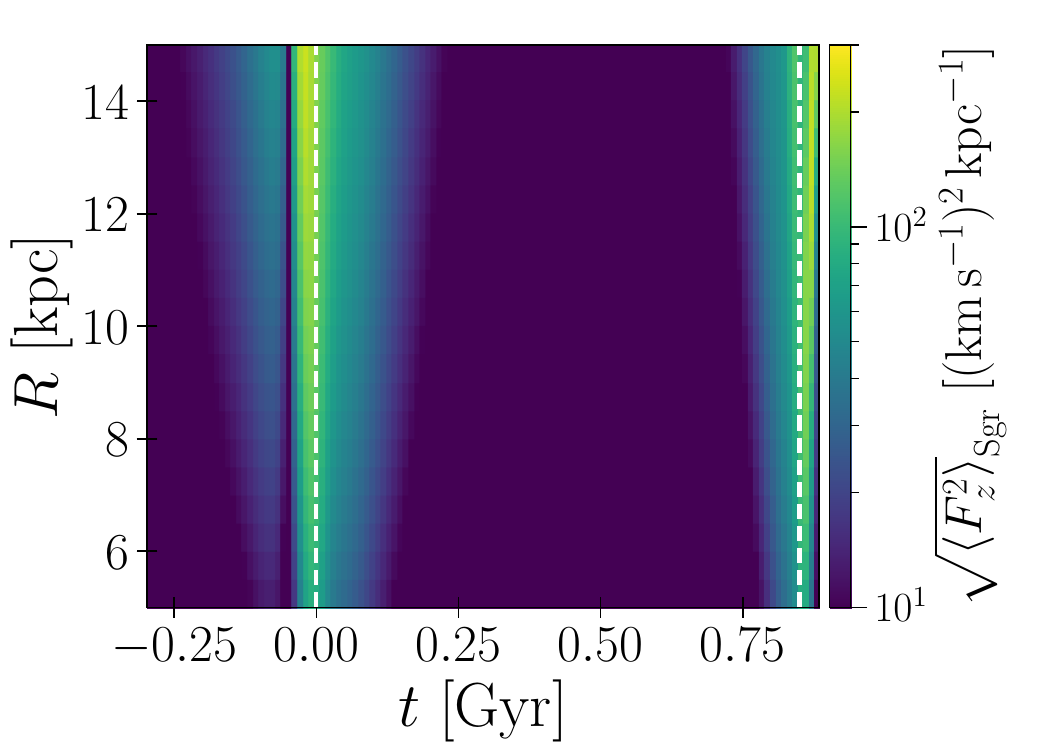}
                \includegraphics[width=0.33\hsize]{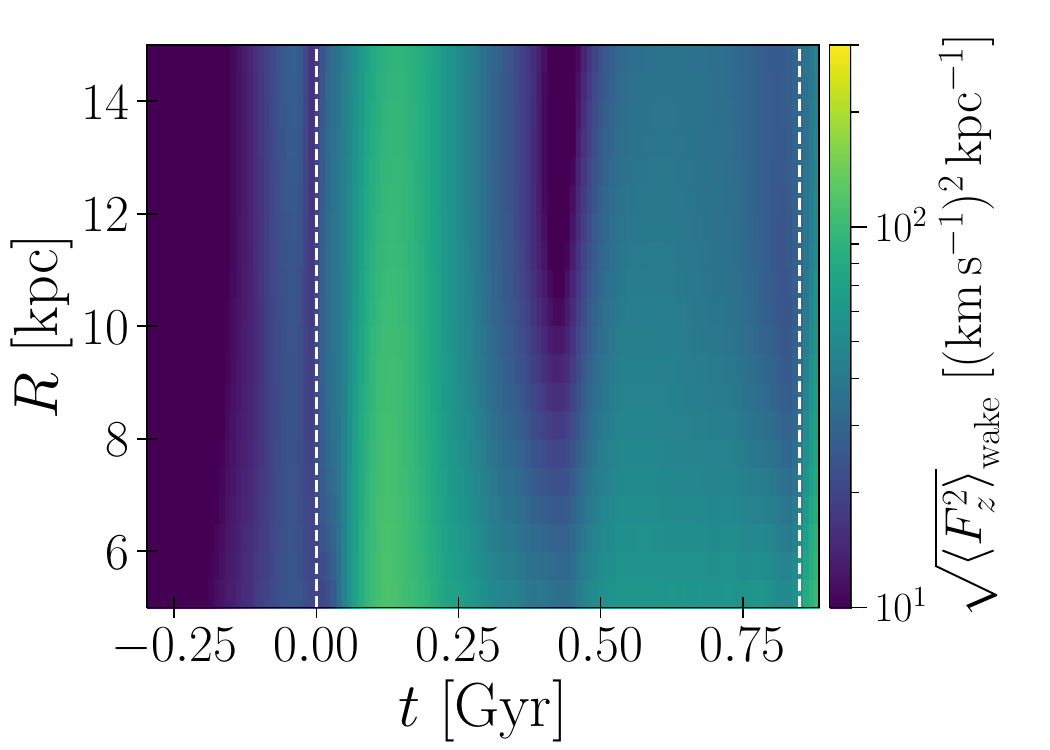}
                \includegraphics[width=0.33\hsize]{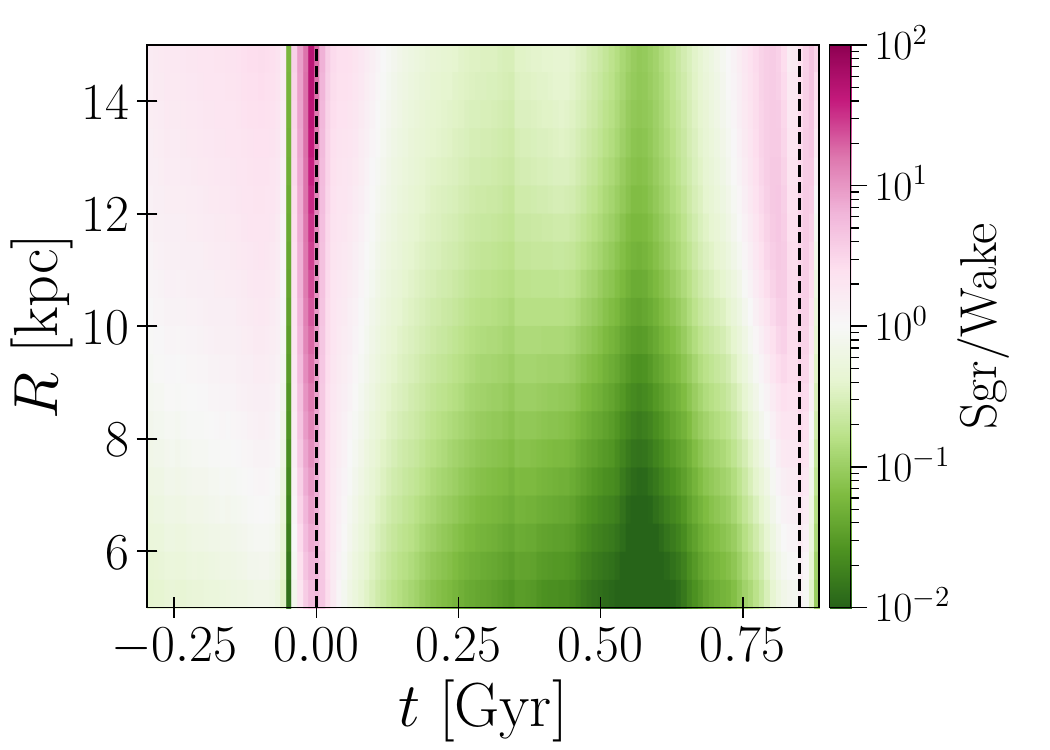}
                \caption{Amplitude of the perturbing force as a function of $R$ and $t$. The colour maps represent the force amplitudes due to Sgr (\textit{first panel}), the DM wake (\textit{second panel}), and their ratio (\textit{third panel}). The vertical dashed lines indicate the times of the Sgr pericentric passages.}\label{fig:Fz}
        \end{center}
\end{figure*}
While Fig.~\ref{fig:Fz_8kpc} illustrates both the amplitude and direction (i.e. positive or negative) of the vertical perturbation along a ring at $R=8$~kpc, Fig.~\ref{fig:Fz} focuses on the amplitudes alone and presents it as a function of radius and time.
We evaluated the amplitude by averaging the square of the vertical force, $F_{z, \mathrm{pert}}^2 (R, \phi, z=0, t)$, over azimuth.
The first panel of Fig.\ref{fig:Fz} shows the amplitude of the Sgr perturbation.
As we saw in Fig.\ref{fig:Fz_8kpc}, Sgr perturbs the disc during short time intervals around the pericentric passages.
The influence of Sgr begins earlier and continues longer in the outer galaxy than in the inner galaxy. The map exhibits a narrow gap at $t=-0.05$~Gyr, when Sgr crosses the disc, and therefore, $F_{z, \mathrm{Sgr}}$ becomes zero.
The second panel of the figure shows the same map for the DM wake.
The amplitude of the wake starts to increase at $t\sim0$, reaches maximum at $t\sim0.1$~Gyr, and then decreases slowly.
After reaching a minimum at $t\sim 0.4$~Gyr, it starts to increase again and becomes almost constant from $t\sim0.5$~Gyr.
In general, the DM halo response to a satellite galaxy can be decomposed into the dynamical friction wake and the collective response (see Appendix~\ref{appendix:potential_models} and \citealt{2021ApJ...919..109G}).
The dynamical friction wake is a transient overdensity located behind the satellite along its orbit, while the collective response manifests as a large-scale dipole asymmetry in the DM halo that persists for a longer duration.
In our model, the peak at $t\sim0.1$~Gyr and the subsequent long-lived component are attributed to the dynamical friction wake and the collective response, respectively.
The third panel shows the ratio of the Sgr force to the wake force.
The impact of the Sgr main body is confined in short time intervals around the pericentric passages, while the wake perturbation  dominates during most of the time between the pericentric passages.
\citet{2021MNRAS.507.2825G} also analysed the force fields of the Sgr and the DM wake in the $N$-body model simulated by \citet{2018MNRAS.481..286L}.
Consistent with our model, the force profile of Sgr exhibits high-amplitude spikes around pericentric passages, whereas the DM force evolves more gradually.

\subsection{Phase spirals}
\begin{figure*}
        \begin{center}
        \includegraphics[width=\hsize]{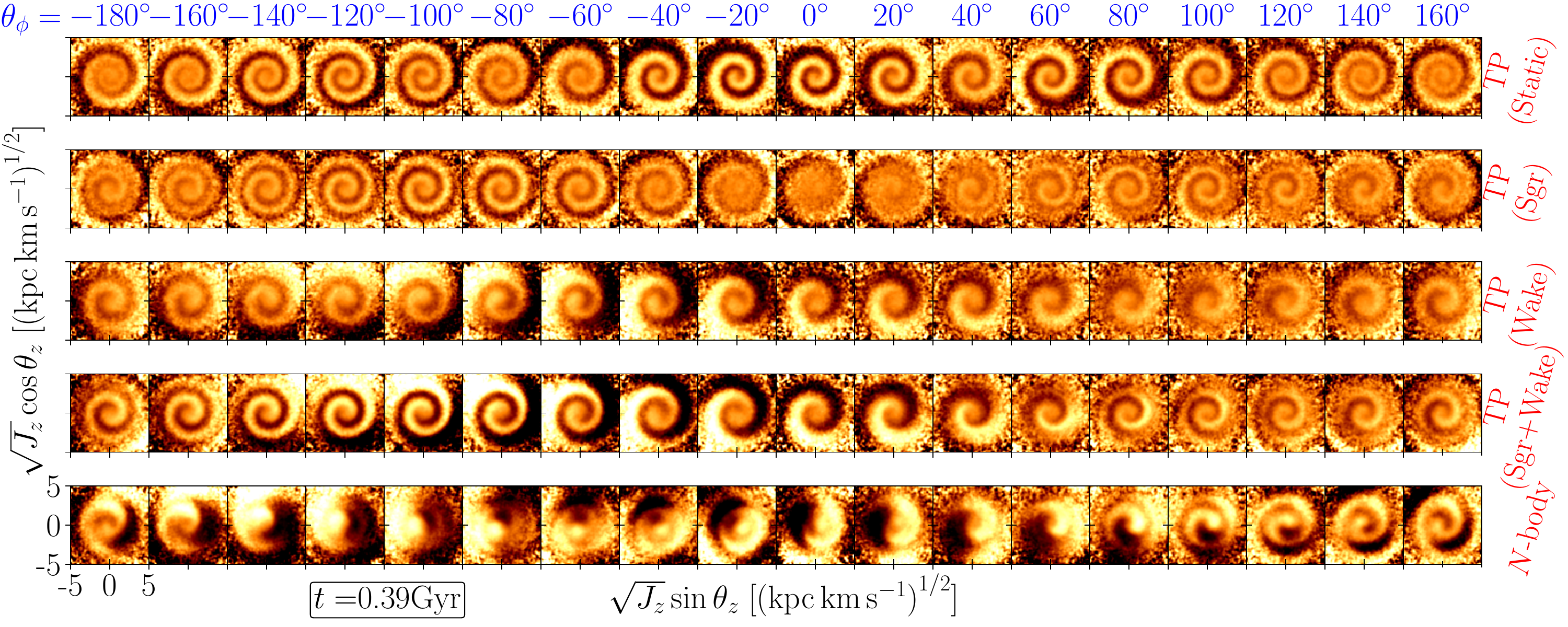}
        \caption{Phase spirals in the TP (static) model (\textit{first row}), TP (Sgr) model (\textit{second row}), TP (wake) model (\textit{third row}), TP (Sgr+wake) model (\textit{fourth row}), and $N$-body model (\textit{fifth row}). Each panel shows the density contrast in the $\sqrt{J_z}\cos\theta_z$-$\sqrt{J_z}\sin\theta_z$ space for 18 bins of $\theta_{\phi}$ at $R_g=8$~kpc.}
        \label{fig:phase_spiral_04}
        \end{center}
\end{figure*}

We first visually inspect the phase spirals in both the $N$-body and test-particle simulations, although we quantify the differences between the models in the next section.
Figure~\ref{fig:phase_spiral_04} presents the particle distributions in the $\sqrt{J_z} \cos \theta_z$--$\sqrt{J_z} \sin \theta_z$ plane in an annulus at $R_g = 8\pm1~\mathrm{kpc}$, with the data binned by azimuthal angle. 
The $\sqrt{J_z}\sin\theta_z$--$\sqrt{J_z}\cos\theta_z$ map is a  projection of vertical action--angle space, which is morphologically analogous to the $z$--$v_z$ phase-space map.
\footnote{For the harmonic oscillator, one has
$z \propto \sqrt{J_z}\sin\theta_z$ and
$v_z \propto \sqrt{J_z}\cos\theta_z$.
In general Galactic potentials, this correspondence is a good approximation only for orbits with small vertical amplitudes, but the resulting map exhibits a phase spiral morphology similar to that in the $z$--$v_z$ plane (see, for example, Fig. 1 of \citealt{2023ApJ...955...74D}).
}
For the present sample, the extent of the distribution is comparable to the $z$--$v_z$ map with $|z|\lesssim1$~kpc and $|v_z|\lesssim50$~km~s$^{-1}$.
From top to bottom, the rows correspond to the TP (static), TP (Sgr), TP (wake), TP (Sgr+wake), and $N$-body models, respectively.
The snapshots are taken at $t \sim0.4~\mathrm{Gyr}$, which corresponds to the epoch at which the phase-spiral patterns can be identified most clearly across all models.
Each panel shows the density contrast relative to the symmetric background density. We estimated the background density by randomising $\theta_z$ of the particles within each azimuthal angle ($\theta_\phi$) bin.

All models exhibit one-arm phase spirals, although their prominence and tightness differ across the models.
In the TP (static) and TP (Sgr) models, the phase spirals are wound tightly in a similar manner, whereas in the test-particle simulations with the DM wake, TP (wake) and TP (Sgr+wake), they appear less tightly wound.
This is because the peak of the wake perturbation force is delayed relative to the Sgr pericentric passage, which introduces a delay in the phase mixing start (see Fig.~\ref{fig:Fz_8kpc}).

Compared to the test-particle models, the phase-space maps of the $N$-body model are notably more complex.
In many panels of the $N$-body model, the phase-space distribution resembles a superposition of a phase spiral and a dipole pattern.
Although weak dipole-like asymmetries are also observed in the TP (wake) and TP (Sgr+wake) models (e.g. panels around $\theta_\phi \sim -120^\circ$ to $-40^\circ$), they are less pronounced than in the $N$-body case.
We attribute this difference to the self-gravitating nature of the disc in the $N$-body simulation.
As suggested by previous studies \citep{2019MNRAS.484.1050D, 2023MNRAS.522..477W}, the phase spiral in the $N$-body model is less tightly wound than in the test-particle models.

\begin{figure}
        \begin{center}
        \includegraphics[width=0.85\hsize]{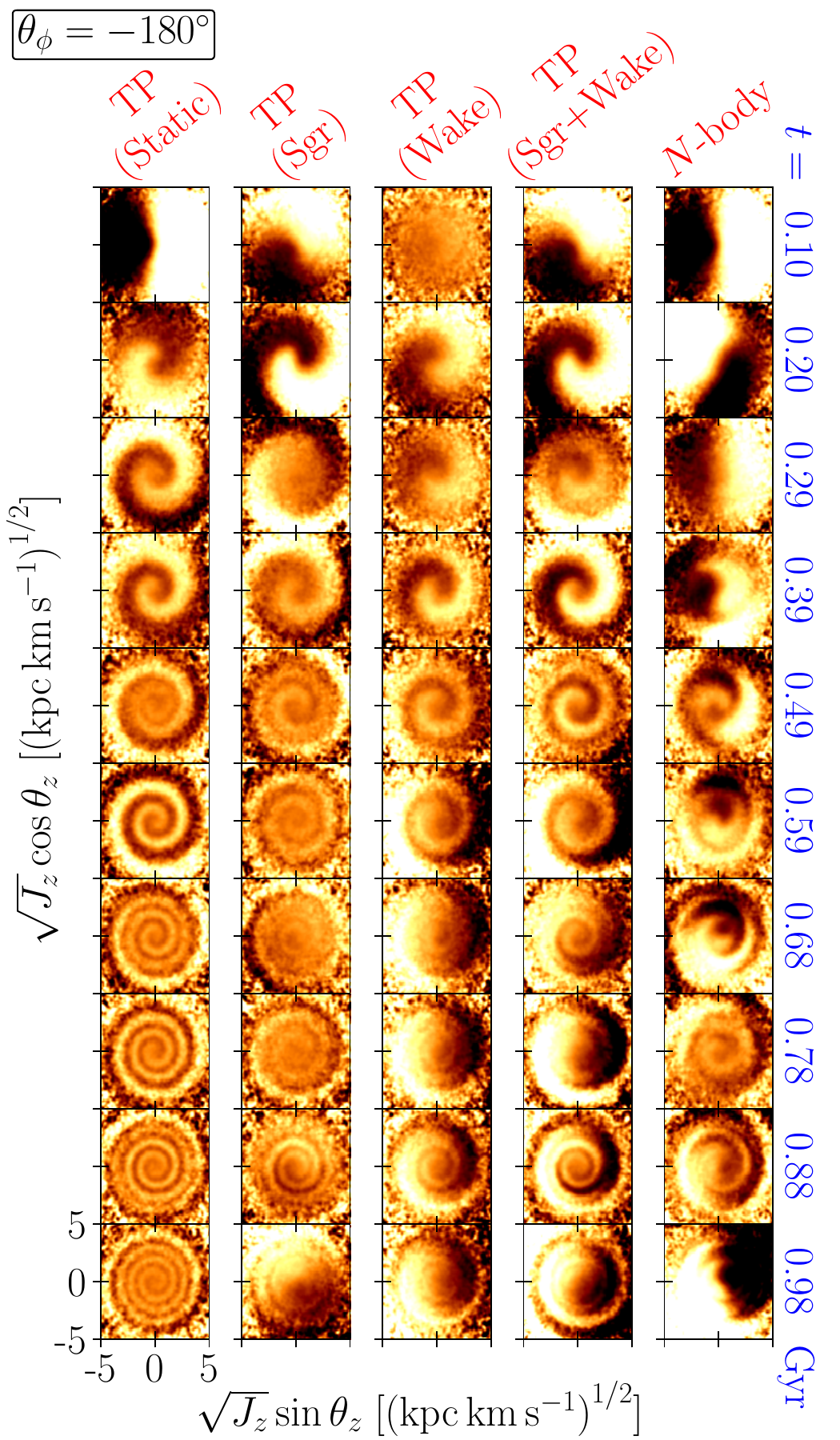}
        \caption{Time evolution of the phase spiral. From top to bottom, the rows correspond to the different models TP (static), TP (Sgr), TP (wake), TP (Sgr+wake), and $N$-body model. From left to right, time increases from $t=0$~Gyr to 0.88~Gyr.}
        \label{fig:phase_spiral_time}
        \end{center}
\end{figure}
We also examine the temporal evolution of the phase spirals in Fig.~\ref{fig:phase_spiral_time}.
The columns represent different models, while the rows correspond to the different times between the first and second pericentric passages.
In the TP (static) model, an initial dipole perturbation is progressively stretched via phase mixing and evolves into a tightly wound phase spiral.
The TP (Sgr) model displays a similar evolution, but at $t = 0.88~\mathrm{Gyr}$, a new dipole perturbation is triggered by Sgr’s second pericentric passage.
In the TP (wake) model, there is no clear asymmetry in the distribution at $t = 0~\mathrm{Gyr}$ because the force of the wake has not yet reached its maximum and is not strong enough to excite the significant bending in the disc.
A phase spiral forms at $t = 0.10~\mathrm{Gyr}$ and becomes increasingly tightly wound until $t = 0.49~\mathrm{Gyr}$.
Unlike the TP (Sgr) model, a new dipole pattern emerges at $t = 0.59~\mathrm{Gyr}$.
        As shown in Appendix~\ref{appendix:potential_models}, the DM wake consists of two distinct components: a transient dynamical friction wake and a collective response.
        These components imprint dipole-like signatures in the vertical phase space at different times.
        The first dipole pattern, which emerges at $t\sim0.1$~Gyr and evolves into the spiral, is triggered by the dynamical friction wake.
        The second dipole pattern appearing at $t \gtrsim 0.5$~Gyr is associated with the collective halo response.
        Since its perturbation is weaker and more slowly varying than that of the dynamical friction wake, it does not fully erase the pre-existing phase spiral, resulting in a superposition of the two features. 

In the TP (Sgr+wake) model, the phase spiral initially resembles that of the TP (Sgr) model: a dipole pattern appears at $t=0$ and develops into a one-arm spiral.
At later times ($0.5 \lesssim t \lesssim 0.8$~Gyr), a new dipole pattern becomes visible that resembles the feature seen in the TP (wake) model, suggesting that the wake-driven perturbation dominates during this epoch.

In the $N$-body model, phase spirals do not form until $t = 0.29~\mathrm{Gyr}$, a behaviour previously reported by \citet{2025A&A...700A.109A}.\footnote{The delayed emergence of phase spirals is also observed in other self-consistent simulations \citep{2021MNRAS.504.3168B, 2025MNRAS.542.1987T}.}
In self-gravitating discs, dipole patterns excited by external perturbations tend to rotate almost rigidly before transforming into phase spirals.
Subsequently, the phase spiral develops and winds up with time.
In comparison to the test-particle models, the phase spirals in the $N$-body model are less tightly wound and more disordered.
In particular, the centre of the spiral does not necessarily coincide with the origin of the adopted phase-space coordinate, and both the strength and width of the spiral show significant variations along it, in contrast to the more regular behaviour seen in the test-particle models.
In the next section we quantitatively examine this non-monotonous phase mixing in self-gravitating systems.

\section{Unwinding phase spirals}
\subsection{Method}\label{subsec:unwinding}
\begin{figure*}
        \begin{center}
        \includegraphics[width=0.85\hsize]{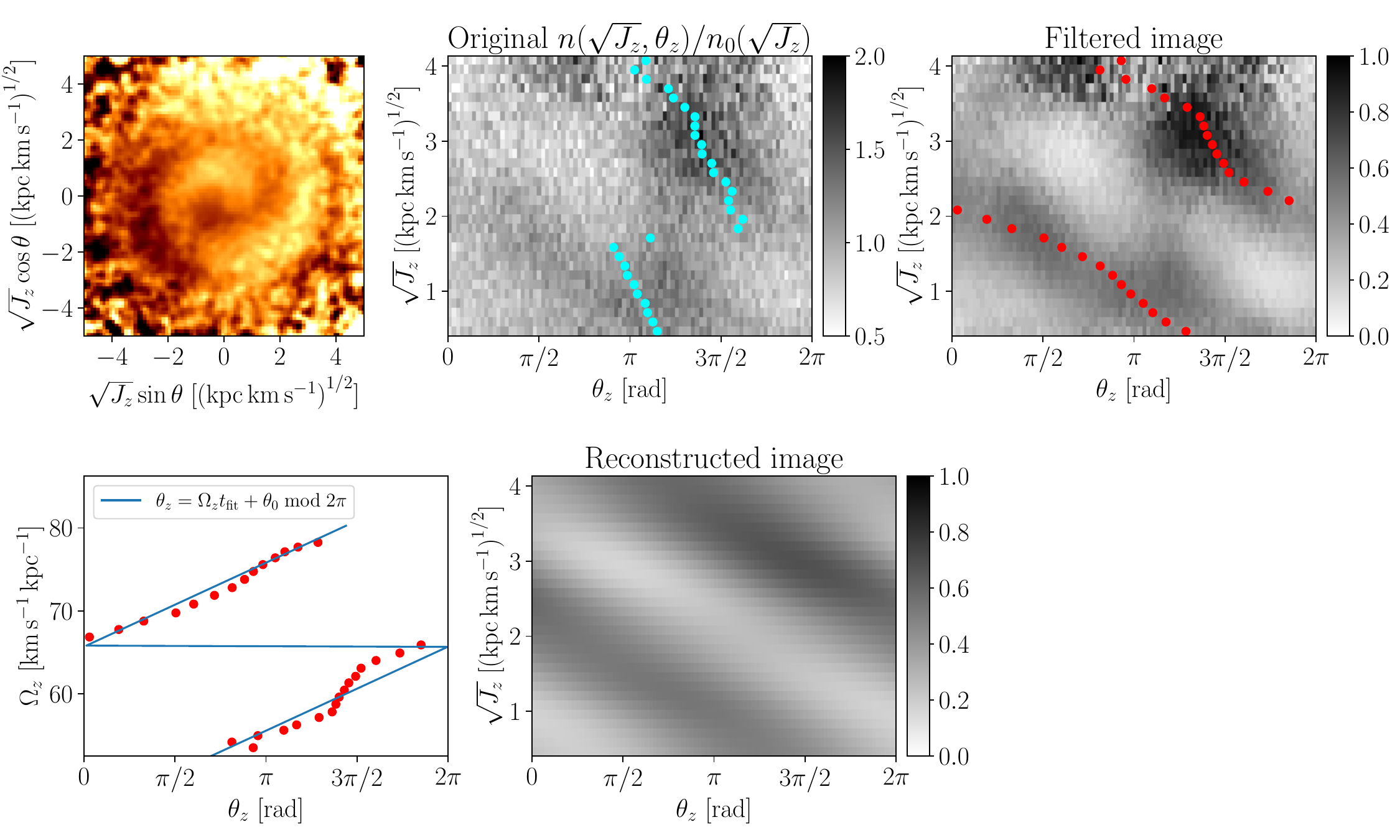}
        \caption{Phase spiral unwinding. \textit{Top left:} Density contrast in the $\sqrt{J_z}\cos\theta_z$--$\sqrt{J_z}\sin\theta_z$ space. \textit{Top middle:} Same data mapped into the $\sqrt{J_z}$--$\theta_z$ space. The cyan dots indicate the phase of the $k=1$ Fourier mode. \textit{Top right:} Filtered $\sqrt{J_z}$--$\theta_z$ map. The red dots indicate the phase of the $k=1$ Fourier mode for the filtered map. \textit{Bottom left:} Phase of the $k=1$ Fourier mode mapped into the $\Omega_z$--$\theta_z$ space, where $\Omega_z$ is the vertical frequency. The blue line shows the best-fit linear model. \textit{Bottom right:} $\sqrt{J_z}$--$\theta_z$ map reconstructed from the fitting result.}\label{fig:fitting_example}
        \end{center}
\end{figure*}
If stars oscillate vertically at frequencies constant with time, which depend only on their (primarily vertical) actions, a phase spiral observed in the $z$--$v_z$ space transforms into a stripe pattern in the $\theta_z$--$\Omega_z$ space, where $\theta_z$ and $\Omega_z$ represent the vertical angle and vertical frequency, respectively.
If a perturbation occurs at $t=0$, the stripe pattern at a later time follows a simple linear relation: $\theta_z = \Omega_z t + \theta_0$.
By fitting this linear model to the data, we can estimate the winding time of the phase spiral, which corresponds to the time elapsed since the perturbation occurred.
This unwinding method was adopted in several previous studies \citep[e.g.][]{2023ApJ...955...74D, 2023MNRAS.521.5917F}.
In our work, we applied a similar approach to both our $N$-body and test-particle simulations.
However, to ensure a robust fit, we first preprocessed the data, as described below.
Here, we present the procedure using a noisy example from the $N$-body model to demonstrate the effectiveness of this processing, and also provide an idealised example from TP (static) model in Appendix~\ref{appendix:phase_spiral_fitting_tp}.

The top left panel of Fig.~\ref{fig:fitting_example} shows an example of the particle distribution in $\sqrt{J_z} \cos \theta_z$--$\sqrt{J_z} \sin \theta_z$ space from the $N$-body simulation at $t=0.39$~Gyr.
A one-arm phase spiral pattern can be identified in the distribution.
In the middle top panel, we map the same distribution into the $\sqrt{J_z}$--$\theta_z$ space.
We divided this space into $30 \times 90$ bins, where the upper and lower boundaries of $\sqrt{J_z}$ correspond to the 5th and 95th percentiles of the $\sqrt{J_z}$ distribution, respectively.
The colour of each pixel indicates the number density $n(\theta_z, \sqrt{J_z})$ normalised by the mean density $n_0(\sqrt{J_z})$ at each $\sqrt{J_z}$.
The diagonal ridge pattern that appears in this projection corresponds to the phase spiral.
In addition to the ridge, we see a vertical band between $\theta_z \sim \pi$ and $\sim 3/2\pi$, which corresponds to the dipole asymmetry that extends across the entire region of the top left panel.
While it is also an interesting feature and might reflect the self-gravitating nature of the disc or the DM wake perturbation, we leave it aside here.

To enhance the ridge pattern for robust fitting, in the following steps, we cleaned the $\sqrt{J_z}$--$\theta_z$ map.
We subtracted the Gaussian-blurred map and applied the bilateral filter \citep{tomasi1998bilateral} and contrast-limited adaptive histogram equalization \citep[CLAHE;][]{PIZER1987355}, using the \texttt{opencv-python} package.
The filtered map is shown in the top right panel, with the pixel values scaled from 0 to 1.
This image processing reduces both small-scale noise and large-scale bias, such as the global dipole asymmetry, and enhances the ridge pattern.

We next applied a Fourier transform to the filtered map along the $\theta_z$ axis,
\begin{align}
\tilde{n}(J_z, \theta_z) = \sum_{k=-\infty}^\infty A_k(J_z) \exp [-i k (\theta_z - \theta_{z,k} (J_z))],
\end{align}
where $\tilde{n}$ is the pixel value in the filtered image, and $A_k$ and $\theta_{z,k}$ are the amplitude and the phase of the $k$-th Fourier mode, respectively.
The red dots in the top right panel indicate the phase of the $k=1$ mode, $\theta_{z,1}$, which traces the ridges (i.e. one-arm phase spiral).\footnote{We can measure $\theta_z$--$\Omega_z$ slope for the $k=2$ mode (i.e. two-arm phase spirals) in the same way as for the $k=1$ mode, but it does not necessarily correspond to the elapsed time since the excitation of the perturbation (see \citealt{2023MNRAS.522..477W} and \citealt{2025MNRAS.543.2159C}).}
We also show the results of Fourier transform for the unfiltered map in the top middle panel with cyan dots. Without filtering, the $k=1$ phase does not trace the ridge, demonstrating the effectiveness of our filtering procedure.

We mapped the detected $k=1$ phase into the $\theta_z$--$\Omega_z$ space shown in the bottom left panel.
The horizontal and vertical coordinates of the dots show the $\theta_{z, 1}$ and mean $\Omega_z$ values in each $\sqrt{J_z}$ bin, respectively.
We fitted these data with the function
\begin{align}
\theta_z = (\Omega_z t_{\mathrm{fit}} + \theta_0) \mod 2\pi
\label{eq:fitting}
\end{align}
by minimising the error function
\begin{align}
        E(t_\mathrm{fit},\theta_0) = 1 - \frac{1}{N}\sum_{j=1}^{N} \cos \left(\theta_{z1, j} - \Omega_{z,j}t_\mathrm{fit} - \theta_0 \right),
        \label{eq:error}
\end{align}
where $N=30$ is the number of the $\sqrt{J_z}$ bins, and $j$ is the index of the bins.
We found the best-fit $(t_\mathrm{fit}, \theta_0)$ using \texttt{scipy.optimize.differential\_evolution}.
The blue line in the bottom left panel shows the fitting result, and the bottom right panel shows the $\sqrt{J_z}$-$\theta_z$ map reconstructed from the fitting result.
The winding time, $t_\mathrm{fit}$, which corresponds to the slope of the ridges in $\theta_z$--$\Omega_z$ space, is equal to the time elapsed since a perturbation is applied if phase mixing proceeds without influence of self-gravity.

As illustrated in Figs.~\ref{fig:phase_spiral_04} and \ref{fig:phase_spiral_time}, phase spirals are not always well defined.
To exclude such cases from the analysis, we evaluated the quality of the Fourier-based pattern recognition and the fitting using two metrics.
The first one is the fitting error defined by Eq.~\eqref{eq:error}.
The second one is the Structural Similarity Index Measure \citep[SSIM;][]{2004ITIP...13..600W}, which quantifies structural similarity between two images in a manner consistent with human visual perception.
We measured the SSIM between the reconstructed map (bottom right panel) and the filtered map (top right panel).
We excluded fitting results from the analysis when the fitting error $E(t_\mathrm{fit}, \theta_0) > 0.2$ and the SSIM $< 0.4$.
These thresholds were selected based on the cases where the ridge detection failed clearly according to visual comparison of the $\sqrt{J_z}$--$\theta_z$ maps, the fitting results, and the corresponding metric values.

\subsection{Results}
\begin{figure}
        \begin{center}
                \includegraphics[width=\hsize]{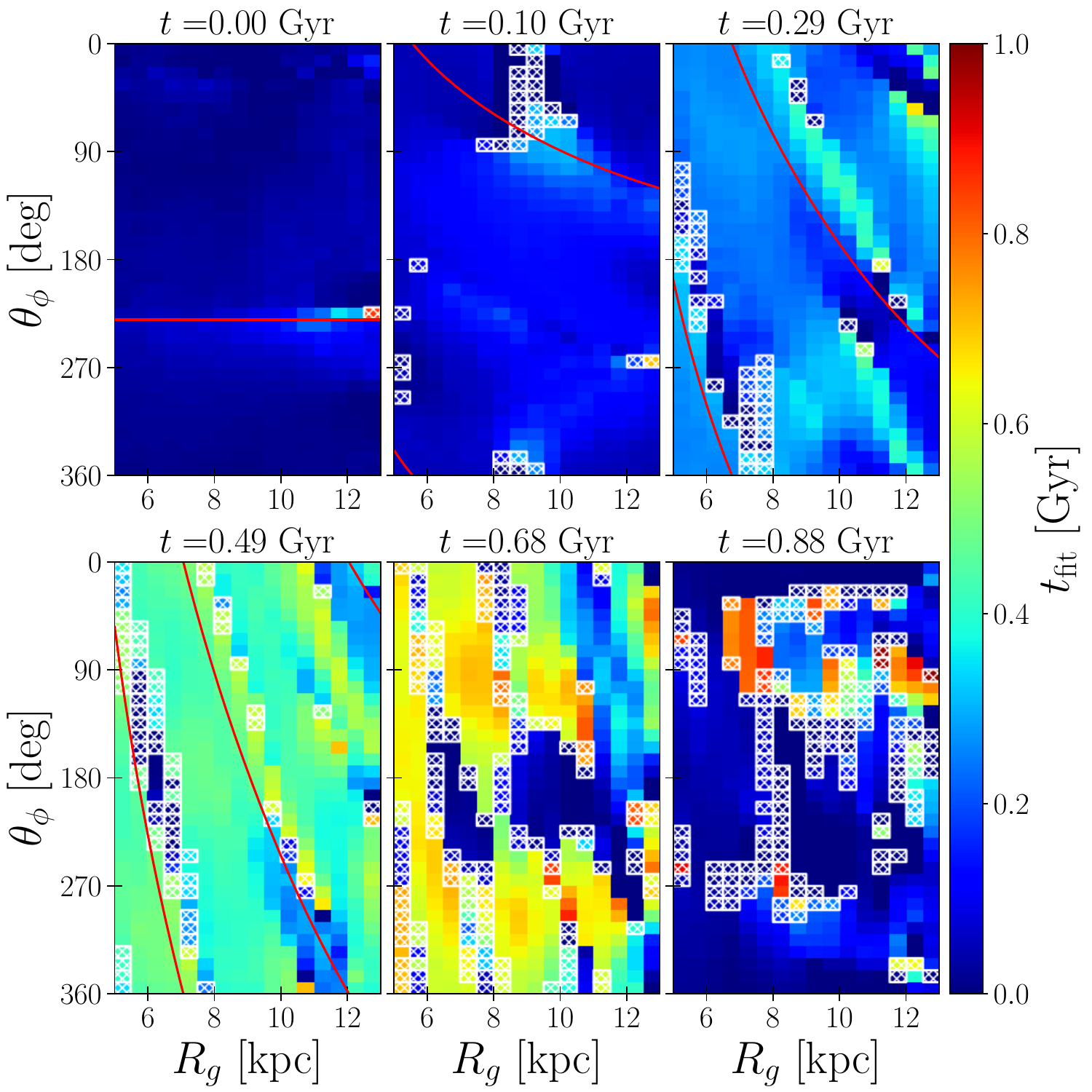}
                \caption{Winding time of the phase spiral as a function of guiding radius, $R_g$, and azimuthal angle, $\theta_\phi$, in the TP (Sgr+wake) model. The six panels correspond to different snapshots from $t = 0$ to 0.88~Gyr. In each panel, bins with poorly defined phase spirals (i.e. $E > 0.2$ and SSIM $< 0.4$) are masked with hatching. The red curves in the panels at $t = 0$, 0.10, 0.29, and 0.49~Gyr indicate $\theta_\phi = \Omega(R_g)t + 230^\circ$.}\label{fig:Rg_theta_tp_sgr_wake}
        \end{center}    
\end{figure}
\begin{figure}
        \begin{center}
                \includegraphics[width=\hsize]{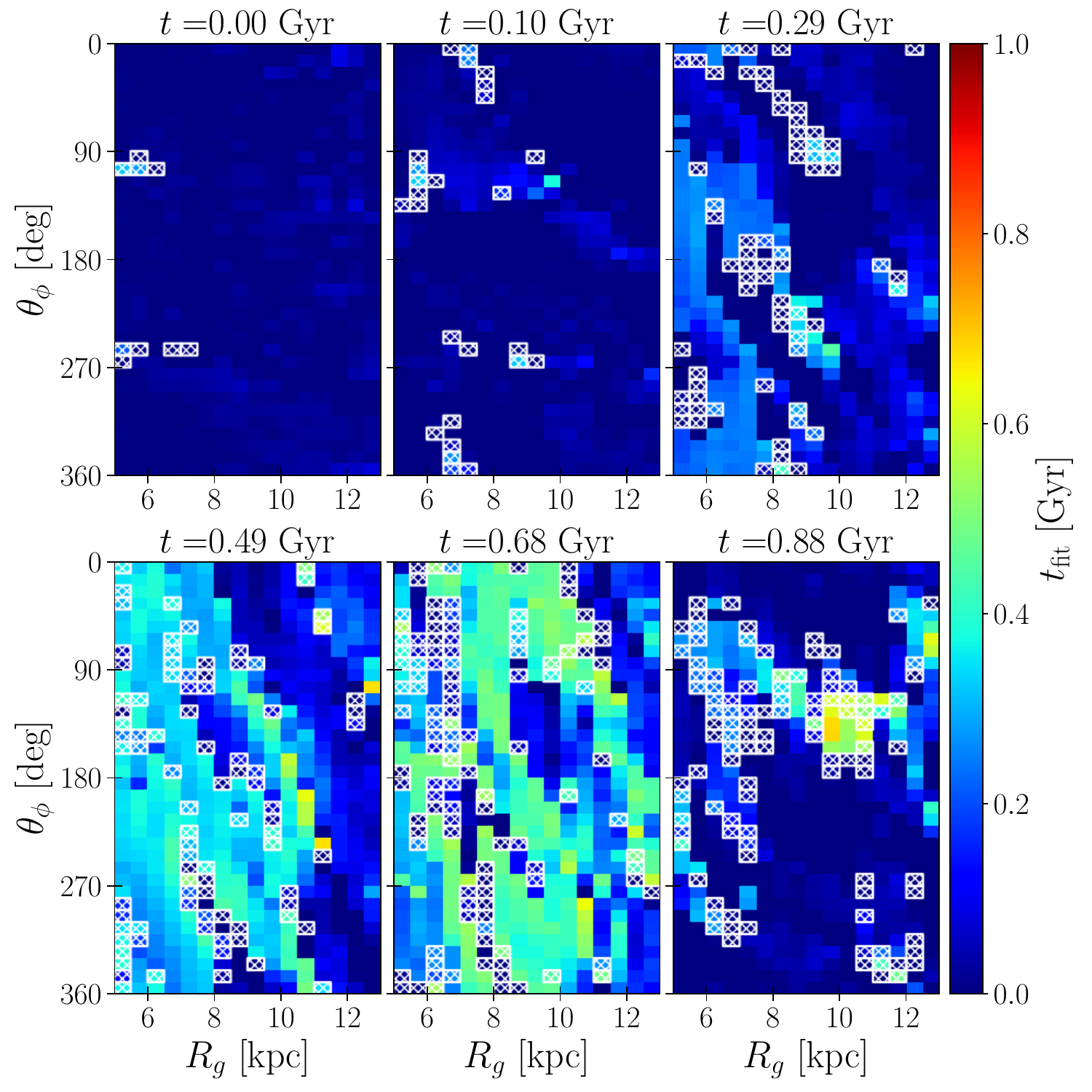}
                \caption{Same as Fig.~\ref{fig:Rg_theta_tp_sgr_wake}, but for the $N$-body model.}\label{fig:Rg_theta_nb}
        \end{center}    
\end{figure}
We first examine the TP (Sgr+wake) model.
We divided the $R_g$--$\theta_\phi$ space into bins of $0.5\,\mathrm{kpc} \times 10^\circ$ and computed the winding time, $t_\mathrm{fit}$, in each bin.
Figure~\ref{fig:Rg_theta_tp_sgr_wake} shows $t_\mathrm{fit}$ across six snapshots, covering the time span from $t = 0$ to 0.9~Gyr.
Equivalent plots from the other test-particle models are shown in Appendix~\ref{appendix:supplementary_figs}.
At $t = 0$ (top left panel), $t_\mathrm{fit}$ is consistent with zero in most bins, except within a narrow range around $\theta_\phi \sim 230^\circ$, which is stronger at outer radii.
This region corresponds to the location in which the effect of Sgr first becomes apparent (see Fig.~\ref{fig:Fz_8kpc}).
As time progresses, the winding time increases almost monotonically from $t = 0$ to 0.49~Gyr. However, diagonal narrow bands emerge, roughly following the curve $-\Omega(R_g)t + 230^\circ$ (indicated by red curves). This pattern reflects the shearing of the initial azimuthal variations due to the differential disc rotation.
The TP (static) model, shown in Fig.~\ref{fig:Rg_theta_tp_static}, exhibits a much more uniform distribution of
        $t_\mathrm{fit}$ within each $R$--$\theta_\phi$ plane.
        This confirms that the azimuthal variations seen in the TP (Sgr+wake) model arise from time-dependent, azimuthally varying perturbations.
At $t = 0.68$~Gyr, regions with $t_\mathrm{fit} \sim 0$~Gyr appear around $\theta_\phi \sim 180^\circ$. We see that this originates from the DM wake perturbation because an equivalent feature is seen in the TP (wake) model but absent in the TP (Sgr) model (see Fig.~\ref{fig:Rg_theta_tp_sgr}).
In the last panel ($t = 0.88$~Gyr),  $t_\mathrm{fit}$ drops to zero across most bins because the impact of the second pericentric passage dominates over that of the first one.
This behaviour indicates that each pericentric passage effectively resets the dynamical clock if the mass of the perturber remains sufficiently large. Similar behaviour has been reported in previous studies \citep{2019MNRAS.485.3134L, 2019MNRAS.486.1167B, 2022MNRAS.510..154G}.

Figure~\ref{fig:Rg_theta_nb} presents analogous maps for the $N$-body model.
As shown in Fig.~\ref{fig:phase_spiral_time}, in this case, phase spirals remain absent for some time following the first pericentric passage of Sgr.
Consistent with this, $t_\mathrm{fit}$ remains near zero not only at $t = 0$~Gyr but also at $t = 0.10$~Gyr.
A rise in $t_\mathrm{fit}$ begins at $t = 0.29$~Gyr, though this increase is not as monotonic as in the TP (Sgr+wake) case: now it progresses more rapidly in the inner disc than in the outer regions.
At $t = 0.49$~Gyr, $t_\mathrm{fit}$ remains systematically lower in the outer disc compared to the inner disc.
By $t = 0.68$~Gyr, $t_\mathrm{fit}$ has reached $\sim0.4$~Gyr across most of the disc between $R_g \sim 7$~kpc and 12~kpc.
At $t = 0.88$~Gyr, $t_\mathrm{fit}$ again drops to zero, reflecting the second pericentric passage of Sgr.

In general, we find that the winding time of the phase spiral in the TP (Sgr+wake) model (Fig.~\ref{fig:Rg_theta_tp_sgr_wake}) increases almost monotonically with time, with values consistent with the elapsed time since the Sgr pericentric passage.
In contrast, in the $N$-body model (Fig.~\ref{fig:Rg_theta_nb}), the winding time evolves non-linearly and stays below the actual time elapsed since the pericentric passage.

\begin{figure}
        \begin{center}
                \includegraphics[width=\hsize]{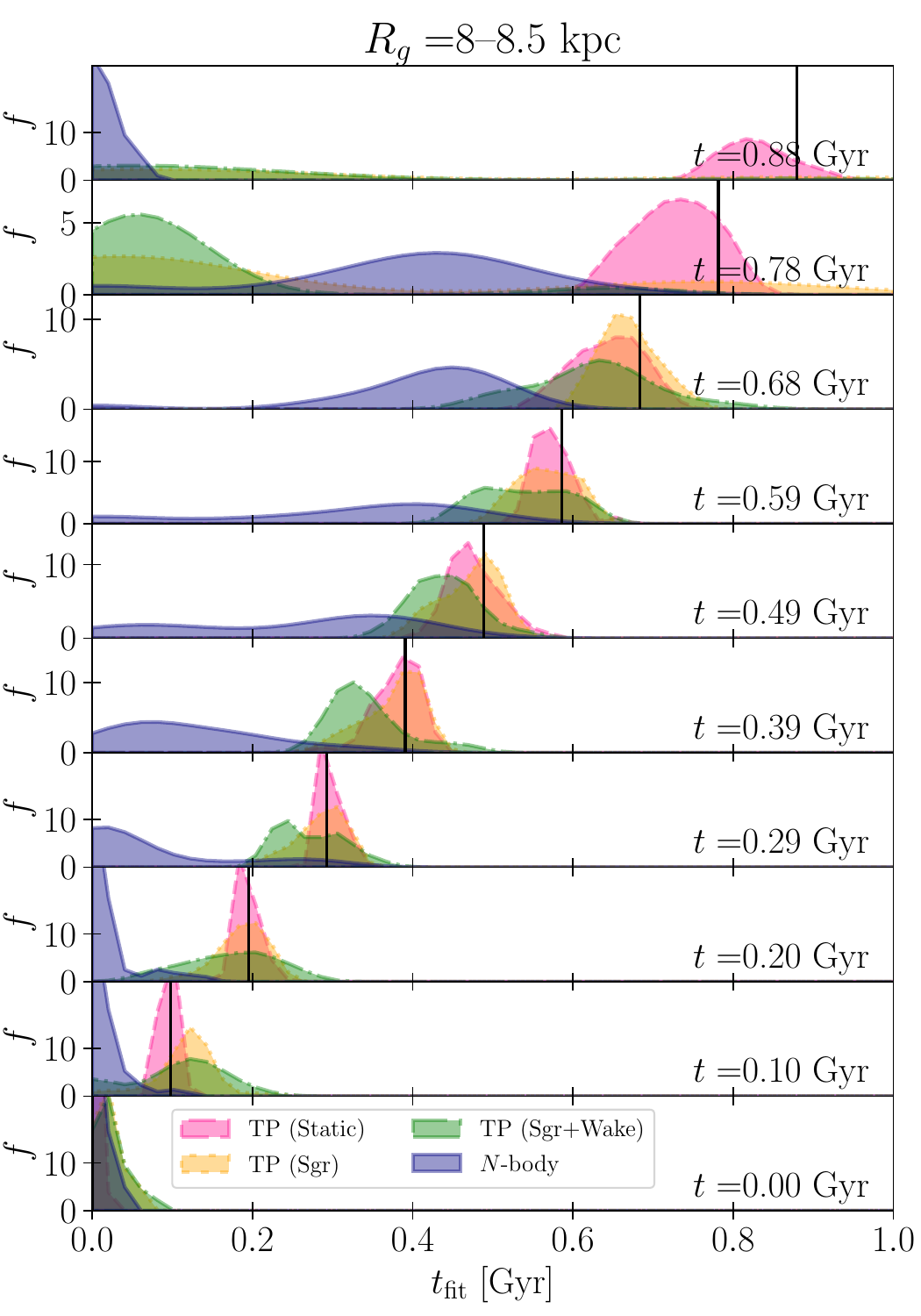}
                \caption{Distribution of the winding time $t_\mathrm{fit}$ in the ring $R_g = 8$--8.5~kpc. The filled pink, yellow, green, and purple areas represent KDE-smoothed distributions for the TP (static), TP (Sgr), TP (Sgr+wake), and $N$-body models, respectively. Each row corresponds to a different time, separated by $\sim0.1$~Gyr. The vertical lines indicate the actual time elapsed since the first pericentric passage.}\label{fig:t_fit_hist}
        \end{center}    
\end{figure}
We next compare the winding time of the phase spiral in the $N$-body and test-particle models more quantitatively.
We extract the columns at $R_g = 8$--8.5~kpc from the same maps shown in Figs.~\ref{fig:Rg_theta_tp_sgr_wake},  \ref{fig:Rg_theta_nb}, \ref{fig:Rg_theta_tp_static}, and \ref{fig:Rg_theta_tp_sgr}, and derive the distributions of $t_\mathrm{fit}$ across the azimuthal angles, $\theta_\phi$.
Figure~\ref{fig:t_fit_hist} shows these distributions for the TP (static), TP (Sgr), TP (Sgr+wake), and $N$-body models.
Here, we do not include the TP (wake) model in the figure because adding it would cause overlap with other histograms and reduce readability.
Each row corresponds to a different time, indicated in the bottom right corner of each panel and marked with black vertical lines.
To obtain smooth histograms, we applied a Gaussian kernel density estimate (KDE) using \texttt{scipy.stats.gaussian\_kde} \citep{2020NatMe..17..261V}.

In the TP (static) model (pink), the distribution peaks of $t_\mathrm{fit}$ coincide with $t$ (vertical lines) up to $t = 0.49$~Gyr, as expected from a pure one-dimensional phase mixing model.
At later times, the peaks shift towards shorter times, but the difference is less than $\sim50$~Myr.
A possible explanation for this small discrepancy is the effect of phase mixing in the radial and azimuthal directions.
The satellite perturbation introduces not only vertical but also in-plane disturbances to disc stars, which are associated with tidally induced spiral arms \citep{2022A&A...668A..61A, 2025A&A...702A.223B} and radial phase spirals \citep{2024MNRAS.52711393H}.
The coupling between vertical and in-plane motions might cause the small deviation of $t_\mathrm{fit}$ from $t$.

In the TP (Sgr) model (yellow), the behaviour is similar to that of the TP (static) model, with peaks of $t_\mathrm{fit}$ remaining close to $t$.
At $t = 0.78$~Gyr and $t = 0.88$~Gyr, the distributions return to $t_\mathrm{fit} \sim 0$ in response to the second pericentric passage of Sgr.

The overall trend in the TP (Sgr+wake) model (green) resembles that of the TP (Sgr) model.
However, the peak values of $t_\mathrm{fit}$ are systematically lower between $t = 0.29$~Gyr and 0.68~Gyr. In this epoch, the DM wake perturbation dominates over that of Sgr.
The distribution in the TP (Sgr+wake) model is also broader than in the TP (Sgr) model.

In contrast, the $N$-body model (purple) exhibits a significantly different evolution.
From $t = 0$ to 0.29~Gyr, the distribution remains concentrated at $t_\mathrm{fit} \sim 0$, with little change in dispersion.
Thereafter, the distribution begins to broaden and shift towards higher $t_\mathrm{fit}$ values.
The peak values of $t_\mathrm{fit}$ in this case are substantially lower than in the test-particle models, and the distributions show higher dispersion, indicating greater azimuthal variation.
In the $N$-body model, $t_\mathrm{fit}$ falls to zero at $t=0.88\ \mathrm{Gyr}$ in response to the second pericentric passage, whereas in the TP (Sgr+wake) model it has already returned to zero by $t=0.78\ \mathrm{Gyr}$.
This offset implies a delayed response of the self-gravitating disc to external perturbations, as the external forces in the two models are essentially identical within the accuracy of the potential expansion.

\begin{figure*}
        \begin{center}
                \includegraphics[width=0.33\hsize]{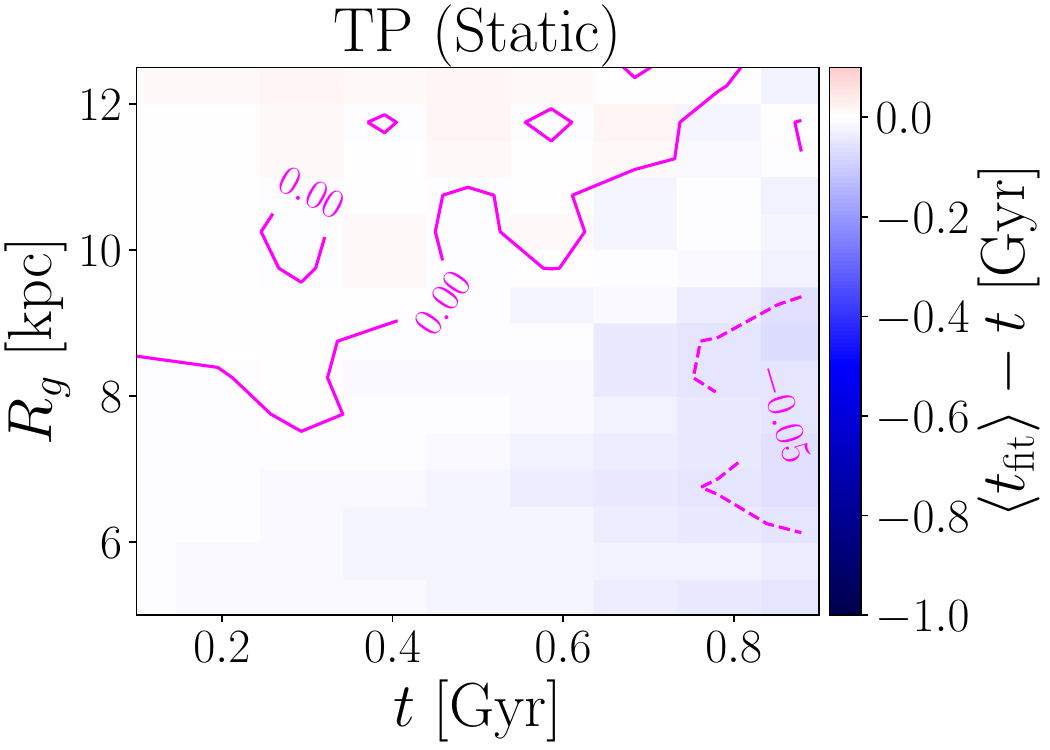}
                \includegraphics[width=0.33\hsize]{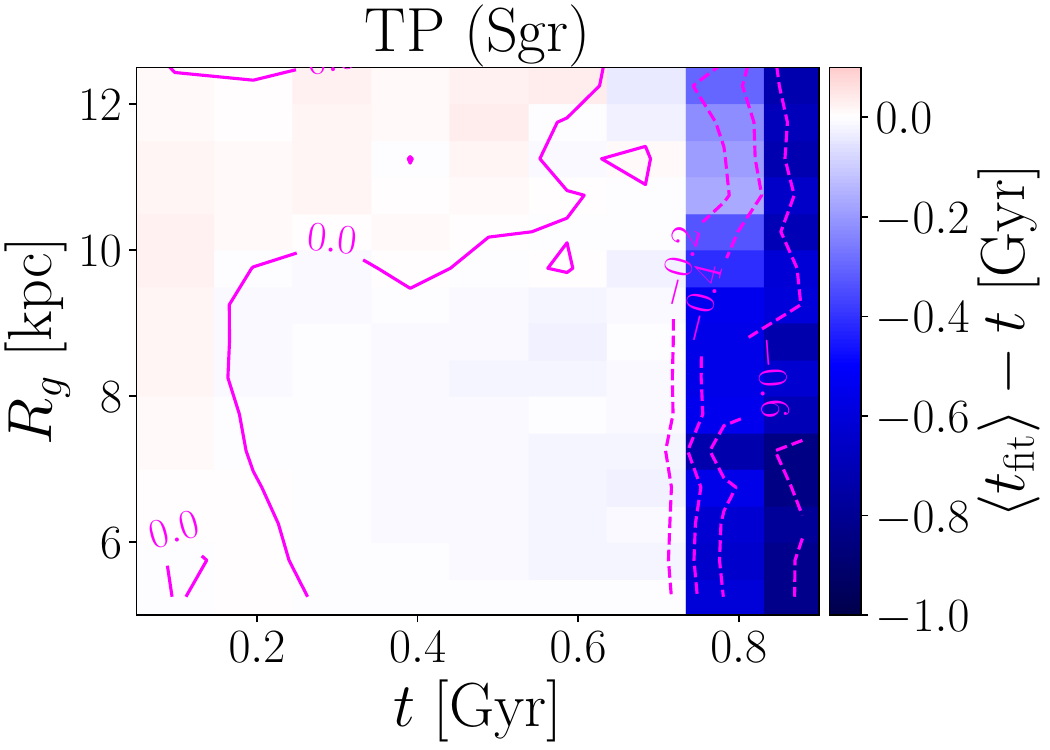}
                \includegraphics[width=0.33\hsize]{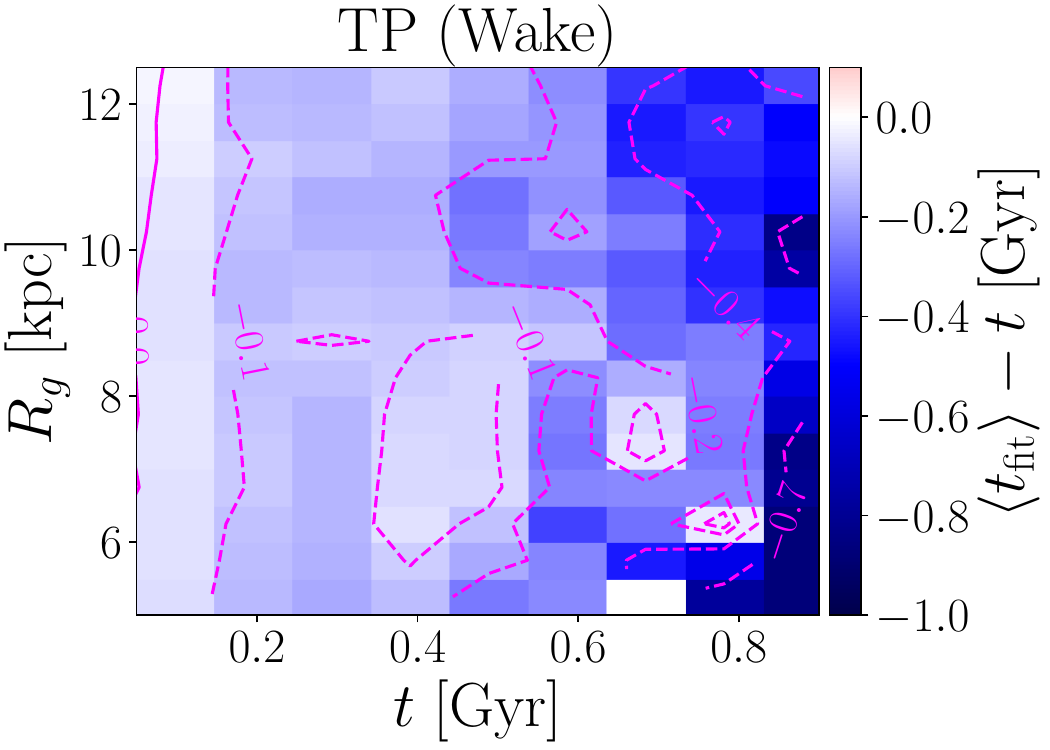}
                \includegraphics[width=0.33\hsize]{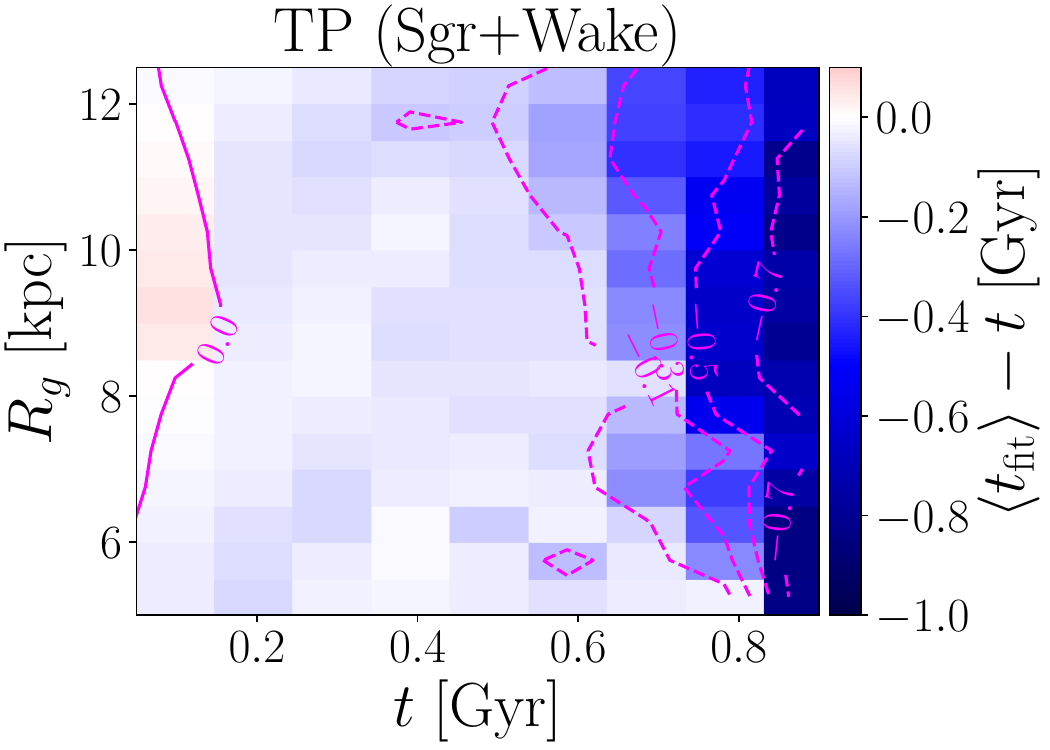}
                \includegraphics[width=0.33\hsize]{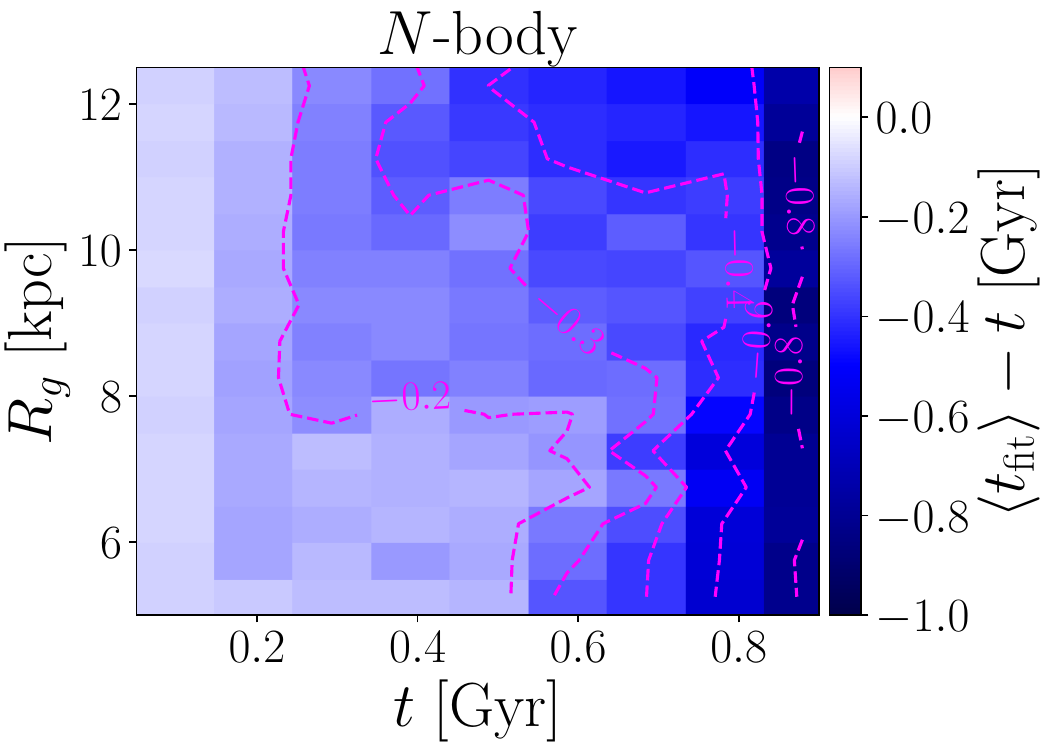}
                \caption{Mean winding time of the phase spiral as a function of $R_g$ and $t$ in the TP (static) (\textit{top left}), TP (Sgr) (\textit{top middle}), TP (wake) (\textit{top right}), TP (Sgr+wake) (\textit{bottom left}), and $N$-body (\textit{bottom right}) models. The colours and contours indicate the mean winding time, $\langle t_\mathrm{fit} \rangle$, relative to the elapsed time since the first pericentric passage of Sgr.}\label{fig:t_fit_t_Rg}
        \end{center}    
\end{figure*}
We finally examine the radial and temporal evolution of the winding time of the phase spiral.
We computed the azimuthal average of $t_\mathrm{fit}$ over $\theta_\phi$ in each bin of $R_g$ and $t$, and show the relative value, $\langle t_\mathrm{fit} \rangle - t$, with respect to the true elapsed time in Fig.~\ref{fig:t_fit_t_Rg}. 

In the TP (static) model (top left panel), $\langle t_\mathrm{fit} \rangle - t$ is nearly zero, although it becomes slightly negative in the inner disc at $t \gtrsim 0.8$Gyr.
As discussed above, this small ($<50$~Myr) deviation is likely due to horizontal mixing.
In the TP (Sgr) model (top middle panel), similar to the TP (static) case, $\langle t_\mathrm{fit} \rangle - t$ stays close to zero until $t = 0.7$~Gyr, after which it drops rapidly to negative values in response to the second pericentric passage of Sgr.
In the TP (wake) model (top right panel), $\langle t_\mathrm{fit} \rangle - t$ is negative across all bins, with differences of $\sim 150$--200~Myr for $t \lesssim 0.6$~Gyr.
Unlike the TP (Sgr) model, the TP (wake) model does not show a radially coherent, sharp drop at the time of the second pericentric passage.
The TP (Sgr+wake) model (bottom left panel) exhibits combined behaviour from both the TP (Sgr) and TP (wake) cases.
Up to $t \sim 0.6$~Gyr, $\langle t_\mathrm{fit} \rangle - t$ is slightly negative, with differences less than 100~Myr.
From $t \sim 0.7$~Gyr, the response to the DM wake becomes evident, starting from the outer disc.
By the final snapshot, $\langle t_\mathrm{fit} \rangle$ drops to zero, and thus $\langle t_\mathrm{fit} \rangle - t$ becomes very negative, reflecting the second pericentric passage of Sgr.
Finally, $\langle t_\mathrm{fit} \rangle - t$ in the $N$-body model (bottom right panel) is also negative across all bins as in the TP (wake) and TP (Sgr+wake) models.
However, while the offset in the previous cases remains below 100~Myr, in the $N$-body model it reaches $\sim 200\text{--}400$~Myr, for instance at $t = 0.5$~Gyr.

To conclude, in the test-particle models, the winding time is slightly shorter than the elapsed time since the Sgr pericentric passage because the peak of the DM wake perturbation lags behind the pericentric passage of the Sgr main body.
The underestimation in the $N$-body model ($\sim 200\text{--}400$~Myr) is even larger than in the TP (Sgr+wake) model ($\lesssim 100$~Myr), suggesting that the less wound phase spirals seen in the $N$-body simulation cannot be explained by the DM wake alone, but are likely the result of self-gravitating effects of the disc.
\citet{2023ApJ...955...74D} also found the same order of offset between the winding time and the peak time of the Sgr perturbation in the $N$-body model by \citet{2021MNRAS.508.1459H}.
Their results also suggest that self-gravity contributes to the less wound phase spirals.

\section{Discussion}
\subsection{Analytical models}\label{subsec:analytic_model}
We observed that phase spirals in the $N$-body model wind up more slowly than those in test-particle models, due to the self-gravitating effects of the disc.
\citet{2023MNRAS.522..477W} predicted this behaviour by using an analytical model, which is an extension of the shearing sheet framework and swing amplification model \citep{1966ApJ...146..810J, 1981seng.proc..111T, 2020MNRAS.496..767B} to three dimensions.
In this  we compare the predictions of the analytical model with the results of our $N$-body simulation.
We adopt model parameters corresponding to those at $R = 8.25$ kpc in the $N$-body model.
Technical details of the model are provided in Appendix~\ref{appendix:analytic_model}.

\begin{figure}
        \begin{center}
                \includegraphics[width=0.95\hsize]{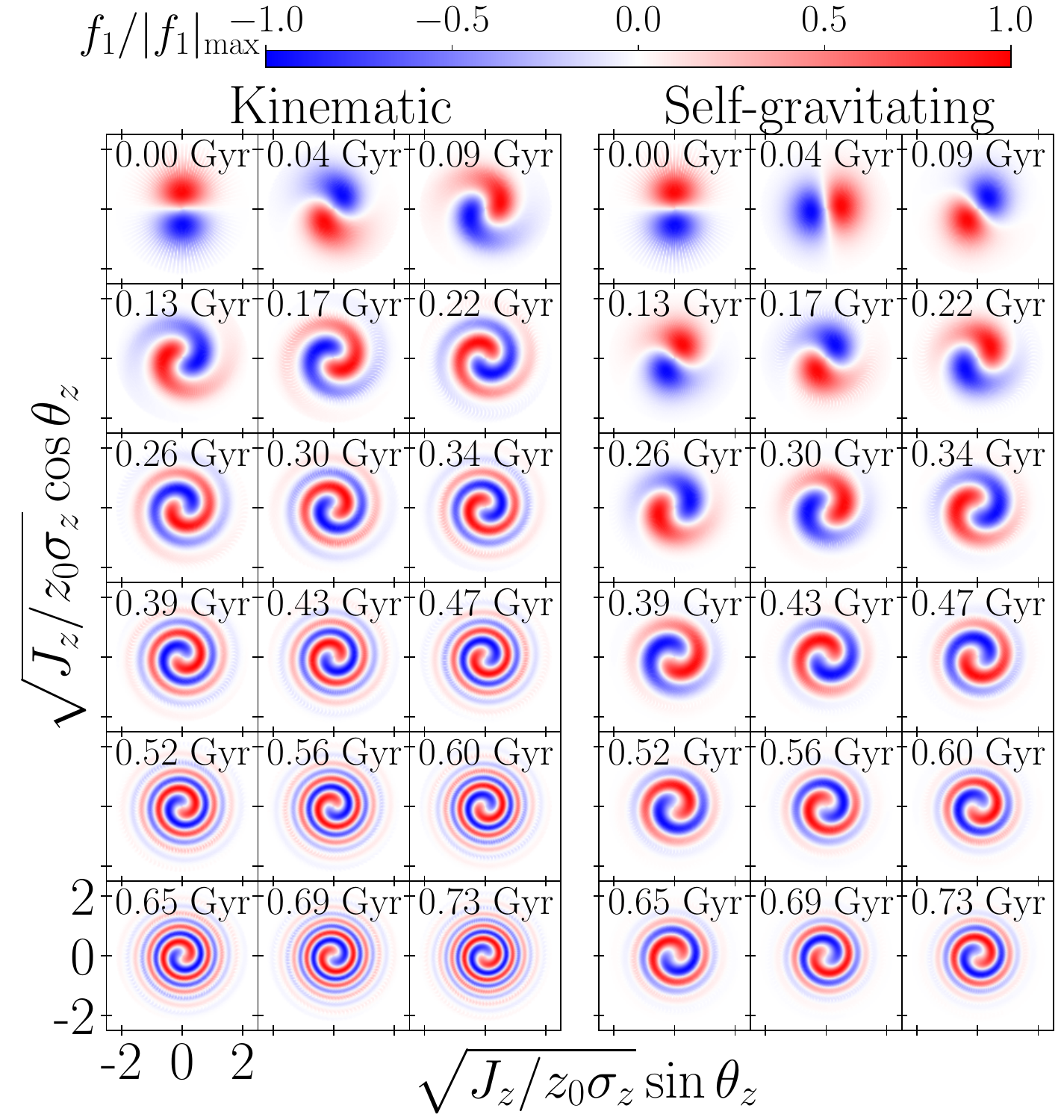}
                \caption{Analytical phase spiral model. Time evolution of the $\sqrt{J_z}\cos\theta_z$--$\sqrt{J_z}\sin\theta_z$ space distribution in the kinematic (\textit{left}) model and in the self-gravitating model (\textit{right}). The colour scales are normalised by the maximum absolute value.}\label{fig:analytic_phase_spiral}
        \end{center}    
\end{figure}
Figure~\ref{fig:analytic_phase_spiral}, which is similar to Figs. 8 and 9 of \citet{2023MNRAS.522..477W}, illustrates the time evolution of the phase spiral in the kinematic (i.e., without self-gravity) and self-gravitating models.
The colour map represents the perturbed term of the distribution function, $f_1(J_z, \theta_z)$, in the $\sqrt{J_z} \cos \theta_z$--$\sqrt{J_z} \sin \theta_z$ space.
As discussed in \citet{2023MNRAS.522..477W}, the amplitudes of $f_1$ in the self-gravitating models are greater than those in the kinematic models.
However, we here focus on differences in the winding rates rather than the amplitudes. Therefore, we normalised the colour scale by the maximum absolute value of $f_1$ for each model at each time.

The external potential excites a dipole moment in the distribution function at $t = 0$, as shown in the top-left panel.
In the kinematic model, this dipole winds up at a constant rate and evolves into a tightly wound phase spiral.
In contrast, in the self-gravitating model, the spiral begins to wind later, at $t \sim 0.15$ Gyr. Prior to that, the dipole moment rotates nearly rigidly.
When comparing the phase spirals in the two models at the same time step, the self-gravitating case appears to be less wound. 
This model successfully reproduces the qualitative evolution of phase spirals observed in the $N$-body simulation.

\begin{figure}
        \begin{center}
                \includegraphics[width=0.8\hsize]{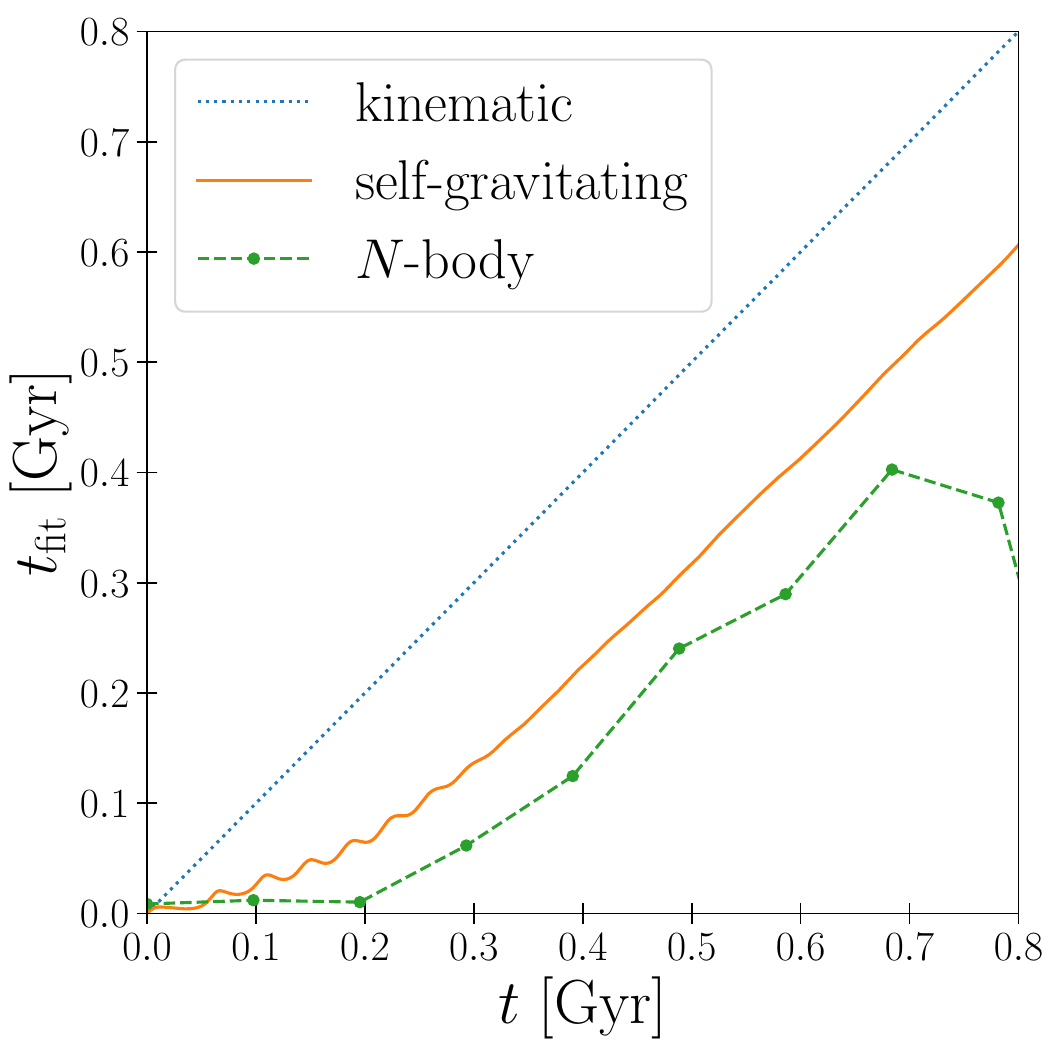}
                \caption{Comparison of the winding times of the phase spirals in the analytical model and the $N$-body model. The dotted blue and solid orange lines represent the values from the kinematic and self-gravitating models, respectively. The green dots connected with dashed lines represent the azimuthally averaged winding time $\langle t_\mathrm{fit} \rangle$ at the bin of $R_g=8$--8.5~kpc in the $N$-body model.
                        The drop in $t_\mathrm{fit}$ at $t \sim 0.8$~Gyr in the $N$-body model is caused by the second pericentric passage, which is not taken into account in the analytical model.
        }\label{fig:analytic_nbody}
        \end{center}    
\end{figure}
We compute the winding time of the phase spiral in the analytical model by measuring the slope of the stripes in $\theta_z$--$\Omega_z$ space, as described in Sect.~\ref{subsec:unwinding}.
Figure~\ref{fig:analytic_nbody} presents the winding time as a function of time.
If self-gravity can be neglected, $t_{\mathrm{fit}}$ is equal to $t$ as indicated by the blue dotted lines.
In the analytical self-gravitating model, $t_{\mathrm{fit}}$  remains approximately zero until $t \sim 0.15$ Gyr.
Thereafter, it increases at a rate comparable to that of the kinematic model (i.e. $\mathrm{d}t_{\mathrm{fit}}/\mathrm{d}t \sim 1$).
The $N$-body model exhibits qualitatively similar behaviour; however, quantitatively, the winding time is shorter than in the self-gravitating analytical model.
In the $N$-body simulation, $t_{\mathrm{fit}}$ begins to increase around $t \sim 0.2\text{--}0.3$ Gyr, which is later than in the analytical model.
The offset between the winding time and the true elapsed time is $\sim0.2$~Gyr in the analytical model, whereas it is $\sim0.3$~Gyr in the $N$-body model at later times.
This difference mainly arises from the variation in the duration of the initial non-winding phase between the two models.

There are several possible reasons for this discrepancy.
Firstly, the analytical model assumes a simple impulsive plane-wave perturbation, whereas the external perturbation in the $N$-body model is more complex.
Secondly, we adopted a simplified potential in the analytical model.
Specifically, a single lowered isothermal model is used to represent the unperturbed vertical potential and density.
In contrast, the vertical potential in the $N$-body model arises from a combination of the stellar disc and DM halo.
Consequently, the potential employed in the analytical model might not precisely correspond to that of the $N$-body model.
Furthermore, the model assumes a separable potential in the analytical model, while in reality, the radial and vertical components of the potential are coupled.
Thirdly, the shearing box framework is a linear model, but non-linear effects might affect the winding rate of the phase spirals in the $N$-body model.

Incorporating such effects is not straightforward and lies beyond the scope of this study.
Most importantly, however, the analytical model---even with its simplified assumptions---qualitatively reproduces the delayed winding of the phase spirals observed in the $N$-body model.

\subsection{Implications for the Milky Way}
\begin{figure}
	\begin{center}
        \includegraphics[width=0.9\hsize]{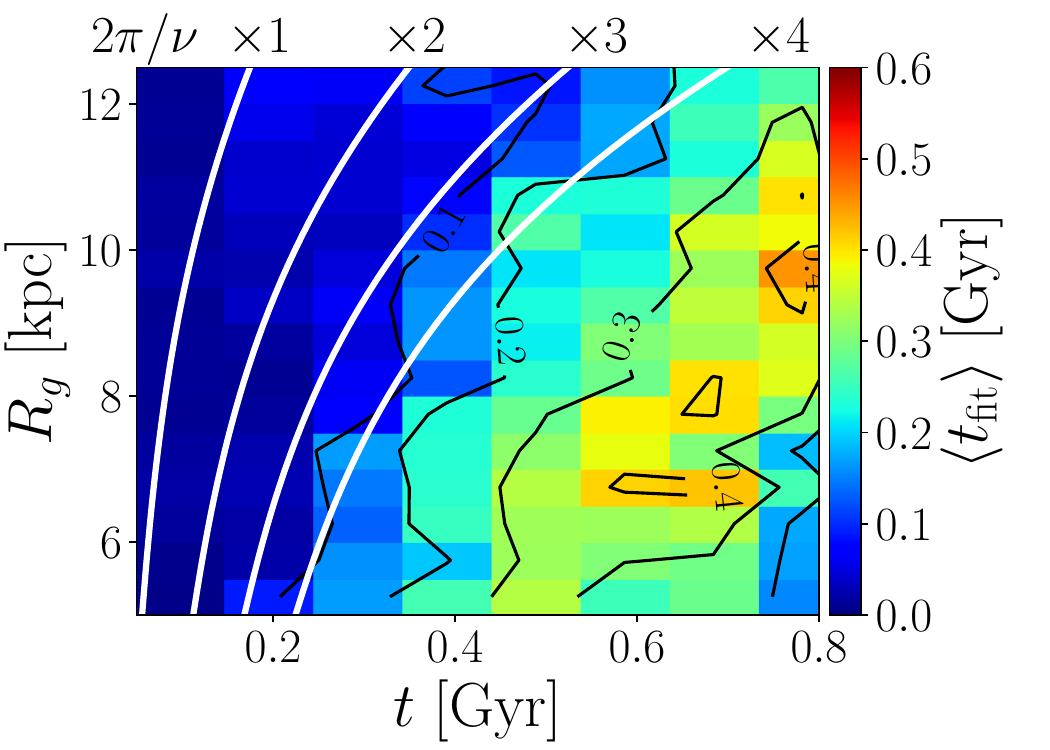}
        \caption{Mean winding time of the phase spiral $\langle t_\mathrm{fit} \rangle$ as a function of guiding radius, $R_g$, and time, $t$, in the $N$-body model. The white curves represent $t=(1,2,3,4) \times 2\pi/\nu(R_g)$.
        }\label{fig:t_fit_t_Rg_abs}
			\end{center}
\end{figure}

We have shown that self-gravity effectively reduces the winding rates of phase spirals, and that the winding time estimated from purely kinematic models is significantly shorter than the actual time elapsed since the disc was perturbed.
To use the observed phase spiral as a reliable dynamical clock, it is necessary to construct a time-evolution model that accounts for the self-gravity of the disc.
The analytical model by \citet{2023MNRAS.522..477W} and our $N$-body model suggest that there is a time lag between the excitation of the bending mode (caused by a close encounter with a dwarf galaxy) and the emergence of the phase spiral.
During this initial stage of the bending mode evolution, the dipole phase-space pattern rotates almost rigidly (i.e. the disc oscillates coherently), and the phase spiral is not well-defined.
Following this non-winding phase, the phase spirals begin to wind at a rate comparable to that predicted by the kinematic model. Based on this observation, we propose that the elapsed time can be approximated by
\begin{align}
        t_{\text{elapsed}} \approx t_{\text{fit}} + \Delta t,
    \label{eq:correction}
\end{align}
where $t_{\text{fit}}$ is the winding time estimated from the slope in the $\theta_z$--$\Omega_z$ space, and $\Delta t$ is the time lag associated with the non-winding phase.

We naively expect that $\Delta t$ scales with a characteristic timescale associated with the stellar vertical oscillations.
To test this expectation, we overplot a line representing the vertical epicycle period, $2\pi/\nu$, and its multiples on the $\langle t_{\mathrm{fit}} \rangle$ map of the $N$-body model in Fig.~\ref{fig:t_fit_t_Rg_abs}.
In this figure, we show the absolute value of $\langle t_{\mathrm{fit}} \rangle$, in contrast to the relative value $\langle t_{\mathrm{fit}}\rangle - t$ shown in Fig.~\ref{fig:t_fit_t_Rg}.
At all radii, the winding time increases slowly until it reaches $\langle t_\mathrm{fit} \rangle \sim 0.1$~Gyr, which corresponds to the non-winding phase, and the time required for $\langle t_\mathrm{fit} \rangle$ to reach this value is equivalent to $\Delta t$.

In our $N$-body model, $\Delta t$ is $\sim3\text{--}4 \times 2\pi/\nu$.
This factor likely depends on dynamical parameters related to self-gravity, such as Toomre's $Q$ and the vertical velocity dispersion.
We cannot directly apply the analytical model to estimate this factor because we saw the discrepancies between the analytical and $N$-body models in the previous subsection.
Qualitatively, the factor is expected to be smaller in hotter discs and larger in colder discs.
Our $N$-body simulation is based on the best-fit MW model of \citet{2019MNRAS.482.1983F}, whose dynamical properties are consistent with recent observations of the MW.
Therefore, we expect the time-lag factor in the real MW to be similarly close to 3--4.
The vertical epicycle frequency in the solar neighbourhood is estimated to be $\nu \sim 70 \, \mathrm{km\,s^{-1}\,kpc^{-1}}$ \citep{2008gady.book.....B}, which gives a time lag of $\Delta t \sim 0.3$~Gyr.

Previous studies have estimated the winding time of the phase spiral in the MW to be approximately 0.3--0.9~Gyr \citep{2018Natur.561..360A, 2023A&A...673A.115A, 2023ApJ...955...74D, 2023MNRAS.521.5917F}. We therefore estimate the total elapsed time since the excitation of the bending mode to be $t_{\text{elapsed}} \sim 0.6$--1.2~Gyr.
Sgr is currently near pericentre, and its previous pericentric passage is estimated to have occurred $\sim 1$--1.5 Gyr ago \citep{2021MNRAS.501.2279V}.
Thus, our estimate of the elapsed time (despite uncertainties in the lag) agrees more closely with the Sgr pericentric passage than previous estimates that did not account for self-gravity effects.

Recently, \citet{2025arXiv250719579W} measured the properties of the phase spiral, including the winding time, across the local disc within $\lesssim 4$~kpc of the Sun using the \textit{Gaia} DR3 dataset.
They found that the observed winding time increases with $R_g$. This contrasts with our $N$-body simulation result, which showed that the azimuthally averaged winding time decreases with $R_g$ (see Fig.~\ref{fig:t_fit_t_Rg_abs}).
This discrepancy might arise from the observational selection function: the \textit{Gaia} data only cover a limited azimuthal range (roughly $\pm 25^{\circ}$ around the Sun-Galaxy centre line).
Figure~\ref{fig:Rg_theta_nb} illustrates a substantial variation in the winding time with azimuthal angle in our $N$-body model, suggesting that the dependence on $R_g$ may change when restricted to a limited azimuthal range rather than the entire disc.
This indicates that both the azimuthal and radial dependence of the time lag $\Delta t$ in Eq.~\eqref{eq:correction} must be taken into account in order to use the phase spiral as a reliable dynamical clock.

Finally, we note the possible effect of the gas disc, which is not included in our $N$-body model.
        The presence of gas alters the gravitational potential of the disc and can also modify the damping rate of disc oscillations through dissipative processes \citep{2022MNRAS.515.5951T}. 
        Due to these effects, the inclusion of a gas disc can change the time lag $\Delta t$.
        Furthermore, $N$-body+gas simulations by \citet{2025MNRAS.542.1987T} demonstrated that phase spirals can be generated by stochastic perturbations in  turbulent gas disc via the mechanism proposed by \citet{2023MNRAS.521..114T}.
        These results suggest that phase spirals in the MW disc might not be attributable to the Sgr impact alone.
        We need further investigations with high-resolution $N$-body+gas simulations to understand the formation and evolution of phase spirals in more realistic discs.

\section{Conclusions}
We investigated the time evolution of vertical phase spirals in the Galactic disc using high-resolution $N$-body and test-particle simulations of the MW-Sgr interaction.
Our goal was to quantify the bias introduced when self-gravitating effects are neglected in the dynamical modelling of vertical phase mixing in the Galactic disc.
\begin{enumerate}
        \item In contrast to test-particle simulations, where phase spirals start to wind up promptly after perturbation, the self-consistent $N$-body model exhibits a delayed response. In the $N$-body model, the dipole pattern in the vertical phase-space distribution, excited by Sgr, initially rotates almost rigidly before transitioning into a winding phase spiral. This non-winding phase (coherent oscillation) causes a systematic underestimation of the excitation time.

        \item  The DM wake induces a more gradual and sustained perturbation than the Sgr impulsive effect.
        Although it affects the phase spiral morphology, it cannot fully account for the delayed and less tightly wound spirals observed in the $N$-body model. This suggests that self-gravity and not the DM wake primarily causes the slower phase mixing observed in the $N$-body disc.

\item The winding time ($t_\mathrm{fit}$) inferred from the slope in the vertical angle--frequency ($\theta_z$--$\Omega_z$) space is consistently shorter than the true elapsed time since the Sgr pericentric passage.
    In the $N$-body model, $t_\mathrm{fit}$ generally increases with the guiding radius, although there is a substantial variation across the azimuthal angle.
    The discrepancy between the winding time and the true elapsed time reaches $\gtrsim 300$~Myr at a guiding radius of $R_g\sim 8$~kpc.

        \item The analytical self-gravitating model of \citet{2023MNRAS.522..477W} reproduces the delayed winding of our model qualitatively, supporting the conclusion that self-gravity plays a central role in shaping the evolution of phase spirals.

        \item In the $N$-body model, the amount of the self-gravity-induced delay is about three to four times the vertical epicycle period ($2\pi/\nu)$.
    When we assume that phase mixing is delayed on the same order in the MW, the time lag in the solar neighbourhood is estimated to be $\sim 0.3$~Gyr.
    When this is combined with the winding time of the observed phase spiral ($\sim$0.3--0.9~Gyr), the inferred elapsed time since the excitation of the bending mode increases to $\sim$0.6--1.2~Gyr.
                Although substantial variations and uncertainties remain, this revised estimate agrees better with the timing of the previous pericentric passage of Sgr and supports the hypothesis that Sgr triggered the observed phase spiral.
\end{enumerate}
We proposed a practical correction for the dynamical clock model of the phase spiral: $t_\mathrm{elapsed} \approx t_\mathrm{fit} + \Delta t$, where $t_\mathrm{fit}$ is the winding time inferred from kinematic analysis, and $\Delta t$ accounts for the non-winding phase caused by self-gravity.
Future work should explore how this time lag depends on the dynamical properties of the Galactic disc and assess the applicability of this correction using $N$-body models under varying conditions. It is essential to incorporate self-gravity for robustly tracing the dynamical history of the MW.

\begin{acknowledgements}
We thank the anonymous referee for constructive comments that improved this manuscript.
This project was developed in part at the ``Winding, Unwinding and Rewinding the Gaia Phase Spiral'' workshop hosted by the Lorentz Center in August 2025, Leiden, The Netherlands.
This research used computational resources of Pegasus and Cygnus provided by Multidisciplinary Cooperative Research Program in Center for Computational Sciences, University of Tsukuba.
We acknowledge the grants PID2021-125451NA-I00 and CNS2022-135232 funded by MICIU/AEI/10.13039/501100011033 and by ``ERDF A way of making Europe’’, by the ``European Union'' and by the ``European Union Next Generation EU/PRTR''.

This work made use of the following software packages: \texttt{astropy} \citep{2013A&A...558A..33A, 2018AJ....156..123A, 2022ApJ...935..167A}, \texttt{Jupyter} \citep{2007CSE.....9c..21P, 2016ppap.book...87K}, \texttt{matplotlib} \citep{2007CSE.....9...90H}, \texttt{numpy} \citep{numpy}, \texttt{pandas} \citep{mckinney-proc-scipy-2010, pandas_10537285}, \texttt{scipy} \citep{2020NatMe..17..261V, scipy_10155614}, \texttt{Agama} \citep{2019MNRAS.482.1525V},
and \texttt{opencv-python} (\url{https://github.com/opencv/opencv-python}).
This research has made use of NASA's Astrophysics Data System. Software citation information aggregated using \texttt{\href{https://www.tomwagg.com/software-citation-station/}{The Software Citation Station}} \citep{2024arXiv240604405W, software-citation-station-zenodo}.
\end{acknowledgements}

\bibliographystyle{aa}
\bibliography{extracted}

@article{2023MNRAS.522..477W,
  author =        {{Widrow}, Lawrence M.},
  journal =       {\mnras},
  month =         jun,
  number =        {1},
  pages =         {477-487},
  title =         {{Swing amplification and the Gaia phase spirals}},
  volume =        {522},
  year =          {2023},
  doi =           {10.1093/mnras/stad973},
}

@article{2018Natur.561..360A,
  author =        {{Antoja}, T. and {Helmi}, A. and
                   {Romero-G{\'o}mez}, M. and {Katz}, D. and
                   {Babusiaux}, C. and {Drimmel}, R. and {Evans}, D.~W. and
                   {Figueras}, F. and {Poggio}, E. and {Reyl{\'e}}, C. and
                   {Robin}, A.~C. and {Seabroke}, G. and {Soubiran}, C.},
  journal =       {\nat},
  month =         sep,
  number =        {7723},
  pages =         {360-362},
  title =         {{A dynamically young and perturbed Milky Way disk}},
  volume =        {561},
  year =          {2018},
  doi =           {10.1038/s41586-018-0510-7},
}

@article{2016A&A...595A...1G,
  author =        {{Gaia Collaboration} and {Prusti}, T. and
                   {de Bruijne}, J.~H.~J. and {Brown}, A.~G.~A. and
                   {Vallenari}, A. and {Babusiaux}, C. and
                   {Bailer-Jones}, C.~A.~L. and {Bastian}, U. and
                   {Biermann}, M. and {Evans}, D.~W. and {Eyer}, L. and
                   {Jansen}, F. and {Jordi}, C. and {Klioner}, S.~A. and
                   {Lammers}, U. and {Lindegren}, L. and {Luri}, X. and
                   {Mignard}, F. and {Milligan}, D.~J. and {Panem}, C. and
                   {Poinsignon}, V. and {Pourbaix}, D. and {Randich}, S. and
                   {Sarri}, G. and {Sartoretti}, P. and
                   {Siddiqui}, H.~I. and {Soubiran}, C. and
                   {Valette}, V. and {van Leeuwen}, F. and
                   {Walton}, N.~A. and {Aerts}, C. and {Arenou}, F. and
                   {Cropper}, M. and {Drimmel}, R. and {H{\o}g}, E. and
                   {Katz}, D. and {Lattanzi}, M.~G. and {O'Mullane}, W. and
                   {Grebel}, E.~K. and {Holland}, A.~D. and {Huc}, C. and
                   {Passot}, X. and {Bramante}, L. and {Cacciari}, C. and
                   {Casta{\~n}eda}, J. and {Chaoul}, L. and {Cheek}, N. and
                   {De Angeli}, F. and {Fabricius}, C. and {Guerra}, R. and
                   {Hern{\'a}ndez}, J. and {Jean-Antoine-Piccolo}, A. and
                   {Masana}, E. and {Messineo}, R. and {Mowlavi}, N. and
                   {Nienartowicz}, K. and {Ord{\'o}{\~n}ez-Blanco}, D. and
                   {Panuzzo}, P. and {Portell}, J. and {Richards}, P.~J. and
                   {Riello}, M. and {Seabroke}, G.~M. and {Tanga}, P. and
                   {Th{\'e}venin}, F. and {Torra}, J. and {Els}, S.~G. and
                   {Gracia-Abril}, G. and {Comoretto}, G. and
                   {Garcia-Reinaldos}, M. and {Lock}, T. and
                   {Mercier}, E. and {Altmann}, M. and {Andrae}, R. and
                   {Astraatmadja}, T.~L. and {Bellas-Velidis}, I. and
                   {Benson}, K. and {Berthier}, J. and {Blomme}, R. and
                   {Busso}, G. and {Carry}, B. and {Cellino}, A. and
                   {Clementini}, G. and {Cowell}, S. and {Creevey}, O. and
                   {Cuypers}, J. and {Davidson}, M. and {De Ridder}, J. and
                   {de Torres}, A. and {Delchambre}, L. and
                   {Dell'Oro}, A. and {Ducourant}, C. and
                   {Fr{\'e}mat}, Y. and {Garc{\'\i}a-Torres}, M. and
                   {Gosset}, E. and {Halbwachs}, J.-L. and
                   {Hambly}, N.~C. and {Harrison}, D.~L. and
                   {Hauser}, M. and {Hestroffer}, D. and
                   {Hodgkin}, S.~T. and {Huckle}, H.~E. and {Hutton}, A. and
                   {Jasniewicz}, G. and {Jordan}, S. and {Kontizas}, M. and
                   {Korn}, A.~J. and {Lanzafame}, A.~C. and
                   {Manteiga}, M. and {Moitinho}, A. and {Muinonen}, K. and
                   {Osinde}, J. and {Pancino}, E. and {Pauwels}, T. and
                   {Petit}, J.-M. and {Recio-Blanco}, A. and
                   {Robin}, A.~C. and {Sarro}, L.~M. and {Siopis}, C. and
                   {Smith}, M. and {Smith}, K.~W. and {Sozzetti}, A. and
                   {Thuillot}, W. and {van Reeven}, W. and {Viala}, Y. and
                   {Abbas}, U. and {Abreu Aramburu}, A. and {Accart}, S. and
                   {Aguado}, J.~J. and {Allan}, P.~M. and {Allasia}, W. and
                   {Altavilla}, G. and {{\'A}lvarez}, M.~A. and
                   {Alves}, J. and {Anderson}, R.~I. and {Andrei}, A.~H. and
                   {Anglada Varela}, E. and {Antiche}, E. and
                   {Antoja}, T. and {Ant{\'o}n}, S. and {Arcay}, B. and
                   {Atzei}, A. and {Ayache}, L. and {Bach}, N. and
                   {Baker}, S.~G. and {Balaguer-N{\'u}{\~n}ez}, L. and
                   {Barache}, C. and {Barata}, C. and {Barbier}, A. and
                   {Barblan}, F. and {Baroni}, M. and
                   {Barrado y Navascu{\'e}s}, D. and {Barros}, M. and
                   {Barstow}, M.~A. and {Becciani}, U. and
                   {Bellazzini}, M. and {Bellei}, G. and
                   {Bello Garc{\'\i}a}, A. and {Belokurov}, V. and
                   {Bendjoya}, P. and {Berihuete}, A. and {Bianchi}, L. and
                   {Bienaym{\'e}}, O. and {Billebaud}, F. and
                   {Blagorodnova}, N. and {Blanco-Cuaresma}, S. and
                   {Boch}, T. and {Bombrun}, A. and {Borrachero}, R. and
                   {Bouquillon}, S. and {Bourda}, G. and {Bouy}, H. and
                   {Bragaglia}, A. and {Breddels}, M.~A. and
                   {Brouillet}, N. and {Br{\"u}semeister}, T. and
                   {Bucciarelli}, B. and {Budnik}, F. and {Burgess}, P. and
                   {Burgon}, R. and {Burlacu}, A. and {Busonero}, D. and
                   {Buzzi}, R. and {Caffau}, E. and {Cambras}, J. and
                   {Campbell}, H. and {Cancelliere}, R. and
                   {Cantat-Gaudin}, T. and {Carlucci}, T. and
                   {Carrasco}, J.~M. and {Castellani}, M. and
                   {Charlot}, P. and {Charnas}, J. and {Charvet}, P. and
                   {Chassat}, F. and {Chiavassa}, A. and {Clotet}, M. and
                   {Cocozza}, G. and {Collins}, R.~S. and {Collins}, P. and
                   {Costigan}, G.},
  journal =       {\aap},
  month =         nov,
  pages =         {A1},
  title =         {{The Gaia mission}},
  volume =        {595},
  year =          {2016},
  doi =           {10.1051/0004-6361/201629272},
  eid =           {A1},
}

@article{2018A&A...616A...1G,
  author =        {{Gaia Collaboration} and {Brown}, A.~G.~A. and
                   {Vallenari}, A. and {Prusti}, T. and
                   {de Bruijne}, J.~H.~J. and {Babusiaux}, C. and
                   {Bailer-Jones}, C.~A.~L. and {Biermann}, M. and
                   {Evans}, D.~W. and {Eyer}, L. and {Jansen}, F. and
                   {Jordi}, C. and {Klioner}, S.~A. and {Lammers}, U. and
                   {Lindegren}, L. and {Luri}, X. and {Mignard}, F. and
                   {Panem}, C. and {Pourbaix}, D. and {Randich}, S. and
                   {Sartoretti}, P. and {Siddiqui}, H.~I. and
                   {Soubiran}, C. and {van Leeuwen}, F. and
                   {Walton}, N.~A. and {Arenou}, F. and {Bastian}, U. and
                   {Cropper}, M. and {Drimmel}, R. and {Katz}, D. and
                   {Lattanzi}, M.~G. and {Bakker}, J. and {Cacciari}, C. and
                   {Casta{\~n}eda}, J. and {Chaoul}, L. and {Cheek}, N. and
                   {De Angeli}, F. and {Fabricius}, C. and {Guerra}, R. and
                   {Holl}, B. and {Masana}, E. and {Messineo}, R. and
                   {Mowlavi}, N. and {Nienartowicz}, K. and
                   {Panuzzo}, P. and {Portell}, J. and {Riello}, M. and
                   {Seabroke}, G.~M. and {Tanga}, P. and
                   {Th{\'e}venin}, F. and {Gracia-Abril}, G. and
                   {Comoretto}, G. and {Garcia-Reinaldos}, M. and
                   {Teyssier}, D. and {Altmann}, M. and {Andrae}, R. and
                   {Audard}, M. and {Bellas-Velidis}, I. and
                   {Benson}, K. and {Berthier}, J. and {Blomme}, R. and
                   {Burgess}, P. and {Busso}, G. and {Carry}, B. and
                   {Cellino}, A. and {Clementini}, G. and {Clotet}, M. and
                   {Creevey}, O. and {Davidson}, M. and {De Ridder}, J. and
                   {Delchambre}, L. and {Dell'Oro}, A. and
                   {Ducourant}, C. and {Fern{\'a}ndez-Hern{\'a}ndez}, J. and
                   {Fouesneau}, M. and {Fr{\'e}mat}, Y. and
                   {Galluccio}, L. and {Garc{\'\i}a-Torres}, M. and
                   {Gonz{\'a}lez-N{\'u}{\~n}ez}, J. and
                   {Gonz{\'a}lez-Vidal}, J.~J. and {Gosset}, E. and
                   {Guy}, L.~P. and {Halbwachs}, J.-L. and
                   {Hambly}, N.~C. and {Harrison}, D.~L. and
                   {Hern{\'a}ndez}, J. and {Hestroffer}, D. and
                   {Hodgkin}, S.~T. and {Hutton}, A. and
                   {Jasniewicz}, G. and {Jean-Antoine-Piccolo}, A. and
                   {Jordan}, S. and {Korn}, A.~J. and
                   {Krone-Martins}, A. and {Lanzafame}, A.~C. and
                   {Lebzelter}, T. and {L{\"o}ffler}, W. and
                   {Manteiga}, M. and {Marrese}, P.~M. and
                   {Mart{\'\i}n-Fleitas}, J.~M. and {Moitinho}, A. and
                   {Mora}, A. and {Muinonen}, K. and {Osinde}, J. and
                   {Pancino}, E. and {Pauwels}, T. and {Petit}, J.-M. and
                   {Recio-Blanco}, A. and {Richards}, P.~J. and
                   {Rimoldini}, L. and {Robin}, A.~C. and {Sarro}, L.~M. and
                   {Siopis}, C. and {Smith}, M. and {Sozzetti}, A. and
                   {S{\"u}veges}, M. and {Torra}, J. and
                   {van Reeven}, W. and {Abbas}, U. and
                   {Abreu Aramburu}, A. and {Accart}, S. and {Aerts}, C. and
                   {Altavilla}, G. and {{\'A}lvarez}, M.~A. and
                   {Alvarez}, R. and {Alves}, J. and {Anderson}, R.~I. and
                   {Andrei}, A.~H. and {Anglada Varela}, E. and
                   {Antiche}, E. and {Antoja}, T. and {Arcay}, B. and
                   {Astraatmadja}, T.~L. and {Bach}, N. and
                   {Baker}, S.~G. and {Balaguer-N{\'u}{\~n}ez}, L. and
                   {Balm}, P. and {Barache}, C. and {Barata}, C. and
                   {Barbato}, D. and {Barblan}, F. and {Barklem}, P.~S. and
                   {Barrado}, D. and {Barros}, M. and {Barstow}, M.~A. and
                   {Bartholom{\'e} Mu{\~n}oz}, S. and {Bassilana}, J.-L. and
                   {Becciani}, U. and {Bellazzini}, M. and
                   {Berihuete}, A. and {Bertone}, S. and {Bianchi}, L. and
                   {Bienaym{\'e}}, O. and {Blanco-Cuaresma}, S. and
                   {Boch}, T. and {Boeche}, C. and {Bombrun}, A. and
                   {Borrachero}, R. and {Bossini}, D. and
                   {Bouquillon}, S. and {Bourda}, G. and {Bragaglia}, A. and
                   {Bramante}, L. and {Breddels}, M.~A. and
                   {Bressan}, A. and {Brouillet}, N. and
                   {Br{\"u}semeister}, T. and {Brugaletta}, E. and
                   {Bucciarelli}, B. and {Burlacu}, A. and
                   {Busonero}, D. and {Butkevich}, A.~G. and {Buzzi}, R. and
                   {Caffau}, E. and {Cancelliere}, R. and
                   {Cannizzaro}, G. and {Cantat-Gaudin}, T. and
                   {Carballo}, R. and {Carlucci}, T. and
                   {Carrasco}, J.~M. and {Casamiquela}, L. and
                   {Castellani}, M. and {Castro-Ginard}, A. and
                   {Charlot}, P. and {Chemin}, L. and {Chiavassa}, A. and
                   {Cocozza}, G. and {Costigan}, G. and {Cowell}, S. and
                   {Crifo}, F. and {Crosta}, M. and {Crowley}, C. and
                   {Cuypers}, J. and {Dafonte}, C. and {Damerdji}, Y. and
                   {Dapergolas}, A. and {David}, P. and {David}, M. and
                   {de Laverny}, P. and {De Luise}, F.},
  journal =       {\aap},
  month =         aug,
  pages =         {A1},
  title =         {{Gaia Data Release 2. Summary of the contents and
                   survey properties}},
  volume =        {616},
  year =          {2018},
  doi =           {10.1051/0004-6361/201833051},
  eid =           {A1},
}

@article{2025NewAR.10001721H,
  author =        {{Hunt}, Jason A.~S. and {Vasiliev}, Eugene},
  journal =       {\nar},
  month =         jun,
  pages =         {101721},
  title =         {{Milky Way dynamics in light of Gaia}},
  volume =        {100},
  year =          {2025},
  doi =           {10.1016/j.newar.2024.101721},
  eid =           {101721},
}

@article{1994Natur.370..194I,
  author =        {{Ibata}, R.~A. and {Gilmore}, G. and {Irwin}, M.~J.},
  journal =       {\nat},
  month =         jul,
  number =        {6486},
  pages =         {194-196},
  title =         {{A dwarf satellite galaxy in Sagittarius}},
  volume =        {370},
  year =          {1994},
  doi =           {10.1038/370194a0},
}

@article{2010ApJ...714..229L,
  author =        {{Law}, David R. and {Majewski}, Steven R.},
  journal =       {\apj},
  month =         may,
  number =        {1},
  pages =         {229-254},
  title =         {{The Sagittarius Dwarf Galaxy: A Model for Evolution
                   in a Triaxial Milky Way Halo}},
  volume =        {714},
  year =          {2010},
  doi =           {10.1088/0004-637X/714/1/229},
}

@article{2011Natur.477..301P,
  author =        {{Purcell}, Chris W. and {Bullock}, James S. and
                   {Tollerud}, Erik J. and {Rocha}, Miguel and
                   {Chakrabarti}, Sukanya},
  journal =       {\nat},
  month =         sep,
  number =        {7364},
  pages =         {301-303},
  title =         {{The Sagittarius impact as an architect of spirality
                   and outer rings in the Milky Way}},
  volume =        {477},
  year =          {2011},
  doi =           {10.1038/nature10417},
}

@article{2017ApJ...836...92D,
  author =        {{Dierickx}, Marion I.~P. and {Loeb}, Abraham},
  journal =       {\apj},
  month =         feb,
  number =        {1},
  pages =         {92},
  title =         {{Predicted Extension of the Sagittarius Stream to the
                   Milky Way Virial Radius}},
  volume =        {836},
  year =          {2017},
  doi =           {10.3847/1538-4357/836/1/92},
  eid =           {92},
}

@article{2021MNRAS.501.2279V,
  author =        {{Vasiliev}, Eugene and {Belokurov}, Vasily and
                   {Erkal}, Denis},
  journal =       {\mnras},
  month =         feb,
  number =        {2},
  pages =         {2279-2304},
  title =         {{Tango for three: Sagittarius, LMC, and the Milky
                   Way}},
  volume =        {501},
  year =          {2021},
  doi =           {10.1093/mnras/staa3673},
}

@article{2018MNRAS.481.1501B,
  author =        {{Binney}, James and {Sch{\"o}nrich}, Ralph},
  journal =       {\mnras},
  month =         dec,
  number =        {2},
  pages =         {1501-1506},
  title =         {{The origin of the Gaia phase-plane spiral}},
  volume =        {481},
  year =          {2018},
  doi =           {10.1093/mnras/sty2378},
}

@article{2021MNRAS.503..376B,
  author =        {{Bennett}, Morgan and {Bovy}, Jo},
  journal =       {\mnras},
  month =         may,
  number =        {1},
  pages =         {376-393},
  title =         {{Did Sgr cause the vertical waves in the solar
                   neighbourhood?}},
  volume =        {503},
  year =          {2021},
  doi =           {10.1093/mnras/stab524},
}

@article{2022ApJ...935..135B,
  author =        {{Banik}, Uddipan and {Weinberg}, Martin D. and
                   {van den Bosch}, Frank C.},
  journal =       {\apj},
  month =         aug,
  number =        {2},
  pages =         {135},
  title =         {{A Comprehensive Perturbative Formalism for Phase
                   Mixing in Perturbed Disks. I. Phase Spirals in an
                   Infinite, Isothermal Slab}},
  volume =        {935},
  year =          {2022},
  doi =           {10.3847/1538-4357/ac7ff9},
  eid =           {135},
}

@article{2023ApJ...952...65B,
  author =        {{Banik}, Uddipan and {van den Bosch}, Frank C. and
                   {Weinberg}, Martin D.},
  journal =       {\apj},
  month =         jul,
  number =        {1},
  pages =         {65},
  title =         {{A Comprehensive Perturbative Formalism for Phase
                   Mixing in Perturbed Disks. II. Phase Spirals in an
                   Inhomogeneous Disk Galaxy with a Nonresponsive Dark
                   Matter Halo}},
  volume =        {952},
  year =          {2023},
  doi =           {10.3847/1538-4357/acd641},
  eid =           {65},
}

@article{2021ApJ...911..107L,
  author =        {{Li}, Zhao-Yu},
  journal =       {\apj},
  month =         apr,
  number =        {2},
  pages =         {107},
  title =         {{Vertical Phase Mixing across the Galactic Disk}},
  volume =        {911},
  year =          {2021},
  doi =           {10.3847/1538-4357/abea17},
  eid =           {107},
}

@article{2022ApJ...928...80G,
  author =        {{Gandhi}, Suroor S. and {Johnston}, Kathryn V. and
                   {Hunt}, Jason A.~S. and {Price-Whelan}, Adrian M. and
                   {Laporte}, Chervin F.~P. and {Hogg}, David W.},
  journal =       {\apj},
  month =         mar,
  number =        {1},
  pages =         {80},
  title =         {{Snails across Scales: Local and Global Phase-mixing
                   Structures as Probes of the Past and Future Milky
                   Way}},
  volume =        {928},
  year =          {2022},
  doi =           {10.3847/1538-4357/ac47f7},
  eid =           {80},
}

@article{2019MNRAS.485.3134L,
  author =        {{Laporte}, Chervin F.~P. and {Minchev}, Ivan and
                   {Johnston}, Kathryn V. and {G{\'o}mez}, Facundo A.},
  journal =       {\mnras},
  month =         may,
  number =        {3},
  pages =         {3134-3152},
  title =         {{Footprints of the Sagittarius dwarf galaxy in the
                   Gaia data set}},
  volume =        {485},
  year =          {2019},
  doi =           {10.1093/mnras/stz583},
}

@article{2019MNRAS.486.1167B,
  author =        {{Bland-Hawthorn}, Joss and {Sharma}, Sanjib and
                   {Tepper-Garcia}, Thor and {Binney}, James and
                   {Freeman}, Ken C. and {Hayden}, Michael R. and
                   {Kos}, Janez and {De Silva}, Gayandhi M. and
                   {Ellis}, Simon and {Lewis}, Geraint F. and
                   {Asplund}, Martin and {Buder}, Sven and
                   {Casey}, Andrew R. and {D'Orazi}, Valentina and
                   {Duong}, Ly and {Khanna}, Shourya and {Lin}, Jane and
                   {Lind}, Karin and {Martell}, Sarah L. and
                   {Ness}, Melissa K. and {Simpson}, Jeffrey D. and
                   {Zucker}, Daniel B. and {Zwitter}, Toma{\v{z}} and
                   {Kafle}, Prajwal R. and {Quillen}, Alice C. and
                   {Ting}, Yuan-Sen and {Wyse}, Rosemary F.~G.},
  journal =       {\mnras},
  month =         jun,
  number =        {1},
  pages =         {1167-1191},
  title =         {{The GALAH survey and Gaia DR2: dissecting the
                   stellar disc's phase space by age, action, chemistry,
                   and location}},
  volume =        {486},
  year =          {2019},
  doi =           {10.1093/mnras/stz217},
}

@article{2021MNRAS.504.3168B,
  author =        {{Bland-Hawthorn}, Joss and
                   {Tepper-Garc{\'\i}a}, Thor},
  journal =       {\mnras},
  month =         jul,
  number =        {3},
  pages =         {3168-3186},
  title =         {{Galactic seismology: the evolving 'phase spiral'
                   after the Sagittarius dwarf impact}},
  volume =        {504},
  year =          {2021},
  doi =           {10.1093/mnras/stab704},
}

@article{2021MNRAS.508.1459H,
  author =        {{Hunt}, Jason A.~S. and {Stelea}, Ioana A. and
                   {Johnston}, Kathryn V. and {Gandhi}, Suroor S. and
                   {Laporte}, Chervin F.~P. and {B{\'e}dorf}, Jeroen},
  journal =       {\mnras},
  month =         nov,
  number =        {1},
  pages =         {1459-1472},
  title =         {{Resolving local and global kinematic signatures of
                   satellite mergers with billion particle simulations}},
  volume =        {508},
  year =          {2021},
  doi =           {10.1093/mnras/stab2580},
}

@article{2022ApJ...927..131B,
  author =        {{Bennett}, Morgan and {Bovy}, Jo and
                   {Hunt}, Jason A.~S.},
  journal =       {\apj},
  month =         mar,
  number =        {1},
  pages =         {131},
  title =         {{Exploring the Sgr-Milky Way-disk Interaction Using
                   High-resolution N-body Simulations}},
  volume =        {927},
  year =          {2022},
  doi =           {10.3847/1538-4357/ac5021},
  eid =           {131},
}

@article{2025A&A...700A.109A,
  author =        {{Asano}, Tetsuro and {Fujii}, Michiko S. and
                   {Baba}, Junichi and {Portegies Zwart}, Simon and
                   {B{\'e}dorf}, Jeroen},
  journal =       {\aap},
  month =         aug,
  pages =         {A109},
  title =         {{Ripples spreading across the Galactic disc:
                   Interplay of direct and indirect effects of the
                   Sagittarius dwarf impact}},
  volume =        {700},
  year =          {2025},
  doi =           {10.1051/0004-6361/202553816},
  eid =           {A109},
}

@article{2025MNRAS.542.1987T,
  author =        {{Tepper-Garc{\'\i}a}, Thor and {Bland-Hawthorn}, Joss and
                   {Bedding}, Timothy R. and {Federrath}, Christoph and
                   {Agertz}, Oscar},
  journal =       {\mnras},
  month =         sep,
  number =        {3},
  pages =         {1987-2003},
  title =         {{Galactic seismology: can the Gaia 'phase spiral'
                   co-exist with a clumpy, turbulent interstellar
                   medium?}},
  volume =        {542},
  year =          {2025},
  doi =           {10.1093/mnras/staf1331},
}

@article{2022MNRAS.510..154G,
  author =        {{Garc{\'\i}a-Conde}, B. and {Roca-F{\`a}brega}, S. and
                   {Antoja}, T. and {Ramos}, P. and {Valenzuela}, O.},
  journal =       {\mnras},
  month =         feb,
  number =        {1},
  pages =         {154-160},
  title =         {{Phase spirals in cosmological simulations of Milky
                   Way-sized galaxies}},
  volume =        {510},
  year =          {2022},
  doi =           {10.1093/mnras/stab3417},
}

@article{2023MNRAS.524..801G,
  author =        {{Grand}, Robert J.~J. and {Pakmor}, R{\"u}diger and
                   {Fragkoudi}, Francesca and {G{\'o}mez}, Facundo A. and
                   {Trick}, Wilma and {Simpson}, Christine M. and
                   {van de Voort}, Freeke and {Bieri}, Rebekka},
  journal =       {\mnras},
  month =         sep,
  number =        {1},
  pages =         {801-816},
  title =         {{An ever-present Gaia snail shell triggered by a dark
                   matter wake}},
  volume =        {524},
  year =          {2023},
  doi =           {10.1093/mnras/stad1969},
}

@article{2023MNRAS.521..114T,
  author =        {{Tremaine}, Scott and {Frankel}, Neige and
                   {Bovy}, Jo},
  journal =       {\mnras},
  month =         may,
  number =        {1},
  pages =         {114-123},
  title =         {{The origin and fate of the Gaia phase-space snail}},
  volume =        {521},
  year =          {2023},
  doi =           {10.1093/mnras/stad577},
}

@article{2025ApJ...980...24G,
  author =        {{Gilman}, Daniel and {Bovy}, Jo and {Frankel}, Neige and
                   {Benson}, Andrew},
  journal =       {\apj},
  month =         feb,
  number =        {1},
  pages =         {24},
  title =         {{Dark Galactic Subhalos and the Gaia Snail}},
  volume =        {980},
  year =          {2025},
  doi =           {10.3847/1538-4357/ada963},
  eid =           {24},
}

@article{2019A&A...622L...6K,
  author =        {{Khoperskov}, Sergey and {Di Matteo}, Paola and
                   {Gerhard}, Ortwin and {Katz}, David and
                   {Haywood}, Misha and {Combes}, Fran{\c{c}}oise and
                   {Berczik}, Peter and {Gomez}, Ana},
  journal =       {\aap},
  month =         feb,
  pages =         {L6},
  title =         {{The echo of the bar buckling: Phase-space spirals in
                   Gaia Data Release 2}},
  volume =        {622},
  year =          {2019},
  doi =           {10.1051/0004-6361/201834707},
  eid =           {L6},
}

@article{1967PhRvL..19..219G,
  author =        {{Gould}, R.~W. and {O'Neil}, T.~M. and
                   {Malmberg}, J.~H.},
  journal =       {\prl},
  month =         jul,
  number =        {5},
  pages =         {219-222},
  title =         {{Plasma Wave Echo}},
  volume =        {19},
  year =          {1967},
  doi =           {10.1103/PhysRevLett.19.219},
}

@article{2025MNRAS.543..190C,
  author =        {{Chiba}, Rimpei and {Ding}, Jupiter and
                   {Hamilton}, Chris and {Kunz}, Matthew W. and
                   {Tremaine}, Scott},
  journal =       {\mnras},
  month =         oct,
  number =        {1},
  pages =         {190-201},
  title =         {{Galactic echoes}},
  volume =        {543},
  year =          {2025},
  doi =           {10.1093/mnras/staf1463},
}

@article{2024A&A...683A..47G,
  author =        {{Garc{\'\i}a-Conde}, B. and {Antoja}, T. and
                   {Roca-F{\`a}brega}, S. and {G{\'o}mez}, F. and
                   {Ramos}, P. and {Garavito-Camargo}, N. and
                   {G{\'o}mez-Flechoso}, M.~A.},
  journal =       {\aap},
  month =         mar,
  pages =         {A47},
  title =         {{Galactoseismology in cosmological simulations.
                   Vertical perturbations by dark matter, satellite
                   galaxies, and gas}},
  volume =        {683},
  year =          {2024},
  doi =           {10.1051/0004-6361/202347446},
  eid =           {A47},
}

@article{2021MNRAS.503.1586L,
  author =        {{Li}, Haochuan and {Widrow}, Lawrence M.},
  journal =       {\mnras},
  month =         may,
  number =        {2},
  pages =         {1586-1598},
  title =         {{The stellar distribution function and local vertical
                   potential from Gaia DR2}},
  volume =        {503},
  year =          {2021},
  doi =           {10.1093/mnras/stab574},
}

@article{2023ApJ...955...74D,
  author =        {{Darragh-Ford}, Elise and {Hunt}, Jason A.~S. and
                   {Price-Whelan}, Adrian M. and {Johnston}, Kathryn V.},
  journal =       {\apj},
  month =         sep,
  number =        {1},
  pages =         {74},
  title =         {{ESCARGOT: Mapping Vertical Phase Spiral
                   Characteristics Throughout the Real and Simulated
                   Milky Way}},
  volume =        {955},
  year =          {2023},
  doi =           {10.3847/1538-4357/acf1fc},
  eid =           {74},
}

@article{2023MNRAS.521.5917F,
  author =        {{Frankel}, Neige and {Bovy}, Jo and {Tremaine}, Scott and
                   {Hogg}, David W.},
  journal =       {\mnras},
  month =         jun,
  number =        {4},
  pages =         {5917-5926},
  title =         {{Vertical motion in the Galactic disc: unwinding the
                   snail}},
  volume =        {521},
  year =          {2023},
  doi =           {10.1093/mnras/stad908},
}

@article{2025ApJ...988..254L,
  author =        {{Lin}, Junxian and {Li}, Zhao-Yu and {Guo}, Rui and
                   {Hunt}, Jason A.~S. and {Antoja}, T. and
                   {Cao}, Chengye},
  journal =       {\apj},
  month =         aug,
  number =        {2},
  pages =         {254},
  title =         {{Formation of the Two-armed Phase Spiral from
                   Multiple External Perturbations}},
  volume =        {988},
  year =          {2025},
  doi =           {10.3847/1538-4357/adea70},
  eid =           {254},
}

@article{2023A&A...673A.115A,
  author =        {{Antoja}, T. and {Ramos}, P. and
                   {Garc{\'\i}a-Conde}, B. and {Bernet}, M. and
                   {Laporte}, C.~F.~P. and {Katz}, D.},
  journal =       {\aap},
  month =         may,
  pages =         {A115},
  title =         {{The phase spiral in Gaia DR3}},
  volume =        {673},
  year =          {2023},
  doi =           {10.1051/0004-6361/202245518},
  eid =           {A115},
}

@article{2025ApJ...987...81F,
  author =        {{Frankel}, Neige and {Hogg}, David W. and
                   {Tremaine}, Scott and {Price-Whelan}, Adrian and
                   {Shen}, Jeff},
  journal =       {\apj},
  month =         jul,
  number =        {1},
  pages =         {81},
  title =         {{Iron Snails: Nonequilibrium Dynamics and Spiral
                   Abundance Patterns}},
  volume =        {987},
  year =          {2025},
  doi =           {10.3847/1538-4357/add5ea},
  eid =           {81},
}

@article{2025arXiv250719579W,
  author =        {{Widmark}, Axel and {Tavangar}, Kiyan and
                   {Kalish}, Josh and {Johnston}, Kathryn V. and
                   {Hunt}, Jason A.~S.},
  journal =       {arXiv e-prints},
  month =         jul,
  pages =         {arXiv:2507.19579},
  title =         {{The phase spiral's origin and evolution: indications
                   from its varying properties across the Milky Way
                   disk}},
  year =          {2025},
  doi =           {10.48550/arXiv.2507.19579},
  eid =           {arXiv:2507.19579},
}

@article{2019MNRAS.484.1050D,
  author =        {{Darling}, Keir and {Widrow}, Lawrence M.},
  journal =       {\mnras},
  month =         mar,
  number =        {1},
  pages =         {1050-1056},
  title =         {{Emergence of the Gaia phase space spirals from
                   bending waves}},
  volume =        {484},
  year =          {2019},
  doi =           {10.1093/mnras/sty3508},
}

@article{1990ApJ...356..359H,
  author =        {{Hernquist}, Lars},
  journal =       {\apj},
  month =         jun,
  pages =         {359},
  title =         {{An Analytical Model for Spherical Galaxies and
                   Bulges}},
  volume =        {356},
  year =          {1990},
  doi =           {10.1086/168845},
}

@article{1997ApJ...490..493N,
  author =        {{Navarro}, Julio F. and {Frenk}, Carlos S. and
                   {White}, Simon D.~M.},
  journal =       {\apj},
  month =         dec,
  number =        {2},
  pages =         {493-508},
  title =         {{A Universal Density Profile from Hierarchical
                   Clustering}},
  volume =        {490},
  year =          {1997},
  doi =           {10.1086/304888},
}

@article{1995MNRAS.277.1341K,
  author =        {{Kuijken}, K. and {Dubinski}, J.},
  journal =       {\mnras},
  month =         dec,
  pages =         {1341},
  title =         {{Nearly Self-Consistent Disc / Bulge / Halo Models
                   for Galaxies}},
  volume =        {277},
  year =          {1995},
  doi =           {10.1093/mnras/277.4.1341},
}

@article{2005ApJ...631..838W,
  author =        {{Widrow}, Lawrence M. and {Dubinski}, John},
  journal =       {\apj},
  month =         oct,
  number =        {2},
  pages =         {838-855},
  title =         {{Equilibrium Disk-Bulge-Halo Models for the Milky Way
                   and Andromeda Galaxies}},
  volume =        {631},
  year =          {2005},
  doi =           {10.1086/432710},
}

@article{2008ApJ...679.1239W,
  author =        {{Widrow}, Lawrence M. and {Pym}, Brent and
                   {Dubinski}, John},
  journal =       {\apj},
  month =         jun,
  number =        {2},
  pages =         {1239-1259},
  title =         {{Dynamical Blueprints for Galaxies}},
  volume =        {679},
  year =          {2008},
  doi =           {10.1086/587636},
}

@article{2019MNRAS.482.1983F,
  author =        {{Fujii}, M.~S. and {B{\'e}dorf}, J. and {Baba}, J. and
                   {Portegies Zwart}, S.},
  journal =       {\mnras},
  month =         jan,
  number =        {2},
  pages =         {1983-2015},
  title =         {{Modelling the Milky Way as a dry Galaxy}},
  volume =        {482},
  year =          {2019},
  doi =           {10.1093/mnras/sty2747},
}

@article{2012JCoPh.231.2825B,
  author =        {{B{\'e}dorf}, Jeroen and {Gaburov}, Evghenii and
                   {Portegies Zwart}, Simon},
  journal =       {Journal of Computational Physics},
  month =         apr,
  number =        {7},
  pages =         {2825-2839},
  title =         {{A sparse octree gravitational N-body code that runs
                   entirely on the GPU processor}},
  volume =        {231},
  year =          {2012},
  doi =           {10.1016/j.jcp.2011.12.024},
}

@inproceedings{2014hpcn.conf...54B,
  author =        {{B{\'e}dorf}, Jeroen and {Gaburov}, Evghenii and
                   {Fujii}, Michiko S. and {Nitadori}, Keigo and
                   {Ishiyama}, Tomoaki and {Portegies Zwart}, Simon},
  booktitle =     {Proceedings of the International Conference for High
                   Performance Computing},
  month =         nov,
  pages =         {54-65},
  title =         {{24.77 Pflops on a Gravitational Tree-Code to
                   Simulate the Milky Way Galaxy with 18600 GPUs}},
  year =          {2014},
  doi =           {10.1109/SC.2014.10},
}

@article{2022MNRAS.515.5951T,
  author =        {{Tepper-Garc{\'\i}a}, Thor and {Bland-Hawthorn}, Joss and
                   {Freeman}, Ken},
  journal =       {\mnras},
  month =         oct,
  number =        {4},
  pages =         {5951-5968},
  title =         {{Galactic seismology: joint evolution of
                   impact-triggered stellar and gaseous disc
                   corrugations}},
  volume =        {515},
  year =          {2022},
  doi =           {10.1093/mnras/stac1926},
}

@article{2019MNRAS.482.1525V,
  author =        {{Vasiliev}, Eugene},
  journal =       {\mnras},
  month =         jan,
  number =        {2},
  pages =         {1525-1544},
  title =         {{AGAMA: action-based galaxy modelling architecture}},
  volume =        {482},
  year =          {2019},
  doi =           {10.1093/mnras/sty2672},
}

@article{2012MNRAS.426.1324B,
  author =        {{Binney}, James},
  journal =       {\mnras},
  month =         oct,
  number =        {2},
  pages =         {1324-1327},
  title =         {{Actions for axisymmetric potentials}},
  volume =        {426},
  year =          {2012},
  doi =           {10.1111/j.1365-2966.2012.21757.x},
}

@article{2025A&A...702A.223B,
  author =        {{Bernet}, Marcel and {Ramos}, Pau and
                   {Antoja}, Teresa and {Price-Whelan}, Adrian and
                   {Brunton}, Steven L. and {Asano}, Tetsuro and
                   {Gir{\'o}n-Soto}, Alexandra},
  journal =       {\aap},
  month =         oct,
  pages =         {A223},
  title =         {{Dynamics of tidal spiral arms: Machine
                   learning-assisted identification of equations and
                   application to the Milky Way}},
  volume =        {702},
  year =          {2025},
  doi =           {10.1051/0004-6361/202556039},
  eid =           {A223},
}

@article{2018MNRAS.481..286L,
  author =        {{Laporte}, Chervin F.~P. and {Johnston}, Kathryn V. and
                   {G{\'o}mez}, Facundo A. and
                   {Garavito-Camargo}, Nicolas and {Besla}, Gurtina},
  journal =       {\mnras},
  month =         nov,
  number =        {1},
  pages =         {286-306},
  title =         {{The influence of Sagittarius and the Large
                   Magellanic Cloud on the stellar disc of the Milky Way
                   Galaxy}},
  volume =        {481},
  year =          {2018},
  doi =           {10.1093/mnras/sty1574},
}

@article{2021MNRAS.507.2825G,
  author =        {{Grion Filho}, Douglas and {Johnston}, Kathryn V. and
                   {Poggio}, Eloisa and {Laporte}, Chervin F.~P. and
                   {Drimmel}, Ronald and {D'Onghia}, Elena},
  journal =       {\mnras},
  month =         oct,
  number =        {2},
  pages =         {2825-2842},
  title =         {{A holistic review of a galactic interaction}},
  volume =        {507},
  year =          {2021},
  doi =           {10.1093/mnras/stab2398},
}

@article{2021ApJ...919..109G,
  author =        {{Garavito-Camargo}, Nicol{\'a}s and {Besla}, Gurtina and
                   {Laporte}, Chervin F.~P. and
                   {Price-Whelan}, Adrian M. and {Cunningham}, Emily C. and
                   {Johnston}, Kathryn V. and {Weinberg}, Martin and
                   {G{\'o}mez}, Facundo A.},
  journal =       {\apj},
  month =         oct,
  number =        {2},
  pages =         {109},
  title =         {{Quantifying the Impact of the Large Magellanic Cloud
                   on the Structure of the Milky Way's Dark Matter Halo
                   Using Basis Function Expansions}},
  volume =        {919},
  year =          {2021},
  doi =           {10.3847/1538-4357/ac0b44},
  eid =           {109},
}

@inproceedings{tomasi1998bilateral,
  author =        {Tomasi, Carlo and Manduchi, Roberto},
  booktitle =     {Sixth international conference on computer vision
                   (IEEE Cat. No. 98CH36271)},
  organization =  {IEEE},
  pages =         {839--846},
  title =         {Bilateral filtering for gray and color images},
  year =          {1998},
}

@article{PIZER1987355,
  author =        {Stephen M. Pizer and E. Philip Amburn and
                   John D. Austin and Robert Cromartie and
                   Ari Geselowitz and Trey Greer and
                   Bart {ter Haar Romeny} and John B. Zimmerman and
                   Karel Zuiderveld},
  journal =       {Computer Vision, Graphics, and Image Processing},
  number =        {3},
  pages =         {355-368},
  title =         {Adaptive histogram equalization and its variations},
  volume =        {39},
  year =          {1987},
  doi =           {https://doi.org/10.1016/S0734-189X(87)80186-X},
  issn =          {0734-189X},
  url =           {https://www.sciencedirect.com/science/article/pii/
                   S0734189X8780186X},
}

@article{2025MNRAS.543.2159C,
  author =        {{Chiba}, Rimpei and {Frankel}, Neige and
                   {Hamilton}, Chris},
  journal =       {\mnras},
  month =         nov,
  number =        {3},
  pages =         {2159-2179},
  title =         {{Origin of the two-armed vertical phase spiral in the
                   inner Galactic disc}},
  volume =        {543},
  year =          {2025},
  doi =           {10.1093/mnras/staf1566},
}

@article{2004ITIP...13..600W,
  author =        {{Wang}, Z. and {Bovik}, A.~C. and {Sheikh}, H.~R. and
                   {Simoncelli}, E.~P.},
  journal =       {IEEE Transactions on Image Processing},
  month =         apr,
  number =        {4},
  pages =         {600-612},
  title =         {{Image Quality Assessment: From Error Visibility to
                   Structural Similarity}},
  volume =        {13},
  year =          {2004},
  doi =           {10.1109/TIP.2003.819861},
}

@article{2020NatMe..17..261V,
  author =        {{Virtanen}, Pauli and {Gommers}, Ralf and
                   {Oliphant}, Travis E. and {Haberland}, Matt and
                   {Reddy}, Tyler and {Cournapeau}, David and
                   {Burovski}, Evgeni and {Peterson}, Pearu and
                   {Weckesser}, Warren and {Bright}, Jonathan and
                   {van der Walt}, St{\'e}fan J. and {Brett}, Matthew and
                   {Wilson}, Joshua and {Millman}, K. Jarrod and
                   {Mayorov}, Nikolay and {Nelson}, Andrew R.~J. and
                   {Jones}, Eric and {Kern}, Robert and {Larson}, Eric and
                   {Carey}, C.~J. and {Polat}, {\.I}lhan and {Feng}, Yu and
                   {Moore}, Eric W. and {VanderPlas}, Jake and
                   {Laxalde}, Denis and {Perktold}, Josef and
                   {Cimrman}, Robert and {Henriksen}, Ian and
                   {Quintero}, E.~A. and {Harris}, Charles R. and
                   {Archibald}, Anne M. and {Ribeiro}, Ant{\^o}nio H. and
                   {Pedregosa}, Fabian and {van Mulbregt}, Paul and
                   {SciPy 1. 0 Contributors}},
  journal =       {Nature Medicine},
  month =         feb,
  pages =         {261-272},
  title =         {{SciPy 1.0: fundamental algorithms for scientific
                   computing in Python}},
  volume =        {17},
  year =          {2020},
  doi =           {10.1038/s41592-019-0686-2},
}

@article{2022A&A...668A..61A,
  author =        {{Antoja}, T. and {Ramos}, P. and
                   {L{\'o}pez-Guitart}, F. and {Anders}, F. and
                   {Bernet}, M. and {Laporte}, C.~F.~P.},
  journal =       {\aap},
  month =         dec,
  pages =         {A61},
  title =         {{Tidally induced spiral arm wraps encoded in phase
                   space}},
  volume =        {668},
  year =          {2022},
  doi =           {10.1051/0004-6361/202244064},
  eid =           {A61},
}

@article{2024MNRAS.52711393H,
  author =        {{Hunt}, Jason A.~S. and {Price-Whelan}, Adrian M. and
                   {Johnston}, Kathryn V. and {McClure}, Rachel L. and
                   {Filion}, Carrie and {Cassese}, Ben and
                   {Horta}, Danny},
  journal =       {\mnras},
  month =         feb,
  number =        {4},
  pages =         {11393-11403},
  title =         {{Radial phase spirals in the Solar neighbourhood}},
  volume =        {527},
  year =          {2024},
  doi =           {10.1093/mnras/stad3918},
}

@article{1966ApJ...146..810J,
  author =        {{Julian}, William H. and {Toomre}, Alar},
  journal =       {\apj},
  month =         dec,
  pages =         {810},
  title =         {{Non-Axisymmetric Responses of Differentially
                   Rotating Disks of Stars}},
  volume =        {146},
  year =          {1966},
  doi =           {10.1086/148957},
}

@inproceedings{1981seng.proc..111T,
  author =        {{Toomre}, A.},
  booktitle =     {Structure and Evolution of Normal Galaxies},
  editor =        {{Fall}, S.~M. and {Lynden-Bell}, D.},
  month =         jan,
  pages =         {111-136},
  title =         {{What amplifies the spirals}},
  year =          {1981},
}

@article{2020MNRAS.496..767B,
  author =        {{Binney}, James},
  journal =       {\mnras},
  month =         jul,
  number =        {1},
  pages =         {767-783},
  title =         {{The shearing sheet and swing amplification
                   revisited}},
  volume =        {496},
  year =          {2020},
  doi =           {10.1093/mnras/staa1485},
}

@book{2008gady.book.....B,
  author =        {{Binney}, James and {Tremaine}, Scott},
  title =         {{Galactic Dynamics: Second Edition}},
  year =          {2008},
}

@article{2013A&A...558A..33A,
  author =        {{Astropy Collaboration} and {Robitaille}, Thomas P. and
                   {Tollerud}, Erik J. and {Greenfield}, Perry and
                   {Droettboom}, Michael and {Bray}, Erik and
                   {Aldcroft}, Tom and {Davis}, Matt and
                   {Ginsburg}, Adam and {Price-Whelan}, Adrian M. and
                   {Kerzendorf}, Wolfgang E. and {Conley}, Alexander and
                   {Crighton}, Neil and {Barbary}, Kyle and
                   {Muna}, Demitri and {Ferguson}, Henry and
                   {Grollier}, Fr{\'e}d{\'e}ric and {Parikh}, Madhura M. and
                   {Nair}, Prasanth H. and {Unther}, Hans M. and
                   {Deil}, Christoph and {Woillez}, Julien and
                   {Conseil}, Simon and {Kramer}, Roban and
                   {Turner}, James E.~H. and {Singer}, Leo and
                   {Fox}, Ryan and {Weaver}, Benjamin A. and
                   {Zabalza}, Victor and {Edwards}, Zachary I. and
                   {Azalee Bostroem}, K. and {Burke}, D.~J. and
                   {Casey}, Andrew R. and {Crawford}, Steven M. and
                   {Dencheva}, Nadia and {Ely}, Justin and
                   {Jenness}, Tim and {Labrie}, Kathleen and
                   {Lim}, Pey Lian and {Pierfederici}, Francesco and
                   {Pontzen}, Andrew and {Ptak}, Andy and
                   {Refsdal}, Brian and {Servillat}, Mathieu and
                   {Streicher}, Ole},
  journal =       {\aap},
  month =         oct,
  pages =         {A33},
  title =         {{Astropy: A community Python package for astronomy}},
  volume =        {558},
  year =          {2013},
  doi =           {10.1051/0004-6361/201322068},
  eid =           {A33},
}

@article{2018AJ....156..123A,
  author =        {{Astropy Collaboration} and {Price-Whelan}, A.~M. and
                   {Sip{\H{o}}cz}, B.~M. and {G{\"u}nther}, H.~M. and
                   {Lim}, P.~L. and {Crawford}, S.~M. and {Conseil}, S. and
                   {Shupe}, D.~L. and {Craig}, M.~W. and {Dencheva}, N. and
                   {Ginsburg}, A. and {VanderPlas}, J.~T. and
                   {Bradley}, L.~D. and {P{\'e}rez-Su{\'a}rez}, D. and
                   {de Val-Borro}, M. and {Aldcroft}, T.~L. and
                   {Cruz}, K.~L. and {Robitaille}, T.~P. and
                   {Tollerud}, E.~J. and {Ardelean}, C. and {Babej}, T. and
                   {Bach}, Y.~P. and {Bachetti}, M. and {Bakanov}, A.~V. and
                   {Bamford}, S.~P. and {Barentsen}, G. and {Barmby}, P. and
                   {Baumbach}, A. and {Berry}, K.~L. and {Biscani}, F. and
                   {Boquien}, M. and {Bostroem}, K.~A. and
                   {Bouma}, L.~G. and {Brammer}, G.~B. and {Bray}, E.~M. and
                   {Breytenbach}, H. and {Buddelmeijer}, H. and
                   {Burke}, D.~J. and {Calderone}, G. and
                   {Cano Rodr{\'\i}guez}, J.~L. and {Cara}, M. and
                   {Cardoso}, J.~V.~M. and {Cheedella}, S. and
                   {Copin}, Y. and {Corrales}, L. and {Crichton}, D. and
                   {D'Avella}, D. and {Deil}, C. and {Depagne}, {\'E}. and
                   {Dietrich}, J.~P. and {Donath}, A. and
                   {Droettboom}, M. and {Earl}, N. and {Erben}, T. and
                   {Fabbro}, S. and {Ferreira}, L.~A. and {Finethy}, T. and
                   {Fox}, R.~T. and {Garrison}, L.~H. and
                   {Gibbons}, S.~L.~J. and {Goldstein}, D.~A. and
                   {Gommers}, R. and {Greco}, J.~P. and {Greenfield}, P. and
                   {Groener}, A.~M. and {Grollier}, F. and {Hagen}, A. and
                   {Hirst}, P. and {Homeier}, D. and {Horton}, A.~J. and
                   {Hosseinzadeh}, G. and {Hu}, L. and {Hunkeler}, J.~S. and
                   {Ivezi{\'c}}, {\v{Z}}. and {Jain}, A. and
                   {Jenness}, T. and {Kanarek}, G. and {Kendrew}, S. and
                   {Kern}, N.~S. and {Kerzendorf}, W.~E. and
                   {Khvalko}, A. and {King}, J. and {Kirkby}, D. and
                   {Kulkarni}, A.~M. and {Kumar}, A. and {Lee}, A. and
                   {Lenz}, D. and {Littlefair}, S.~P. and {Ma}, Z. and
                   {Macleod}, D.~M. and {Mastropietro}, M. and
                   {McCully}, C. and {Montagnac}, S. and {Morris}, B.~M. and
                   {Mueller}, M. and {Mumford}, S.~J. and {Muna}, D. and
                   {Murphy}, N.~A. and {Nelson}, S. and {Nguyen}, G.~H. and
                   {Ninan}, J.~P. and {N{\"o}the}, M. and {Ogaz}, S. and
                   {Oh}, S. and {Parejko}, J.~K. and {Parley}, N. and
                   {Pascual}, S. and {Patil}, R. and {Patil}, A.~A. and
                   {Plunkett}, A.~L. and {Prochaska}, J.~X. and
                   {Rastogi}, T. and {Reddy Janga}, V. and {Sabater}, J. and
                   {Sakurikar}, P. and {Seifert}, M. and
                   {Sherbert}, L.~E. and {Sherwood-Taylor}, H. and
                   {Shih}, A.~Y. and {Sick}, J. and {Silbiger}, M.~T. and
                   {Singanamalla}, S. and {Singer}, L.~P. and
                   {Sladen}, P.~H. and {Sooley}, K.~A. and
                   {Sornarajah}, S. and {Streicher}, O. and {Teuben}, P. and
                   {Thomas}, S.~W. and {Tremblay}, G.~R. and
                   {Turner}, J.~E.~H. and {Terr{\'o}n}, V. and
                   {van Kerkwijk}, M.~H. and {de la Vega}, A. and
                   {Watkins}, L.~L. and {Weaver}, B.~A. and
                   {Whitmore}, J.~B. and {Woillez}, J. and {Zabalza}, V. and
                   {Astropy Contributors}},
  journal =       {\aj},
  month =         sep,
  number =        {3},
  pages =         {123},
  title =         {{The Astropy Project: Building an Open-science
                   Project and Status of the v2.0 Core Package}},
  volume =        {156},
  year =          {2018},
  doi =           {10.3847/1538-3881/aabc4f},
  eid =           {123},
}

@article{2022ApJ...935..167A,
  author =        {{Astropy Collaboration} and {Price-Whelan}, Adrian M. and
                   {Lim}, Pey Lian and {Earl}, Nicholas and
                   {Starkman}, Nathaniel and {Bradley}, Larry and
                   {Shupe}, David L. and {Patil}, Aarya A. and
                   {Corrales}, Lia and {Brasseur}, C.~E. and
                   {N{\"o}the}, Maximilian and {Donath}, Axel and
                   {Tollerud}, Erik and {Morris}, Brett M. and
                   {Ginsburg}, Adam and {Vaher}, Eero and
                   {Weaver}, Benjamin A. and {Tocknell}, James and
                   {Jamieson}, William and {van Kerkwijk}, Marten H. and
                   {Robitaille}, Thomas P. and {Merry}, Bruce and
                   {Bachetti}, Matteo and {G{\"u}nther}, H. Moritz and
                   {Aldcroft}, Thomas L. and {Alvarado-Montes}, Jaime A. and
                   {Archibald}, Anne M. and {B{\'o}di}, Attila and
                   {Bapat}, Shreyas and {Barentsen}, Geert and
                   {Baz{\'a}n}, Juanjo and {Biswas}, Manish and
                   {Boquien}, M{\'e}d{\'e}ric and {Burke}, D.~J. and
                   {Cara}, Daria and {Cara}, Mihai and {Conroy}, Kyle E. and
                   {Conseil}, Simon and {Craig}, Matthew W. and
                   {Cross}, Robert M. and {Cruz}, Kelle L. and
                   {D'Eugenio}, Francesco and {Dencheva}, Nadia and
                   {Devillepoix}, Hadrien A.~R. and
                   {Dietrich}, J{\"o}rg P. and {Eigenbrot}, Arthur Davis and
                   {Erben}, Thomas and {Ferreira}, Leonardo and
                   {Foreman-Mackey}, Daniel and {Fox}, Ryan and
                   {Freij}, Nabil and {Garg}, Suyog and {Geda}, Robel and
                   {Glattly}, Lauren and {Gondhalekar}, Yash and
                   {Gordon}, Karl D. and {Grant}, David and
                   {Greenfield}, Perry and {Groener}, Austen M. and
                   {Guest}, Steve and {Gurovich}, Sebastian and
                   {Handberg}, Rasmus and {Hart}, Akeem and
                   {Hatfield-Dodds}, Zac and {Homeier}, Derek and
                   {Hosseinzadeh}, Griffin and {Jenness}, Tim and
                   {Jones}, Craig K. and {Joseph}, Prajwel and
                   {Kalmbach}, J. Bryce and {Karamehmetoglu}, Emir and
                   {Ka{\l}uszy{\'n}ski}, Miko{\l}aj and
                   {Kelley}, Michael S.~P. and {Kern}, Nicholas and
                   {Kerzendorf}, Wolfgang E. and {Koch}, Eric W. and
                   {Kulumani}, Shankar and {Lee}, Antony and {Ly}, Chun and
                   {Ma}, Zhiyuan and {MacBride}, Conor and
                   {Maljaars}, Jakob M. and {Muna}, Demitri and
                   {Murphy}, N.~A. and {Norman}, Henrik and
                   {O'Steen}, Richard and {Oman}, Kyle A. and
                   {Pacifici}, Camilla and {Pascual}, Sergio and
                   {Pascual-Granado}, J. and {Patil}, Rohit R. and
                   {Perren}, Gabriel I. and {Pickering}, Timothy E. and
                   {Rastogi}, Tanuj and {Roulston}, Benjamin R. and
                   {Ryan}, Daniel F. and {Rykoff}, Eli S. and
                   {Sabater}, Jose and {Sakurikar}, Parikshit and
                   {Salgado}, Jes{\'u}s and {Sanghi}, Aniket and
                   {Saunders}, Nicholas and {Savchenko}, Volodymyr and
                   {Schwardt}, Ludwig and {Seifert-Eckert}, Michael and
                   {Shih}, Albert Y. and {Jain}, Anany Shrey and
                   {Shukla}, Gyanendra and {Sick}, Jonathan and
                   {Simpson}, Chris and {Singanamalla}, Sudheesh and
                   {Singer}, Leo P. and {Singhal}, Jaladh and
                   {Sinha}, Manodeep and {Sip{\H{o}}cz}, Brigitta M. and
                   {Spitler}, Lee R. and {Stansby}, David and
                   {Streicher}, Ole and {{\v{S}}umak}, Jani and
                   {Swinbank}, John D. and {Taranu}, Dan S. and
                   {Tewary}, Nikita and {Tremblay}, Grant R. and
                   {de Val-Borro}, Miguel and {Van Kooten}, Samuel J. and
                   {Vasovi{\'c}}, Zlatan and {Verma}, Shresth and
                   {de Miranda Cardoso}, Jos{\'e} Vin{\'\i}cius and
                   {Williams}, Peter K.~G. and {Wilson}, Tom J. and
                   {Winkel}, Benjamin and {Wood-Vasey}, W.~M. and
                   {Xue}, Rui and {Yoachim}, Peter and {Zhang}, Chen and
                   {Zonca}, Andrea and {Astropy Project Contributors}},
  journal =       {\apj},
  month =         aug,
  number =        {2},
  pages =         {167},
  title =         {{The Astropy Project: Sustaining and Growing a
                   Community-oriented Open-source Project and the Latest
                   Major Release (v5.0) of the Core Package}},
  volume =        {935},
  year =          {2022},
  doi =           {10.3847/1538-4357/ac7c74},
  eid =           {167},
}

@article{2007CSE.....9c..21P,
  author =        {{Perez}, Fernando and {Granger}, Brian E.},
  journal =       {Computing in Science and Engineering},
  month =         jan,
  number =        {3},
  pages =         {21-29},
  title =         {{IPython: A System for Interactive Scientific
                   Computing}},
  volume =        {9},
  year =          {2007},
  doi =           {10.1109/MCSE.2007.53},
}

@incollection{2016ppap.book...87K,
  author =        {{Kluyver}, Thomas and {Ragan-Kelley}, Benjain and
                   {P{\'e}rez}, Fernando and {Granger}, Brian and
                   {Bussonnier}, Matthias and {Frederic}, Jonathan and
                   {Kelley}, Kyle and {Hamrick}, Jessica and
                   {Grout}, Jason and {Corlay}, Sylvain and
                   {Ivanov}, Paul and {Avila}, Dami{\'a}n and
                   {Abdalla}, Safia and {Willing}, Carol and
                   {Jupyter Development Team}},
  booktitle =     {IOS Press},
  pages =         {87-90},
  title =         {{Jupyter Notebooks{\textemdash}a publishing format
                   for reproducible computational workflows}},
  year =          {2016},
  doi =           {10.3233/978-1-61499-649-1-87},
}

@article{2007CSE.....9...90H,
  author =        {{Hunter}, John D.},
  journal =       {Computing in Science and Engineering},
  month =         jan,
  number =        {3},
  pages =         {90-95},
  title =         {{Matplotlib: A 2D Graphics Environment}},
  volume =        {9},
  year =          {2007},
  doi =           {10.1109/MCSE.2007.55},
}

@article{numpy,
  author =        {Charles R. Harris and K. Jarrod Millman and
                   St{\'{e}}fan J. van der Walt and Ralf Gommers and
                   Pauli Virtanen and David Cournapeau and Eric Wieser and
                   Julian Taylor and Sebastian Berg and
                   Nathaniel J. Smith and Robert Kern and Matti Picus and
                   Stephan Hoyer and Marten H. van Kerkwijk and
                   Matthew Brett and Allan Haldane and
                   Jaime Fern{\'{a}}ndez del R{\'{i}}o and Mark Wiebe and
                   Pearu Peterson and Pierre G{\'{e}}rard-Marchant and
                   Kevin Sheppard and Tyler Reddy and Warren Weckesser and
                   Hameer Abbasi and Christoph Gohlke and
                   Travis E. Oliphant},
  journal =       {Nature},
  month =         sep,
  number =        {7825},
  pages =         {357--362},
  publisher =     {Springer Science and Business Media {LLC}},
  title =         {Array programming with {NumPy}},
  volume =        {585},
  year =          {2020},
  doi =           {10.1038/s41586-020-2649-2},
  url =           {https://doi.org/10.1038/s41586-020-2649-2},
}

@inproceedings{mckinney-proc-scipy-2010,
  author =        {{W}es {M}c{K}inney},
  booktitle =     {{P}roceedings of the 9th {P}ython in {S}cience
                   {C}onference},
  editor =        {{S}t\'efan van der {W}alt and {J}arrod {M}illman},
  pages =         {56 - 61},
  title =         {{D}ata {S}tructures for {S}tatistical {C}omputing in
                   {P}ython},
  year =          {2010},
  doi =           {10.25080/Majora-92bf1922-00a},
}

@misc{pandas_10537285,
  author =        {The pandas development team},
  month =         jan,
  publisher =     {Zenodo},
  title =         {pandas-dev/pandas: Pandas},
  year =          {2024},
  doi =           {10.5281/zenodo.10537285},
  url =           {https://doi.org/10.5281/zenodo.10537285},
}

@misc{scipy_10155614,
  author =        {Ralf Gommers and Pauli Virtanen and Matt Haberland and
                   Evgeni Burovski and Warren Weckesser and Tyler Reddy and
                   Travis E. Oliphant and David Cournapeau and
                   Andrew Nelson and alexbrc and Pamphile Roy and
                   Pearu Peterson and Josh Wilson and Ilhan Polat and
                   endolith and Nikolay Mayorov and Stefan van der Walt and
                   Matthew Brett and Denis Laxalde and Eric Larson and
                   Jarrod Millman and Atsushi Sakai and Lars and
                   peterbell10 and CJ Carey and Paul van Mulbregt and
                   eric-jones and Nicholas McKibben and Robert Kern and
                   Kai},
  month =         nov,
  publisher =     {Zenodo},
  title =         {scipy/scipy: SciPy 1.11.4},
  year =          {2023},
  doi =           {10.5281/zenodo.10155614},
  url =           {https://doi.org/10.5281/zenodo.10155614},
}

@article{2024arXiv240604405W,
  author =        {{Wagg}, Tom and {Broekgaarden}, Floor S.},
  journal =       {arXiv e-prints},
  month =         jun,
  pages =         {arXiv:2406.04405},
  title =         {{Streamlining and standardizing software citations
                   with The Software Citation Station}},
  year =          {2024},
  doi =           {10.48550/arXiv.2406.04405},
  eid =           {arXiv:2406.04405},
}

@misc{software-citation-station-zenodo,
  author =        {Tom Wagg and Floor Broekgaarden and Kayhan Gültekin},
  month =         aug,
  publisher =     {Zenodo},
  title =         {TomWagg/software-citation-station: v1.2},
  year =          {2024},
  doi =           {10.5281/zenodo.13225824},
  url =           {https://doi.org/10.5281/zenodo.13225824},
}

@article{2023Galax..11...59V,
  author =        {{Vasiliev}, Eugene},
  journal =       {Galaxies},
  month =         apr,
  number =        {2},
  pages =         {59},
  title =         {{The Effect of the LMC on the Milky Way System}},
  volume =        {11},
  year =          {2023},
  doi =           {10.3390/galaxies11020059},
  eid =           {59},
}

@article{1983A&A...117....9M,
  author =        {{Mulder}, W.~A.},
  journal =       {\aap},
  month =         jan,
  number =        {1},
  pages =         {9-16},
  title =         {{Dynamical friction on extended objects}},
  volume =        {117},
  year =          {1983},
}

@article{1986ApJ...300...93W,
  author =        {{Weinberg}, M.~D.},
  journal =       {\apj},
  month =         jan,
  pages =         {93},
  title =         {{Orbital Decay of Satellite Galaxies in Spherical
                   Systems}},
  volume =        {300},
  year =          {1986},
  doi =           {10.1086/163785},
}

@article{1989MNRAS.239..549W,
  author =        {{Weinberg}, Martin D.},
  journal =       {\mnras},
  month =         aug,
  pages =         {549-569},
  title =         {{Self-gravitating response of a spherical galaxy to
                   sinking satellites}},
  volume =        {239},
  year =          {1989},
  doi =           {10.1093/mnras/239.2.549},
}

\begin{appendix}
\section{Potential models}\label{appendix:potential_models}
Here we provide details of the construction of the gravitational potential models.
They were derived from $N$-body snapshots using the \texttt{Agama} library \citep{2019MNRAS.482.1525V}.
Table ~\ref{table:potential_parameters} summarises the \texttt{Agama}'s potential expansion parameters adopted for each component.
For all parameters not listed, we used the default values.

For the static axisymmetric potentials (MW disc, MW bulge, and the DM halo without the wake), we applied the potential expansion to the snapshot at $t=-300$~Myr, which is sufficiently prior to the first pericentric passage of Sgr.
In contrast, for the MW halo model including the DM wake, we used potential expansions from all snapshots.
To account for temporal evolution, we employed the \texttt{Evolving} modifier, which linearly interpolates the potential between adjacent time stamps.
The gravitational potential of Sgr was constructed in the same way, but with the coordinate system shifted to the Sgr-centric frame.
The centre of Sgr was defined as the density peak of its stellar component.
When performing test-particle simulations, the Sgr potential was transformed back into the MW-centric frame and evolved in time using the \texttt{Shifted} and \texttt{Evolving} modifiers.

\begin{table}[h!]
        \caption{Parameters of the potential expansion.}\label{table:potential_parameters}
        \centering
        \begin{tabular}{l c}
                \hline\hline
                Parameter & Value \\
                \hline
                MW disc & \\
                \hline
                \texttt{type} & CylSpline \\
                \texttt{symmetry} & Axisymmetric \\
                \texttt{gridSizeR} & 20 \\
                \texttt{gridSizez} & 20 \\
                \texttt{Rmin}& 0.2 kpc \\
                \texttt{Rmax}& 100 kpc \\
                \texttt{zmin}& 0.05 kpc \\
                \texttt{zmax}& 20 kpc \\
                \hline
                MW bulge & \\
                \hline
                \texttt{type} & Multipole \\
                \texttt{symmetry} & Axisymmetric \\
                \texttt{lmax} & 10 \\
                \hline
                MW halo (without DM wake) & \\
                \hline
                \texttt{type} & Multipole \\
                \texttt{symmetry} & Axisymmetric \\
                \texttt{lmax} & 20 \\
                \texttt{rmax} & 200 kpc \\
                \hline
                MW halo (with DM wake) & \\
                \hline
                \texttt{type} & Multipole \\
                \texttt{lmax} & 20 \\
                \texttt{mmax} & 20 \\
                \texttt{rmax} & 200 kpc \\
                \hline
                Sgr & \\
                \hline
                \texttt{type} & Multipole \\
                \texttt{lmax} & 10 \\
                \texttt{mmax} & 10 \\
                \texttt{rmax} & 50 kpc \\
                \hline
        \end{tabular}
\end{table}

\begin{figure}[!ht]
        \begin{center}
                \includegraphics[width=\hsize]{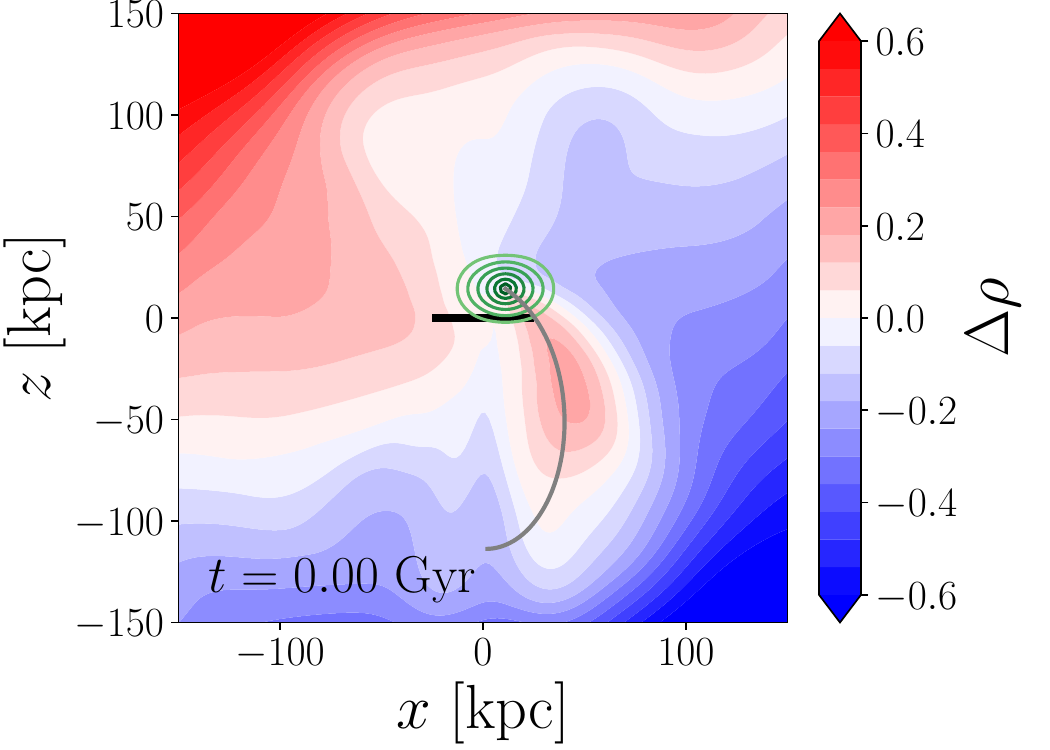}
                \caption{Milky Way halo density distribution reconstructed from the potential model. The background colour map indicates the density contrast defined by Eq.~\eqref{eq:density_contrast} at $t = 0$~Gyr.
                        The green contours represent the projected density distribution of Sgr, showing in logarithmic scale between $10^{-2}$ and $10^0$ of the maximum density. The horizontal black line and grey curve indicate the MW disc position and the orbital trajectory of Sgr, respectively.
                }\label{fig:halo_wake_092}
        \end{center}
\end{figure}
\begin{figure}[!ht]
        \begin{center}
                \includegraphics[width=\hsize]{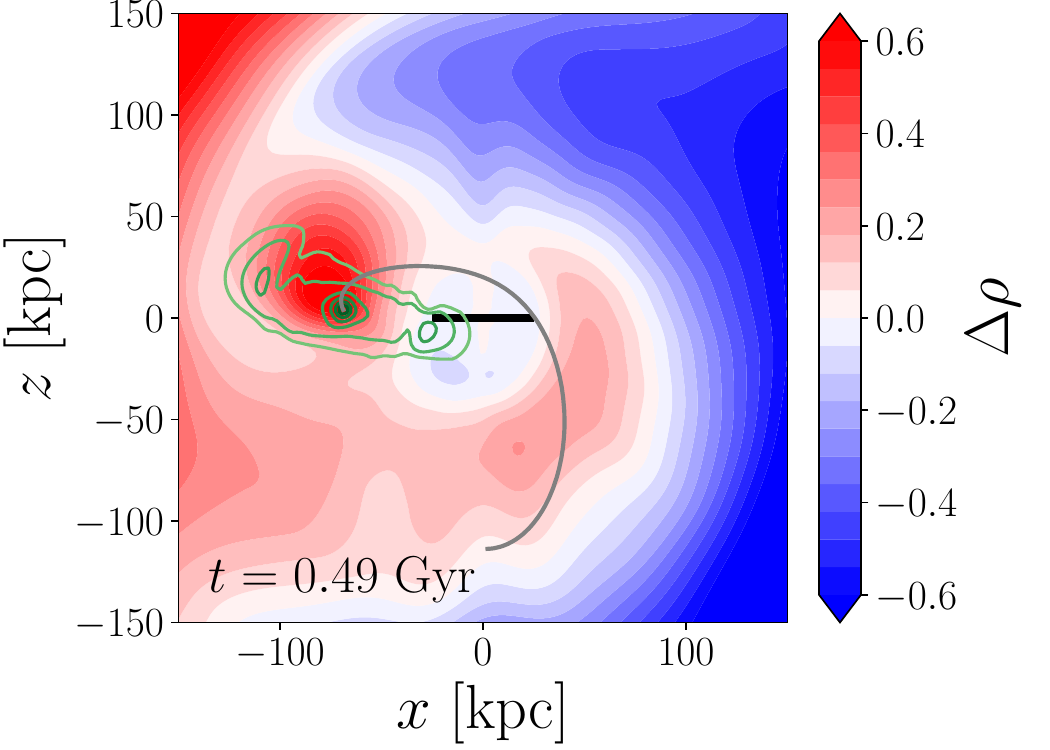}
                \caption{Same as Fig.~\ref{fig:halo_wake_092} but at $t=0.49$~Gyr.
                }\label{fig:halo_wake_142}
        \end{center}
\end{figure}
To verify that the potential model adequately captures halo substructures, we reconstructed the density field of the MW halo from the potential expansion and computed the density contrast,
\begin{align}
        \Delta \rho(x, z) = \frac{\int_{|y| <20\,\mathrm{kpc}}\rho(x, y, z)}{\int_{|y| <20\,\mathrm{kpc}}\rho_0(x, y, z)}-1, \label{eq:density_contrast}
\end{align}
where $\rho$ and $\rho_0$ are the densities of the MW halo models with and without the DM wake, respectively.
The integration over $y$ from $-20$~kpc to 20~kpc was chosen because the orbital plane of Sgr is approximately aligned with the $x$--$z$ plane.
Figure~\ref{fig:halo_wake_092} shows the projected density contrast at $t=0$~Gyr, together with the projected density distribution of Sgr, also reconstructed from the potential expansion.
The overall morphology of the density contrast is consistent with the expected response of a DM halo to a massive satellite.
Similar features have been reported in previous studies of the MW halo response to the Large Magellanic Cloud (LMC) (see \citealt{2023Galax..11...59V} and references therein).
In particular, \citet{2021ApJ...919..109G} presented a comparable density map based on their $N$-body simulations of the MW-LMC interaction.
In the figure, two characteristic features can be clearly identified. The first is an overdense region trailing the orbit of Sgr, which corresponds to a transient wake induced by dynamical friction \citep{1983A&A...117....9M, 1986ApJ...300...93W}.
The second is a large-scale dipole density pattern, characterised by an overdensity in the upper left and an underdensity in the lower right of the panel.
This dipole arises from the reflex motion of the MW halo and can persist longer than the transient wake, as it is sustained by resonant effects and self-gravity \citep{1989MNRAS.239..549W}.

In Fig.~\ref{fig:halo_wake_142} we also present the same plot at $t=0.49$~Gyr, when Sgr is close to its apocentre.
We can identify the dynamical friction wake at $(x, z)\sim (-100, 50)$~kpc inside the collective response overdensity formed during the first infall.
At this time, the centre of the dynamical friction wake is distant from the disc and its dynamical impact on the disc is considered to be relatively weaker than that of the collective response.

\section{Phase spiral fitting in the test particle model}\label{appendix:phase_spiral_fitting_tp}
\begin{figure*}
        \begin{center}
                \includegraphics[width=0.85\hsize]{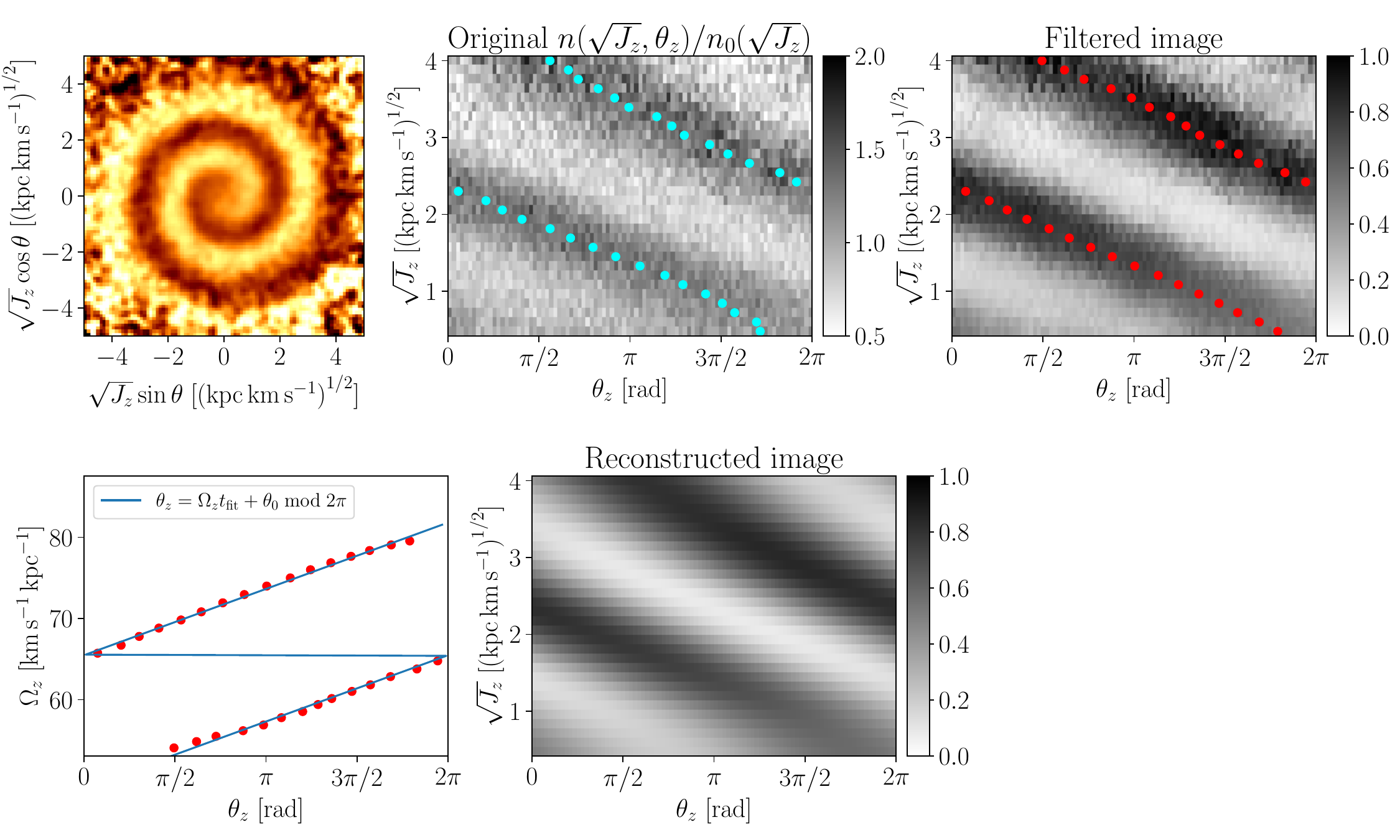}
                \caption{
                        Phase spiral fitting example in the TP (static) model. Figure format is the same as Fig.~\ref{fig:fitting_example}.
                }\label{fig:fitting_example_static}
        \end{center}
\end{figure*}

We demonstrate that our phase spiral fitting method (Sect.~\ref{subsec:unwinding}) works accurately in ideal cases.
We selected a particle sample from a bin with $R_g = 7\text{--}7.5$~kpc and $\theta_\phi = 150 \text{--}160^\circ$ in the snapshot at $t = 0.39$~Gyr from the TP (static) model, where the phase spiral is clearly visible.
Figure~\ref{fig:fitting_example_static} illustrates the fitting procedure for this sample, following the same format as Fig.~\ref{fig:fitting_example}.
In this case, the $k = 1$ Fourier mode of the original $\sqrt{J_z}$--$\theta_z$ distribution (top middle panel) already captures the diagonal ridge structure well, while the filtering further suppresses noise (top right panel).
In the $\Omega_z$--$\theta_z$ space (bottom left panel), the extracted $k = 1$ mode forms straight lines.
The winding time inferred from the fitting is $t_\mathrm{fit} = 0.38$~Gyr, in good agreement with the true elapsed time $t = 0.39$~Gyr.

\FloatBarrier

\section{Supplementary figures from the test particle models}\label{appendix:supplementary_figs}
\begin{figure}[!ht]
        \begin{center}
                \includegraphics[width=\hsize]{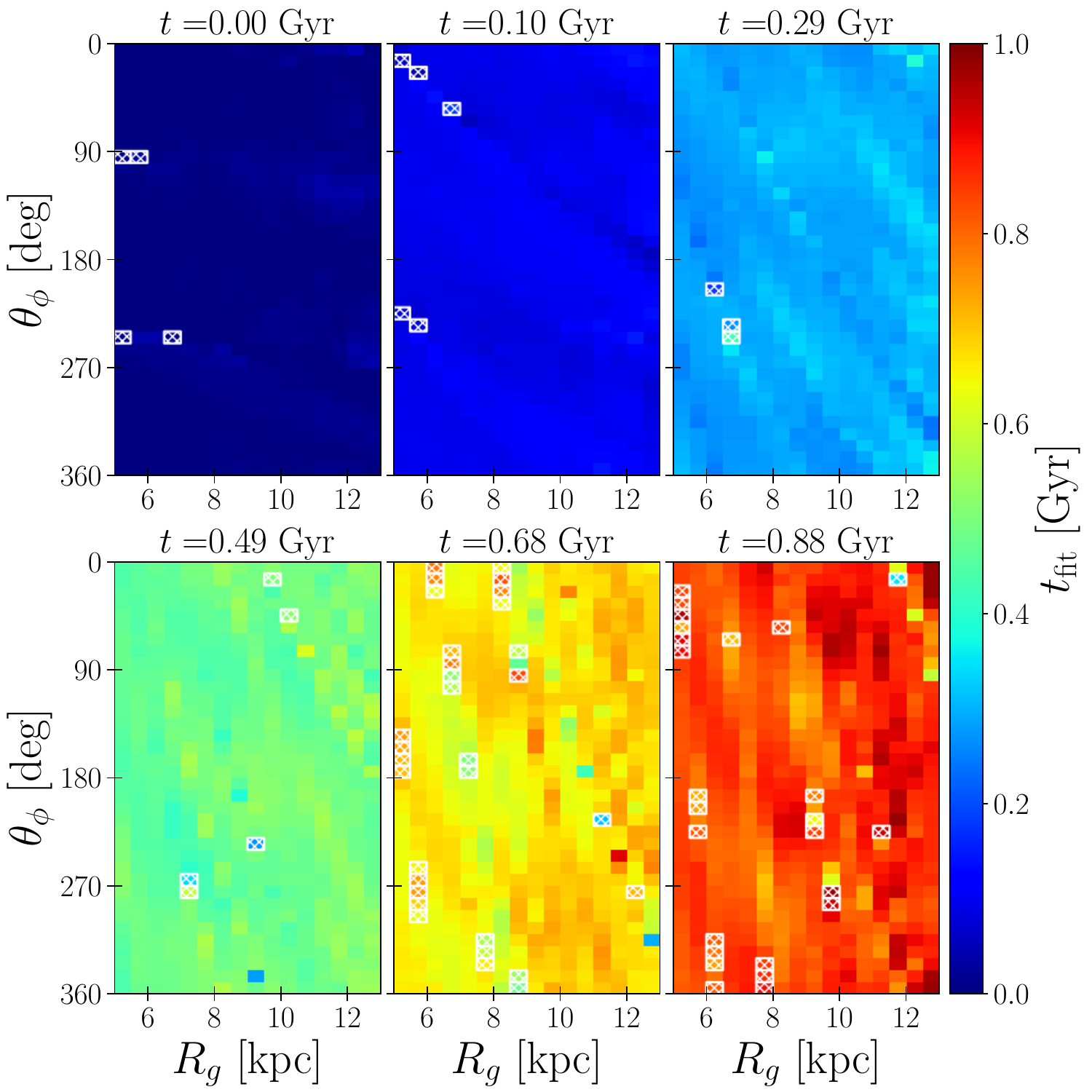}
                \caption{Same as Fig.~\ref{fig:Rg_theta_tp_sgr_wake} but for the TP (static) model.}\label{fig:Rg_theta_tp_static}
        \end{center}
\end{figure}

\begin{figure}
        \begin{center}
                \includegraphics[width=\hsize]{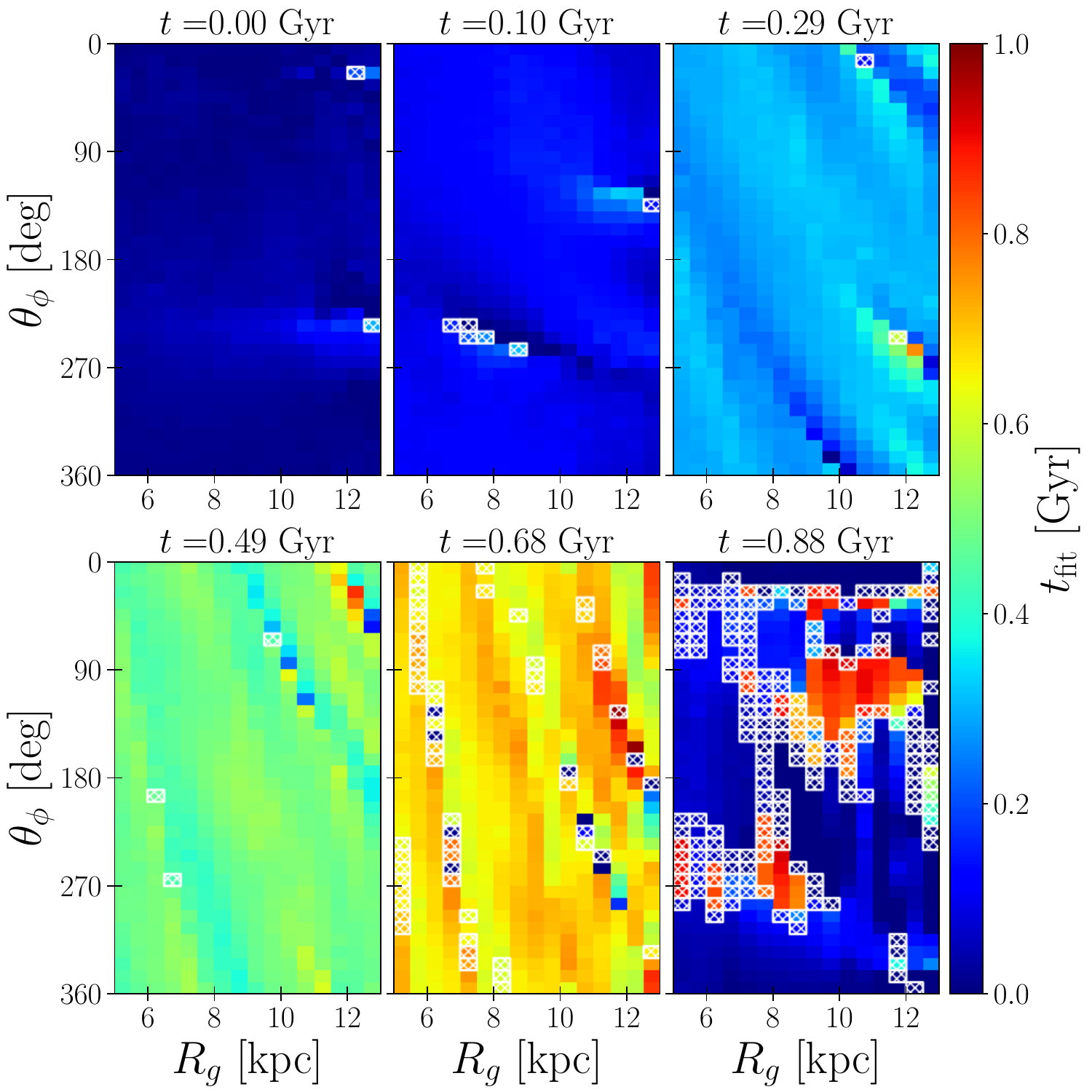}
                \caption{Same as Fig.~\ref{fig:Rg_theta_tp_sgr_wake} but for the TP (Sgr) model.}\label{fig:Rg_theta_tp_sgr}
        \end{center}
\end{figure}

\begin{figure}
        \begin{center}
                \includegraphics[width=\hsize]{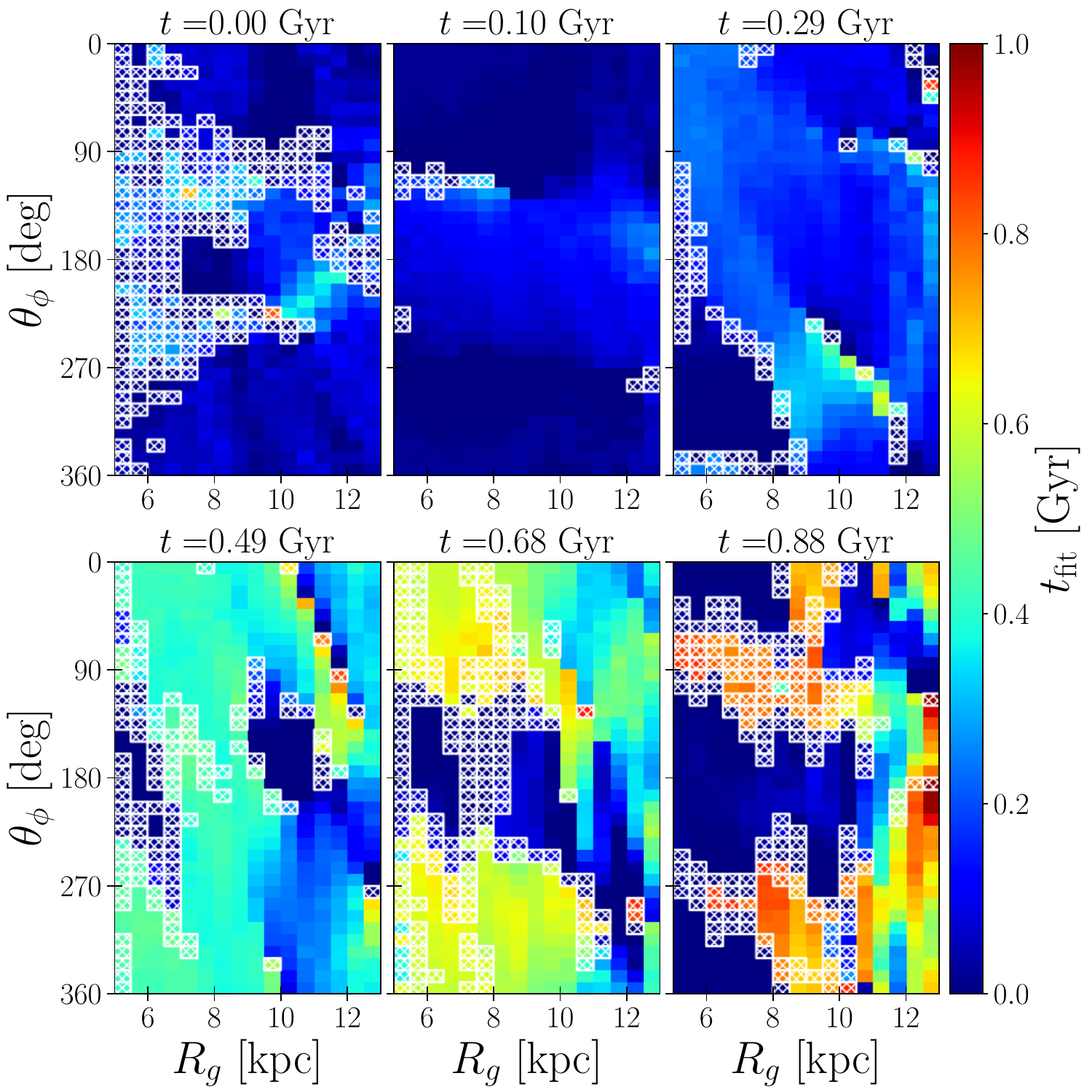}
                \caption{Same as Fig.~\ref{fig:Rg_theta_tp_sgr_wake} but for the TP (wake) model.}\label{fig:Rg_theta_tp_wake}
        \end{center}
\end{figure}

Figures~\ref{fig:Rg_theta_tp_static}, \ref{fig:Rg_theta_tp_sgr}, and \ref{fig:Rg_theta_tp_wake} show the winding time of the phase spiral in the TP (static), TP (Sgr), TP (wake), and TP (Sgr+wake) models in the same manner as in Fig.~\ref{fig:Rg_theta_tp_sgr_wake}.

\FloatBarrier

\section{Shearing box model}\label{appendix:analytic_model}
The analytical framework developed by \citet{2023MNRAS.522..477W} is based on linear perturbation theory within the shearing box approximation. This extends the classical shearing sheet formalism \citep{1966ApJ...146..810J, 2020MNRAS.496..767B} by incorporating vertical structure and self-gravity.

The shearing box is a local Cartesian coordinate system, $\bm{x}=(x, y, z)$, centred on a reference point at $R=R_0$ and rotating with angular velocity $\Omega$, where $x$ denotes the radial direction, $y$ the azimuthal direction, and $z$ the vertical direction. The model assumes that the distribution function (DF) and gravitational potential can be decomposed into unperturbed and perturbed components:
\begin{align}
f = f_0 + f_1, \quad \Phi = \Phi_0 + \Phi_1,
\end{align}
where $f_0$ and $\Phi_0$ represent the time-independent unperturbed DF and potential, and $f_1$ and $\Phi_1$ are linear perturbations. The perturbed potential $\Phi_1$ includes contributions from both external disturbances and the disc's self-gravitating response.

Assuming epicyclic motion, the unperturbed Hamiltonian is expressed in separable form, $H(\bm{x}, \bm{p}) = H_x + H_z$, where $\bm{p} = (p_x, p_y, p_z)$ is the momentum vector. The in-plane and vertical components of the Hamiltonian are defined as
\begin{align}
H_x &= \frac{1}{2} p_x^2 + \kappa^2 (x - \bar{x})^2, \\
H_z &= \frac{1}{2} p_z^2 + \Phi_z(z),
\end{align}
where $\bar{x} = 2\Omega(p_y - R_0\Omega)/\kappa^2$ is the guiding centre in the shearing box, $\kappa$ is the epicyclic frequency, and $\Phi_z(z)$ is the vertical potential.

We adopt an unperturbed DF in which the in-plane and vertical components follow a Maxwell--Boltzmann and a lowered isothermal form, respectively:
\begin{align}
f_0(\bm{x}, \bm{p}) &= \frac{\Omega \Sigma_0}{(2\pi)^{3/2} \kappa z_0 \sigma_x^2 \sigma_z} e^{-H_x/\sigma_x^2} F_z(H_z), \\
F_z(H_z) &=
\begin{cases}
N_z \left( e^{-H_z/\sigma_z^2} - e^{-E_z/\sigma_z^2} \right), & \text{for } 0 < H_z < E_z, \\
0, & \text{for } H_z \geq E_z,
\end{cases} \label{eq:F_z}
\end{align}
where $\Sigma_0$ is the surface density, $\sigma_x$ and $\sigma_z$ are the in-plane and vertical velocity dispersions, and $z_0 = \sigma_z^2 / \pi G \Sigma_0$ represents the thickness of the disc.
The vertical density profile can be obtained by integrating Eq.~\eqref{eq:F_z} over $p_z$, and the corresponding potential $\Phi_z(z)$ is obtained by solving Poisson’s equation.

The time evolution of the perturbed DF, $f_1$, is governed by the linearised collisionless Boltzmann equation:
\begin{align}
\frac{df_1}{dt} = \frac{\partial f_1}{\partial t} + \left[ f_1, H \right] = \left[ \Phi_1, f_0 \right],
\end{align}
where the square brackets denote Poisson brackets. We integrate this equation along unperturbed orbits. By decomposing $f_1$ into in-plane and vertical responses, $J_p$ and $J_z$, we obtain
\begin{align}
f_1(\bm{x}, \bm{p}, t) &= J_p(\bm{x}, \bm{p}, t) + J_z(\bm{x}, \bm{p}, t), \\
J_p &= \int_{t_i}^{t} dt'\, \frac{\partial \Phi_1}{\partial \bm{x}_p} \cdot \frac{\partial f_0}{\partial \bm{p}_p}, \\
J_z &= \int_{t_i}^{t} dt'\, \frac{\partial \Phi_1}{\partial z} \frac{\partial f_0}{\partial p_z},
\end{align}
where $\bm{x}_p$ and $\bm{p}_p$ are the in-plane position and momentum vectors, and $t_i$ is the time when the perturbation is initially applied.
Assuming a plane-wave perturbation, the potential takes the form:
\begin{align}
\Phi_1(\bm{x}, t) = e^{i\bm{k}_p(t) \cdot \bm{x}_p} \widetilde{\Phi}_1(z, t),
\end{align}
where $\widetilde{\Phi}_1(z, t)$ is the vertical component of the potential and $\bm{k}_p(t) = (k_x, k_y)$ is the in-plane wavevector. 
The phase $\bm{k}_p \cdot \bm{x}_p$ is constant along circular orbits, and if the shearing box is centred at the wave's corotation radius, then
\begin{align}
k_x(t) = 2 A k_y t,
\end{align}
where $A$ is Oort's first constant. We assume that the wavevector is aligned with the azimuthal direction at $t = 0$. The perturbed potential is related to the perturbed density $e^{i\bm{k}_p \cdot \bm{x}_p} \widetilde{\rho}_1(z, t)$ via a Green's function solution of Poisson's equation:
\begin{align}
\widetilde{\Phi}_1(z, t) = -\frac{2\pi G}{k_p} P(z, t), \quad
P(z, t) = \int_{-\infty}^{\infty} \widetilde{\rho}_1(\zeta, t) e^{-k_p |z - \zeta|} d\zeta,
\end{align}
and the vertical force is given by
\begin{align}
\frac{\partial \widetilde{\Phi}_1}{\partial z} = 2\pi G Q(z, t), \quad
Q(z, t) = \int_{-\infty}^{\infty} \widetilde{\rho}_1(\zeta, t) e^{-k_p|z - \zeta|} \mathrm{sgn}(z - \zeta) d\zeta.
\end{align}

The vertical DF is defined by integrating $f_1 = e^{i \bm{k}_p \cdot \bm{x}_p} \widetilde{f}_1$ over $\bm{p}_p$:
\begin{align}
        f_{1z}(x, y, z, p_z, t) = e^{i\bm{k}_p \cdot \bm{x}_p} \int d^2 \bm{p}_p \, \widetilde{f}_1
= e^{i\bm{k}_p \cdot \bm{x}_p} (\widetilde{\mathcal{J}}_p + \widetilde{\mathcal{J}}_z).
\end{align}
Using variable changes and Gaussian integrals (see Sect.~3 of \citealt{2020MNRAS.496..767B}), we obtained a Volterra integral equation,
\begin{align}
\widetilde{\mathcal{J}}_p(z, p_z, t) = \frac{\kappa F_z(H_z)}{\sqrt{8\pi} z_0 \sigma_z} \int_{t_i}^{t} dt' \, K_p(t, t') P(z', t'),
\label{eq:volterra_Jp}
\end{align}
where the kernel $K_p(t, t')$ is given by
\begin{align}
K_p(t, t') = \frac{4 \bm{c} \cdot \hat{\bm{b}} \, \exp(-0.572 Q_T^2 \hat{b}^2)}{\sqrt{1 + 4 A^2 t'^2}},
\end{align}
with the Toomre parameter $Q_T$, and
\begin{align}
\hat{b}_x &= \frac{k_y}{k_\mathrm{crit}} \left[ A(t' \sin \kappa t' - t \sin \kappa t) + \frac{\Omega}{\kappa} (\cos \kappa t' - \cos \kappa t) \right], \\
\hat{b}_y &= \frac{k_y}{k_\mathrm{crit}} \left[ A(t' \cos \kappa t' - t \cos \kappa t) - \frac{\Omega}{\kappa} (\sin \kappa t' - \sin \kappa t) \right],
\end{align}
where $k_\mathrm{crit} = \kappa^2 / (2\pi G \Sigma_0)$. The vector $\bm{c}$ is defined as
\begin{align}
c_x &= -A t' \cos \kappa t' + \frac{\Omega}{\kappa} \sin \kappa t', \\
c_y &= A t' \sin \kappa t' + \frac{\Omega}{\kappa} \cos \kappa t'.
\end{align}
The vertical response is given by
\begin{align}
\widetilde{\mathcal{J}}_z(z, p_z, t) = -\frac{N_z e^{-H_z/\sigma_z^2}}{\sqrt{2\pi} z_0^2 \sigma_z} \int_{t_i}^{t} dt'\, p_z' K_z(t, t') Q(z', t'),
\label{eq:volterra_Jz}
\end{align}
where the vertical kernel is
\begin{align}
K_z(t, t') = \exp\left(-0.572 Q_T^2 \hat{b}^2\right).
\end{align}

Up to this point, we have not assumed a specific vertical form for the external perturbation.
If the perturbation kicks the north and south parts of the galactic disc in the same (opposite) direction, a bending (breathing) mode is excited.
Since this study focuses on the one-arm phase spiral, in Sect.~\ref{subsec:analytic_model} we used the following antisymmetric external density perturbation following Sect.~3.2 of \citet{2023MNRAS.522..477W}:
\begin{align}
\tilde{\rho}_e(z, t) = \frac{\Sigma_e}{\kappa \Delta^2} \delta(t) z \exp\left(-\frac{z^2}{\Delta^2}\right),
\end{align}
where $\Sigma_e$ and $\Delta$ correspond to the strength and vertical scale length of the perturbation, respectively.
In the linear approximation, the winding rate of the phase spiral is independent of the perturbation strength $\Sigma_e$, which only affects the amplitude of the response.
        In the kinematic (non-self-gravitating) case, the perturbed density is $\tilde{\rho}_1 = \tilde{\rho}_e$, while in the self-gravitating case, $\tilde{\rho}_1$ consists of $\tilde{\rho}_e$ and the self-gravitating term, $\int dp_z (\widetilde{\mathcal{J}_p} + \widetilde{\mathcal{J}}_z)$.

We numerically solve the Volterra integral equations Eqs.~\eqref{eq:volterra_Jp} and \eqref{eq:volterra_Jz}.
We adopt the parameters listed in Table~\ref{table:shearing_box_parameters}, which are chosen to match the local properties of our $N$-body model at $R_g = 8.25$~kpc.

\begin{table}[h!]
        \caption{Free parameters of the shearing box model.}\label{table:shearing_box_parameters}
        \centering
        \begin{tabular}{l c}
                \hline\hline
                Parameter & Value \\
                \hline
                Surface density, $\Sigma_0$ & $66\,M_\odot\,\mathrm{pc}^{-2}$ \\
                In-plane velocity dispersion, $\sigma_x$ & $38\,\mathrm{km\,s^{-1}}$ \\
                Vertical velocity dispersion, $\sigma_z$ & $15\,\mathrm{km\,s^{-1}}$ \\
                Circular frequency, $\Omega$ & $29\,\mathrm{km\,s^{-1}\,kpc^{-1}}$ \\
                Epicyclic frequency, $\kappa$ & $44\,\mathrm{km\,s^{-1}\,kpc^{-1}}$ \\
                Azimuthal wavenumber, $k_y$ & $1/(8.25\,\mathrm{kpc})$ \\
                Upper energy cutoff, $E_z$ & $5\sigma_z^2$ \\
                Perturbation scale length, $\Delta$ & $3z_0$ \\
                \hline
        \end{tabular}
\end{table}

\end{appendix}

\end{document}